\documentclass[num-refs]{wiley-article}

\usepackage{siunitx}

\usepackage{tikz}
\usepackage{mathtools}
\usepackage{pgfplots}
\usetikzlibrary{shapes,arrows}
\usetikzlibrary{decorations.pathreplacing,bending,calc,shadows,angles, arrows.meta, quotes,shapes.geometric,decorations.text,hobby}
\usepackage{subcaption}
\usepackage{caption}
\usepackage{booktabs}

\DeclareMathAlphabet{\pazocal}{OMS}{zplm}{m}{n}

\newcommand{\Pb}{\pazocal{P}}
\DeclareMathOperator{\tr}{tr}

\papertype{Original Article}
\paperfield{Journal Section}

\title{Free surface flows using inverse smoothed particle hydrodynamics}

\author[1\authfn{1}]{Kalale Chola PhD}
\author[2\authfn{2}]{Tsumoru Shintake PhD}

\contrib[\authfn{1}]{Equally contributing authors.}

\affil[1]{Quantum wave spectroscopy and ocean power, Okinawa Institute of Science and Technology Graduate University, Tancha 1919-1, Onna-son, Okinawa, 904-0495, Japan}
\affil[2]{Department, Institution, City, State or Province, Postal Code, Country}

\corraddress{Quantum wave spectroscopy and ocean power, Okinawa Institute of Science and Technology Graduate University, Tancha 1919-1, Onna-son, Okinawa, 904-0495, Japan}
\corremail{kalale.chola@oist.jp}

\presentadd[\authfn{2}]{Quantum wave spectroscopy and ocean power, Okinawa Institute of Science and Technology Graduate University, Tancha 1919-1, Onna-son, Okinawa, 904-0495, Japan}

\runningauthor{Kalale Chola et al.}

\begin{document}

\maketitle

\begin{abstract}
Free surface flow problems including mixing processes in dam break flows and wave breaking phenomena are characterized by large deformation of the free surface. As such, they cannot be easily investigated by analytical and numerical mesh based models. In this paper a new method called un-smoothed particle hydrodynamics (SPH-$i$) solver is tested to study how well it can capture wave breaking and mixing of water bodies in 2D dam-break flows over wet beds. The model captures the breaking and mixing processes relatively well when compared against experimental results. Therefore, it can be inferred that the capability of the proposed model to predict the development and evolution of breaking waves as well as the interface at the water-water interface in dam-break mixing processes has been demonstrated. 

\keywords{ particle method, meshfree, free surface, partition-of-unity, hydrodynamics, large eddy simulation, turbulence models}
\end{abstract}

\section{Introduction}
Simulation of free surface flow problems with large deformation remains a challenging task in computational fluid dynamics (CFD). This is particularly difficult for mesh based models since a fixed mesh inadequate, adaptive mesh refinement becomes a necessity. Furthermore, it is difficult to develop highly efficient free-surface tracking algorithms. A method that overcomes this problems is called smoothed particle hydrodynamics (SPH). The method was independently developed by Lucy \cite{Lucy1977} and Gingold and Monaghan \cite{Gingold1977} for astrophysical problems. The method was later extended to study continuum fluid and solid mechanics by Monaghan \cite{Monaghan1992,Monaghan1994}. Since then several versions of the SPH method have been proposed \cite{Monaghan2005,Price2012} to address aspects of the method ranging from stability, accuracy \cite{Das2015}, consistency \cite{LIU2006} and convergence \cite{Dehnen2012}.

Recent research has shown that SPH has tremendous potential in the simulation of multiphase flows and fluid-structure interaction problems \cite{Dalrymple2006,Violeau2007,Hu2006,HU2007}. As the method matures into a standard CFD approach, several studies have demonstrated that SPH can handle free-surface, multi-phase, fluid structure and problems with moving boundaries with relatively good accuracy.

A particularly interesting benchmark problem for investigating rapid and interfacial flows is the dam-break flow over dry and wet beds. This is a complex flow phenomena that involves the formation of shock waves due to step changes at the initial water level. Owing to its Lagrangian nature, the SPH method can handle this problem in both two-dimensional and three-dimensional configurations. Dam-break flows over dry beds \cite{Monaghan1994} and their interaction with structures were successfully demonstrated \cite{Colagrossi2010,Gomez2004,Gomez2010}. The mixing process arising from the dam-break over wet beds is a much more complicated process that involves the formation of breaking waves and has been studied both experimentally and in \cite{Crespo2008}. Understanding the mixing process has both theoretical and practical significance in water and environmental engineering in predicting the the impact of floods on the ecosystem and structures.
The SPH method has further been successfully applied to the simulation of breaking water waves and wave-body interaction problems \cite{Gomez2004,Dalrymple2006,KHAYYER2008,KHAYYER2009,REN2014}.

In this paper numerical modeling of dam-break flow over dry and wet beds and breaking waves will be performed. Emphasis is placed on the capability of the proposed SPH-$i$ \cite{2018arXiv180711244C,2018arXiv180711728C} to capture the mixing process, along with the pre- and post-wave breaking phenomena. Unlike mesh methods, due to its Lagrangian nature, it is a robust method as no free-surface tracking for imposing the dynamic free surface boundary condition is required. Thus the evolution of the interface and mixing processes can be easily handled.    


\section{The SPH-$i$ Model}
We have proposed an SPH model called SPH-$i$ that implicitly models turbulence and is a high order model. For the reader, full details about model development and theory can be found in \cite{2018arXiv180711244C,2018arXiv180711728C}. Unlike standard SPH models which involves only a filtering or convolution process, the SPH-$i$ models inolves both a filter/convolution process and a de-filtering/deconvolution process. 

To study hydrodynamics of a system, we generally start with the Lagrangian form of the compressible Navier-Stokes equations (CNSEs) for a continuum. 
\begin{align}
\frac{d\rho}{dt} &= -\rho\bm{\nabla}\cdot\mathbf{u} \label{eq2016:1}\\
\rho\frac{d\mathbf{u}}{dt}&= -\bm{\nabla} p + \bm{\nabla}\cdot\underline{\underline{\sigma}} + \rho\mathbf{b} \label{eq2016:2}\\ 
\kappa_{s}\frac{dp}{dt} &= -\bm{\nabla}\cdot\mathbf{u} + \gamma\alpha\bm{\nabla}\cdot(\kappa_{s}\bm{\nabla} p) -\alpha\bm{\nabla}\cdot\left(\frac{1}{\rho}\bm{\nabla}\rho\right)+\frac{\alpha\beta}{k}\Phi\label{eq2016:2aa}
\end{align}
with fluid velocity $\mathbf{u}$, fluid mass-density $\rho$ , fluid pressure $p$, viscous stress tensor $\underline{\underline{\sigma}}$, viscous dissipation $\Phi =\underline{\underline{\sigma}}:\bm{\nabla}\mathbf{u}$, thermal diffusivity $\alpha$, volumetric thermal expansivity $\beta$, thermal conductivity $k$, adiabatic index $\gamma=c_p/c_v=K_{S}/K_{T}$, adiabatic incompressibility modulus $K_{S}$, isothermal incompressible modulus $K_{T}$, specific heat capacity at constant pressure $c_{p}$, specific heat capacity at constant volume $c_{v}$. For water and air, $\gamma$ is 7.0 and 1.4 respectively. We further assume that the adiabatic incompressibility modulus varies linearly with pressure so that 
\begin{align}
K_{S}= K_{S,0} + p\equiv\frac{1}{\kappa_{s}}
\end{align}
For an ideal gas, it then follows that $K_{S}$ is given by,
\begin{align}
K_{S}= \gamma K_{T}=-\gamma v\left(\frac{\partial p}{\partial v}\right)_{T}=\gamma p
\end{align}
meaning that for an ideal gas, $K_{S,0}=0$. For gases and liquids in general, we have that the incompressibility modulus under standard conditions $K_{S,0}$ is related to the standard speed of sound $c_0$ as $K_{S,0} = \rho_{0}c_{0}^{2}$.

For a Newtonian fluid, the Cauchy stress tensor $\underline{\underline{\tau}}$ can be expressed in terms of a deviatoric stress and a normal stress as
\begin{eqnarray}
\underline{\underline{\tau}}&= -p\underline{\underline{1}}+ \underline{\underline{\sigma}}=-p\underline{\underline{1}}+\nu\rho (\bm{\nabla}\mathbf{u}+\bm{\nabla}\mathbf{u}^{\textup{T}}-\frac{2}{d}\bm{\nabla}\cdot\mathbf{u}\underline{\underline{1}})\label{eqn:201619}
\end{eqnarray}
where $\nu$ is the kinematic viscosity and $d$ is the space dimension. Note that the Stokes hypothesis for a zero bulk viscosity has been assumed for simplicity.
The PDEs (\ref{eq2016:1}), (\ref{eq2016:2}) (\ref{eq2016:2aa}) are valid over a continuum where the turbulent field $\{\rho,p,\mathbf{u}\}$ is assumed to be sufficiently smooth and continuous.

\subsection{Filtering Process}
The convolution or filtering problem can be stated formally as:
Given the continuum field\{$\rho(\mathbf{r})$, $p(\mathbf{r})$, $\mathbf{u}(\mathbf{r})$\} defined on a domain $\Omega$, compute local approximations \{$\langle\rho_{h}(\mathbf{r})\rangle$, $\langle p_{h}(\mathbf{r})\rangle$,$\widetilde{\mathbf{u}}_{h}(\mathbf{r})$\} which faithfully represent the behavior of the continuum field on scales above some, user defined, filter length (here denoted $h$) and which truncates scales smaller than $\mathcal{O}(h)$. 

The filtering procedure is chosen so as to derive a filtered form of the compressible Navier-Stokes equations (CNSEs) that are consistent with the explicit LES model. This is defined as the filtering integral transform (FIT) and its application to the CNSEs is discussed in \cite{2018arXiv180711244C}.  
	
Let $\Omega_{h}(\mathbf{r})$ be a locally compact space within the fluid domain $\Omega$. Then the filtered mass density, momentum density and pressure are given by the FIT; for each $w_{h}\in C^{\infty}_{c}(\Omega_{h})$
\begin{align}
\langle\rho_{h}(\mathbf{r})\rangle &=\int_{\Omega_{h}(\mathbf{r})}\rho(\mathbf{r}^{\prime})w_{h}(\mathbf{r}-\mathbf{r}^{\prime})d^{\nu}\mathbf{r}^{\prime}\label{deq:l1}\\
\langle\rho_{h}(\mathbf{r})\rangle\widetilde{\mathbf{u}}_{h}(\mathbf{r}) &=\int_{\Omega_{h}(\mathbf{r})}\rho(\mathbf{r}^{\prime})\mathbf{u}(\mathbf{r}^{\prime})w_{h}(\mathbf{r}-\mathbf{r}^{\prime})d^{\nu}\mathbf{r}^{\prime}\label{deq:1pp}\\
\langle p_{h}(\mathbf{r})\rangle &=\int_{\Omega_{h}(\mathbf{r})}p(\mathbf{r}^{\prime})w_{h}(\mathbf{r}-\mathbf{r}^{\prime})d^{\nu}\mathbf{r}^{\prime}\label{deq:pp1}
\end{align} 

The smoothed field \{$\langle\rho_{h}(\mathbf{r})\rangle$, $\langle p_{h}(\mathbf{r})\rangle$, $\widetilde{\mathbf{u}}_{h}(\mathbf{r})$\} represents the interaction of fluid particles located at $\mathbf{r}$, $\mathbf{r}^{\prime}\in\Omega_{h}(\mathbf{r})$. Furthermore, the choice of the velocity smoothing here arises from the physical consideration that the smoothed velocity $\widetilde{\mathbf{u}}_{h}:=\langle\mathbf{P}_{h}\rangle/\langle\rho_{h}\rangle$ where $\mathbf{P}$ is the momentum density.

If the FIT is applied to (\ref{eq2016:1}), (\ref{eq2016:2}) and (\ref{eq2016:2aa}) we obtain the following set of filtered equations.
\begin{eqnarray}
\frac{d}{dt}\langle\rho_{h}(\mathbf{r})\rangle&=&-\langle\rho_{h}(\mathbf{r})\rangle\bm{\nabla}\cdot\widetilde{\mathbf{u}}_{h}(\mathbf{r})\label{eqn:2017cc2}\\
\frac{d }{dt}\langle p_{h}(\mathbf{r})\rangle&=&-\langle K_{S}\bm{\nabla}\cdot\mathbf{u},w_{h}\rangle + \gamma\alpha \langle K_{S}\bm{\nabla}\cdot(\kappa_{s}\bm{\nabla} p),w_{h}\rangle\label{eqn:2017c3}\\
\langle\rho_{h}\rangle\frac{d}{dt}\widetilde{\mathbf{u}}_{h}&=&\langle\bm{\nabla}\cdot \underline{\underline{\tau}},w_{h}\rangle-\bm{\nabla}\cdot\langle \underline{\underline{\mathcal{H}}}_{h}\rangle+ \langle\rho_{h}\rangle\widetilde{\mathbf{b}}_{h}\label{eqn:2017cc4}\\
\frac{d\mathbf{r}}{dt} &=& \widetilde{\mathbf{u}}_{h}(\mathbf{r})\label{eqn:2017c5}
\end{eqnarray}
where the material derivative after the filtering becomes
\begin{align}
\frac{d}{dt}&=\frac{\partial}{\partial t} + \widetilde{\mathbf{u}}_{h}\cdot\bm{\nabla}
\end{align}
The sub-particle stress (SPS) tensor arising from the filtering process is given by the following definition.

Application of the FIT is applied to the momentum equation introduces momentum transfer due to small scale motion. The SPS  represents the effect of the unresolved small scales on the local approximations. This is defined by the following 
\begin{align}
\langle\underline{\underline{\mathcal{H}}}_{h}(\mathbf{r})\rangle&=\int_{\Omega(\mathbf{r})}\rho(\mathbf{r}^{\prime})(\mathbf{u}(\mathbf{r}^{\prime})-\widetilde{\mathbf{u}}_{h}(\mathbf{r}))\otimes(\mathbf{u}(\mathbf{r}^{\prime})-\widetilde{\mathbf{u}}_{h}(\mathbf{r}))w_{h}d\Omega(\mathbf{r}^{\prime})\nonumber\\&=\langle\rho_{h}(\mathbf{r})\rangle\left(\widetilde{(\mathbf{u}\otimes\mathbf{u})}_{h}(\mathbf{r})-\widetilde{\mathbf{u}}_{h}(\mathbf{r})\otimes\widetilde{\mathbf{u}}_{h}(\mathbf{r})\right)\label{deq:2018d11}\quad\text{by the FIT}
\end{align}

The main task now is to de-filter the filtered equations (\ref{eqn:2017cc2}), (\ref{eqn:2017c3}), (\ref{eqn:2017cc4}) and (\ref{eqn:2017c5}). To this end, an inverse filtering procedure is necessary.

\subsection{De-filtering process}
Consider a fluid particle located at $\mathbf{r}$ and has a test space $\Omega_{h}(\mathbf{r})$ within the fluid domain $\Omega$. Given the locally averaged mass density, momentum density and pressure on $\Omega_{h}(\mathbf{r})$, we can reconstruct the continuum field by de-filtering the filtered mass density, momentum density and pressure as defined by the FIT above. Mathematically, for each $w_{h}\in C^{\infty}_{c}(\Omega_{h})$, there exists a $\varphi_{h}\in C^{\infty}_{c}(\Omega_{h})$ such that
\begin{align}
\rho(\mathbf{r}) &=\int_{\Omega_{h}(\mathbf{r})}\langle\rho_{h}(\mathbf{r}^{\prime})\rangle\varphi_{h}(\mathbf{r}-\mathbf{r}^{\prime})d^{\nu}\mathbf{r}^{\prime}\label{deq:3}\\
\rho(\mathbf{r})\mathbf{u}(\mathbf{r}) &=\int_{\Omega_{h}(\mathbf{r})}\langle\rho_{h}(\mathbf{r}^{\prime})\rangle\widetilde{\mathbf{u}}_{h}(\mathbf{r}^{\prime})\varphi_{h}(\mathbf{r}-\mathbf{r}^{\prime})d^{\nu}\mathbf{r}^{\prime}\label{deq:3e}\\
p(\mathbf{r}) &=\int_{\Omega_{h}(\mathbf{r})}\langle p_{h}(\mathbf{r}^{\prime})\rangle\varphi_{h}(\mathbf{r}-\mathbf{r}^{\prime})d^{\nu}\mathbf{r}^{\prime}\label{deq:3g}
\end{align} 
We then call $w_{h}$ as the convolution filter and $\varphi_{h}$ as the deconvolution filter.

The de-filtered SPH or SPH$-i$ model for short, is a complete model resulting from the application of the DIT to the filtered CNSEs. Unlike the SPH which uses the zeroth order deconvolution method, SPH$-i$ is based on the general deconvolution method. The mathematical procedure is shown below; steps [1]$\sim$[3] is the convolution operation on the fields $\{\rho, p, \mathbf{u}\}$ to produce local approximations $\{\langle\rho_{h}\rangle, \langle p_{h}\rangle, \widetilde{\mathbf{u}}_{h}\}$. For completeness, in steps [4]$\sim$[6] a deconvolution operation is dynamically performed on the local approximations to reconstruct the original continuum field $\{\rho, p, \mathbf{u}\}$.

\begin{enumerate}
	\item smoothed mass density
	\begin{align}
	\langle\rho_{h}(\mathbf{r})\rangle&=\int_{\Omega_{h}(\mathbf{r})}\rho(\mathbf{r}^{\prime})w_{h}(\mathbf{r}-\mathbf{r}^{\prime})d^{\nu}\mathbf{r}^{\prime}\label{deq:d12}\nonumber\\
	&\stackrel{.}{=} \rho(\mathbf{r})-\int_{\Omega_{h}(\mathbf{r})}\bigg(\rho(\mathbf{r})-\rho(\mathbf{r}^{\prime})\bigg)w_{h}(\mathbf{r}-\mathbf{r}^{\prime})d^{\nu}\mathbf{r}^{\prime}
	\end{align}
	\item smoothed pressure
	\begin{align}
	\langle p_{h}(\mathbf{r})\rangle&=\int_{\Omega_{h}(\mathbf{r})}p(\mathbf{r}^{\prime})w_{h}(\mathbf{r}-\mathbf{r}^{\prime})d\Omega(\mathbf{r}^{\prime})\label{deq:d16}\\
	&\stackrel{.}{=}p(\mathbf{r})-\int_{\Omega_{h}(\mathbf{r})}\bigg(p(\mathbf{r})-p(\mathbf{r}^{\prime})\bigg)w_{h}(\mathbf{r}-\mathbf{r}^{\prime})d^{\nu}\mathbf{r}^{\prime}
	\end{align}
	\item smoothed velocity
	\begin{align}
	\widetilde{\mathbf{u}}_{h}(\mathbf{r})&=\frac{1}{\langle\rho_{h}(\mathbf{r})\rangle}\int_{\Omega_{h}(\mathbf{r})}\rho(\mathbf{r}^{\prime})\mathbf{u}(\mathbf{r}^{\prime})w_{h}(\mathbf{r}-\mathbf{r}^{\prime})d\Omega(\mathbf{r}^{\prime})\label{deq:d13}\\
	&\stackrel{.}{=}\mathbf{u}(\mathbf{r})-\frac{1}{\langle\rho_{h}(\mathbf{r})\rangle}\int_{\Omega_{h}(\mathbf{r})}\rho(\mathbf{r}^{\prime})\bigg(\mathbf{u}(\mathbf{r})-\mathbf{u}(\mathbf{r}^{\prime})\bigg)w_{h}(\mathbf{r}-\mathbf{r}^{\prime})d^{\nu}\mathbf{r}^{\prime}
	\end{align}
	\item de-filtered continuity equation
	\begin{align}
	\frac{d\rho}{dt}&=-\rho D(\mathbf{u}\vert\varphi)\label{deq:d14}
	\end{align}
	\item de-filtered pressure equation
	\begin{align}
	\kappa_{s}(p)\frac{dp}{dt} &=
	-D(\mathbf{u}\vert\varphi)\nonumber \\
	& \qquad{} +\gamma\alpha L(\kappa_{s}(p),p\vert\varphi)-\alpha L\left(\frac{1}{\rho},\rho\vert\varphi\right)
	\end{align}
	\item de-filtered momentum equation
	\begin{align}
	\rho\frac{d\mathbf{u}}{dt}&=-G(p\vert\varphi)+D(\mu,\underline{\underline{\sigma}}\vert\varphi)+\rho\mathbf{b}\label{deq:d15}
	\end{align}
	\item moving the un-smoothed particles
	\begin{align}
	\frac{d\mathbf{r}}{dt} &= \mathbf{u}(\mathbf{r})
	\end{align}
\end{enumerate}

where the momentum conserving operators are defined as follows.
\begin{align}
G(p\vert\varphi) &=\bm{\nabla} p= \int_{\Omega_{h}(\mathbf{r})}\left(\frac{p(\mathbf{r})}{\rho(\mathbf{r})}\langle\rho_{h}(\mathbf{r}^{\prime})\rangle +\frac{p(\mathbf{r}^{\prime})}{\rho(\mathbf{r}^{\prime})}\langle\rho_{h}(\mathbf{r})\rangle\right)\bm{\nabla}\varphi_{h} d^{\nu}\mathbf{r}^{\prime}\label{eqn:2017d11}\\
D(\mathbf{u}\vert\varphi)&=\bm{\nabla}\cdot\mathbf{u}(\mathbf{r})=-\frac{1}{\rho(\mathbf{r})}\int_{\Omega_{h}(\mathbf{r})}\langle\rho_{h}(\mathbf{r}^{\prime})\rangle \bigg(\mathbf{u}(\mathbf{r})-\widetilde{\mathbf{u}}_{h}(\mathbf{r}^{\prime})\bigg)\cdot\bm{\nabla}\varphi_{h}d^{\nu}\mathbf{r}^{\prime}\label{eqn:2017d12}
\end{align}
\begin{align}
D(\mu,\underline{\underline{\sigma}}\vert\varphi) &:=\bm{\nabla}\cdot\underline{\underline{\sigma}}\nonumber\\
&=\int_{\Omega_{h}(\mathbf{r})}\left(\frac{\underline{\underline{\sigma}}(\mathbf{r})}{\rho(\mathbf{r})}\langle\rho_{h}(\mathbf{r}^{\prime})\rangle +\frac{\underline{\underline{\sigma}}(\mathbf{r}^{\prime})}{\rho(\mathbf{r}^{\prime})}\langle\rho_{h}(\mathbf{r})\rangle\right)\cdot\bm{\nabla}\varphi_{h} d^{\nu}\mathbf{r}^{\prime}\label{eqn:2017dd11}\\
L(\kappa^{s},p\vert\varphi)&:=\bm{\nabla}\cdot(\kappa^{s}\bm{\nabla} p)\nonumber\\
&=\frac{1}{2}\int_{\Omega_{h}(\mathbf{r})}\bigg[\bigg(\langle\kappa^{s}_{h}(\mathbf{r})\rangle+\langle\kappa^{s}_{h}(\mathbf{r}^{\prime})\rangle\bigg)\bigg(p(\mathbf{r})-p(\mathbf{r}^{\prime})\bigg)\nonumber\\
&+\bigg(\kappa^{s}(\mathbf{r})+\kappa^{s}(\mathbf{r}^{\prime})\bigg)\bigg(\langle p_{h}(\mathbf{r})\rangle-\langle p_{h}(\mathbf{r}^{\prime})\rangle\bigg)\bigg]\frac{(\mathbf{r}-\mathbf{r}^{\prime})\cdot\bm{\nabla}\varphi_{h}}{\vert\vert\mathbf{r}-\mathbf{r}^{\prime}\vert\vert^{2}}d^{\nu}\mathbf{r}^{\prime}\label{eqn:2017d13}
\end{align} 
 
To get the discrete forms we just replace integrals by summations. The reader must also see that the differential forms of the above are the original compressible Navier-Stokes equations.

For the work presented in this paper we choose the following convolution filter for stability reasons.
\begin{align}
w_{h}&=\left(1-\frac{q^2}{4}\right)^{8}\bigg(2048 - 3072q^{2} + 2816q^{4} - 1984q^{6}\nonumber\\
&\qquad+ 1184q^{8} - 628q^{10} + 305q^{12}\bigg)\label{deq:2018d2}
\end{align}
Its associated deconvolution filter on $\mathbb{R}^{2}$ is given by a truncated series as
\begin{align}
\varphi_{h,n}(\mathbf{r}-\mathbf{r}^{\prime\prime})&=\sum_{k=0}^{n}\widetilde{M}_{k}\frac{h^{2k}}{(2k)!}\nabla^{2k}w_{h}(\mathbf{r}-\mathbf{r}^{\prime\prime})\equiv\sum_{k=0}^{n}f_{k}(h)\psi_{h,2k}(\mathbf{r}-\mathbf{r}^{\prime\prime})\label{deq:2017k7}
\end{align}
which is a convergent series. Refer to \cite{2018arXiv180711728C} for full details about this model.

\section{Model validation}
\subsection{Hydrostatic pressure in a water tank}
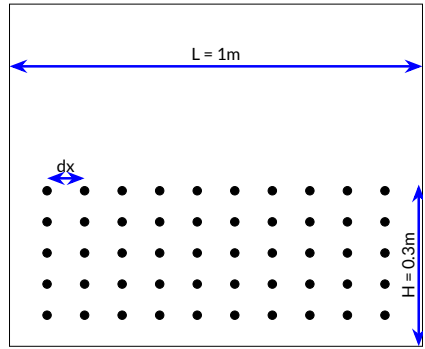
\begin{figure}
	\centering
	\begin{tikzpicture}[scale=0.8]
	\begin{axis}[ticks=none,xmax=10,ymax=10]
	\addplot[draw=none,no marks] coordinates {(0,0) (10,10)};
	\foreach \x in {0,1,...,9}
	{
		\foreach \y in {0,1,...,9}
		{\x --["\ifthenelse{\x<10 \AND \y>4}{}{	\begingroup\edef\temp{\endgroup\noexpand\draw[fill]  (axis cs:\x,\y) circle (2pt);}\temp}"] \y;
		}
	}
	\draw[blue, very thick, {Stealth}-{Stealth}, postaction={decoration={raise=3pt, text along path, text={dx},text align=center}, decorate}] (0,44) -- (10,44);
	\draw[blue, very thick, {Stealth}-{Stealth},
	postaction={decoration={raise=3pt, text along path, text={L = 1m},text align=center}, decorate}] (-10,80) -- (100,80);
	\draw[blue, very thick, {Stealth}-{Stealth}, postaction={decoration={raise=2pt, text along path, text={H = 0.3m},text align=center}, decorate}] (99,-10) -- (99,42);
	\end{axis}
	\end{tikzpicture}
	\caption[Water tank schematic]{numerical water tank setup.}
	\label{fig:tankSchematic}
\end{figure}

The hydrostatic pressure problem is one of the fundamental benchmark test cases in SPH. For many traditional SPH approaches it is a challenging problem to obtain a stable, regularized pressure field \cite{Chen2013,Oger2007}. The set for this numerical simulation is as shown in figure \ref{fig:tankSchematic}.  

The numerical water tank has a square shape of side $1\textup{m}$. It is filled with fluid particles to a height of $h_{0}=0.3\textup{m}$ and a pressure probe point P is placed $y_{p}=0.06\textup{m}$ above the bottom. With a zero reference pressure on the free surface, the hydrostatic pressure at the probe point P is given by $p=\rho_{0}\mathnormal{g}(h_{0}-y_{p})=2354.4\textup{Pa}$. The initial particle spacing is set at $dx=dy=0.01\textup{m}$ so that the total number of fluid particles is $N=100\times30=3000$ while the average number of nearest neighbor is fixed at $N_{n}=19$. The density is initialized to the rest value $\rho_{0}=1000\textup{Kgm}^{-3}$ whereas the pressure and velocity are initially set to zero. The compact support radius in units of $h$ is then $\xi=\sqrt{N_{n}/4\pi}=1.23$. This results from the mass conservation law and details can be found in \cite{Price2012}. The smoothing length is thus $h=\xi\Delta x=0.0123$m. 

The solid boundary treatment is enforced by the proposed damped boundary force model with a damping coefficient of $k^{d}=0.10606\textup{Nsm}^{-1}$ and a stiffness coefficient of $k^{s}=0.06065\textup{Nm}^{-1}$ for each particle.  
The thermal diffusivity and kinematic viscosity were respectively $\alpha=0.015\textup{m}^{2}\textup{s}^{-1}$ and $\nu=1.0\times10^{-6}\textup{m}^{2}\textup{s}^{-1}$. A suitable time step for the time integration was $\Delta t=1.0\times10^{-5}\textup{s}$

\begin{figure}
	\centering
	\includegraphics[width=8cm]{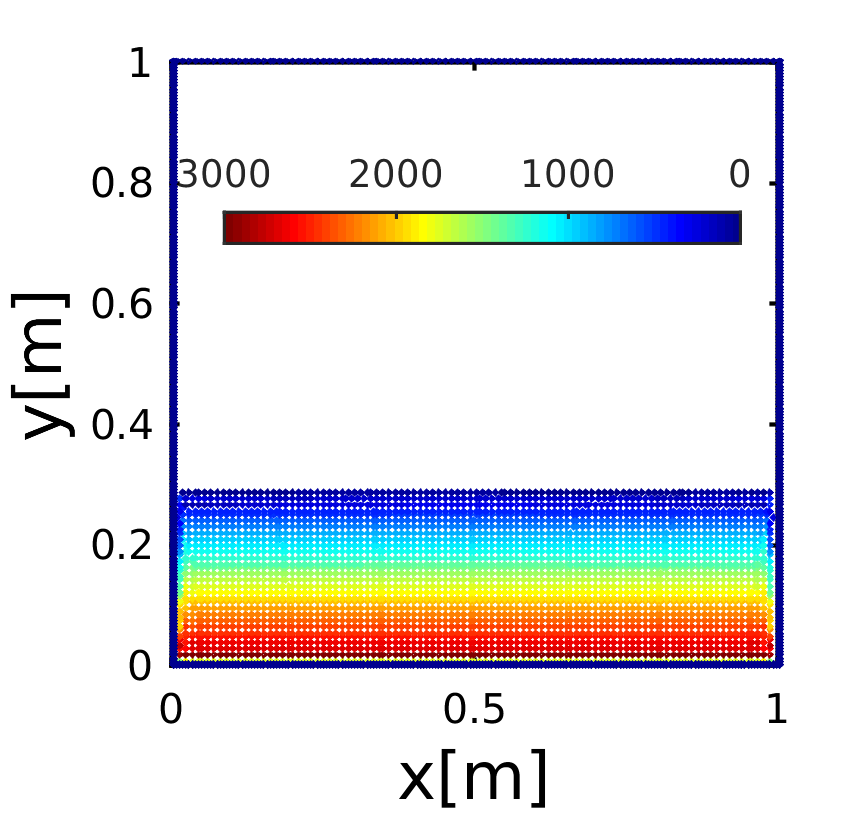}
	\caption[Quasi-static pressure field at time $t=15\textup{s}$]{Snapshot of the pressure field at time $t=15\textup{s}$ obtained by the SPH$-i$ model.}
	\label{fig:staticSnapshot}
\end{figure}

Figure \ref{fig:staticSnapshot} shows a snapshot of the pressure filed obtained by the SPH$-i$ model at time $t=15\textup{s}$.  
In order to check the long term numerical stability, the simulation was allowed to run for an extended time of $t=15\textup{s}$. It is observed that SPH$-i$ model gives a smooth pressure distribution.

\begin{figure}
	\centering
	\begin{subfigure}[b]{0.7\textwidth}
		\includegraphics[width=0.85\linewidth]{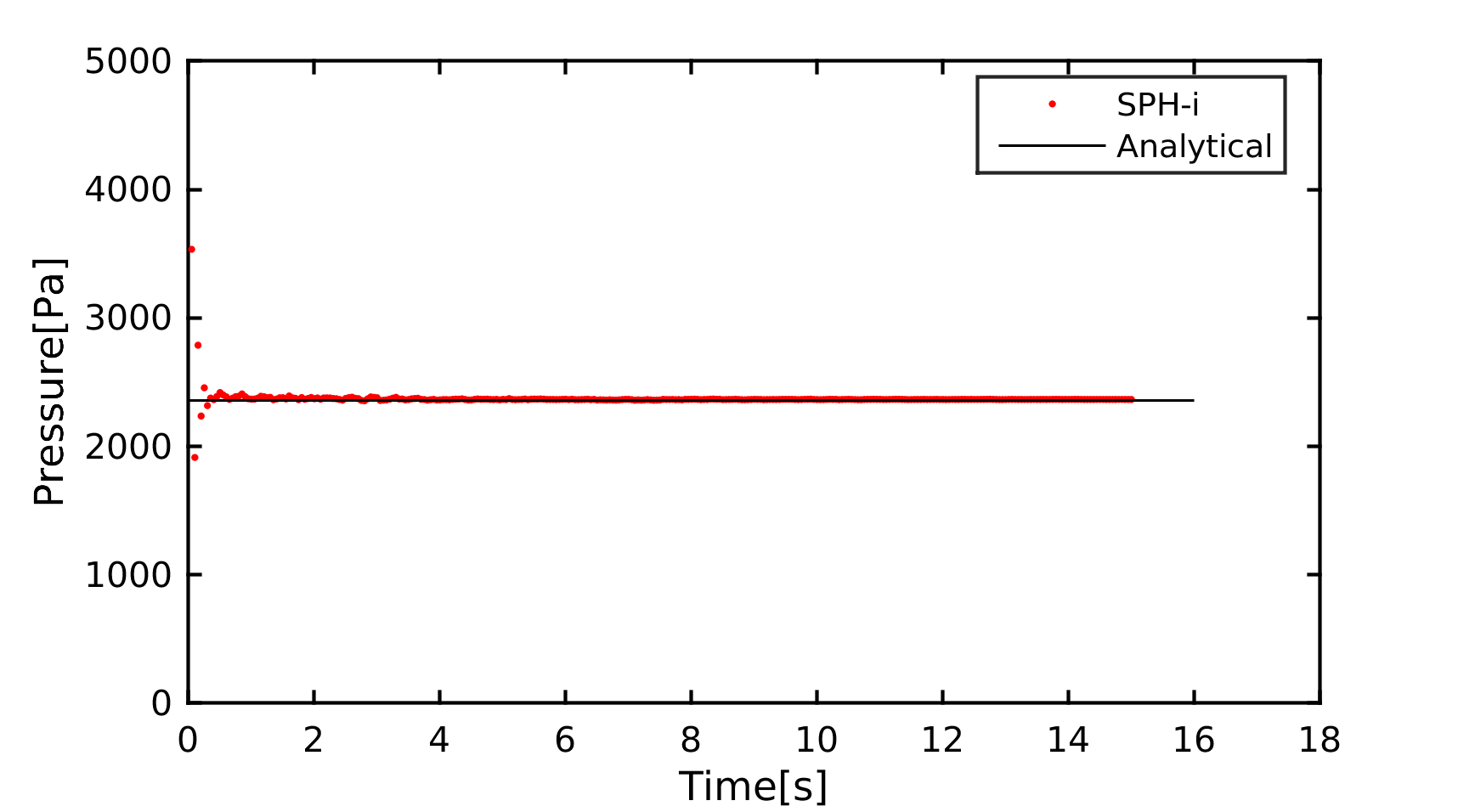}
		\caption[Pressure time history]{Pressure time history}
		\label{fig:PressureHistory}
	\end{subfigure}%
	\begin{subfigure}[b]{0.4\textwidth}
		\includegraphics[width=0.6\linewidth]{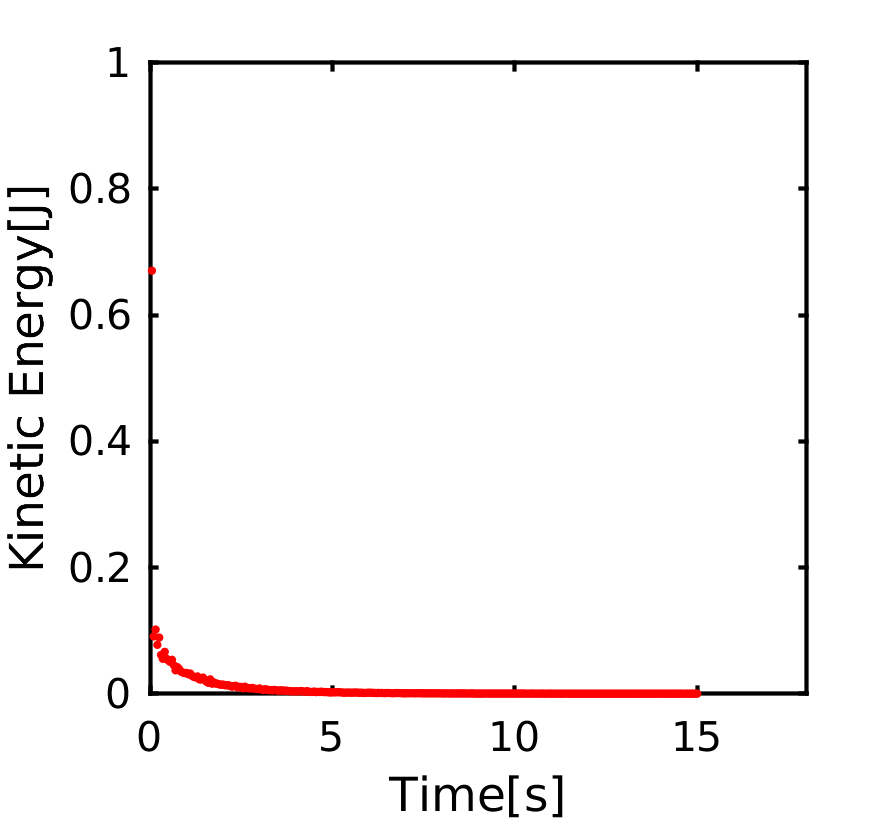}
		\caption[Total kinetic energy]{Total kinetic energy}
		\label{fig:Kinetic}
	\end{subfigure}
	\caption[Quasi-hydrostatatic equilibrium]{Time history of the pressure at the probe point P $(y=0.06\textup{m})$ obtained by the SPH$-i$ model in \ref{fig:PressureHistory}. Here \ref{fig:Kinetic} is the total kinetic energy of the fluid.}
	\label{fig:staticHistory}
\end{figure}
\begin{figure}
	\centering
	\begin{subfigure}[b]{0.6\textwidth}
		\includegraphics[width=0.85\linewidth]{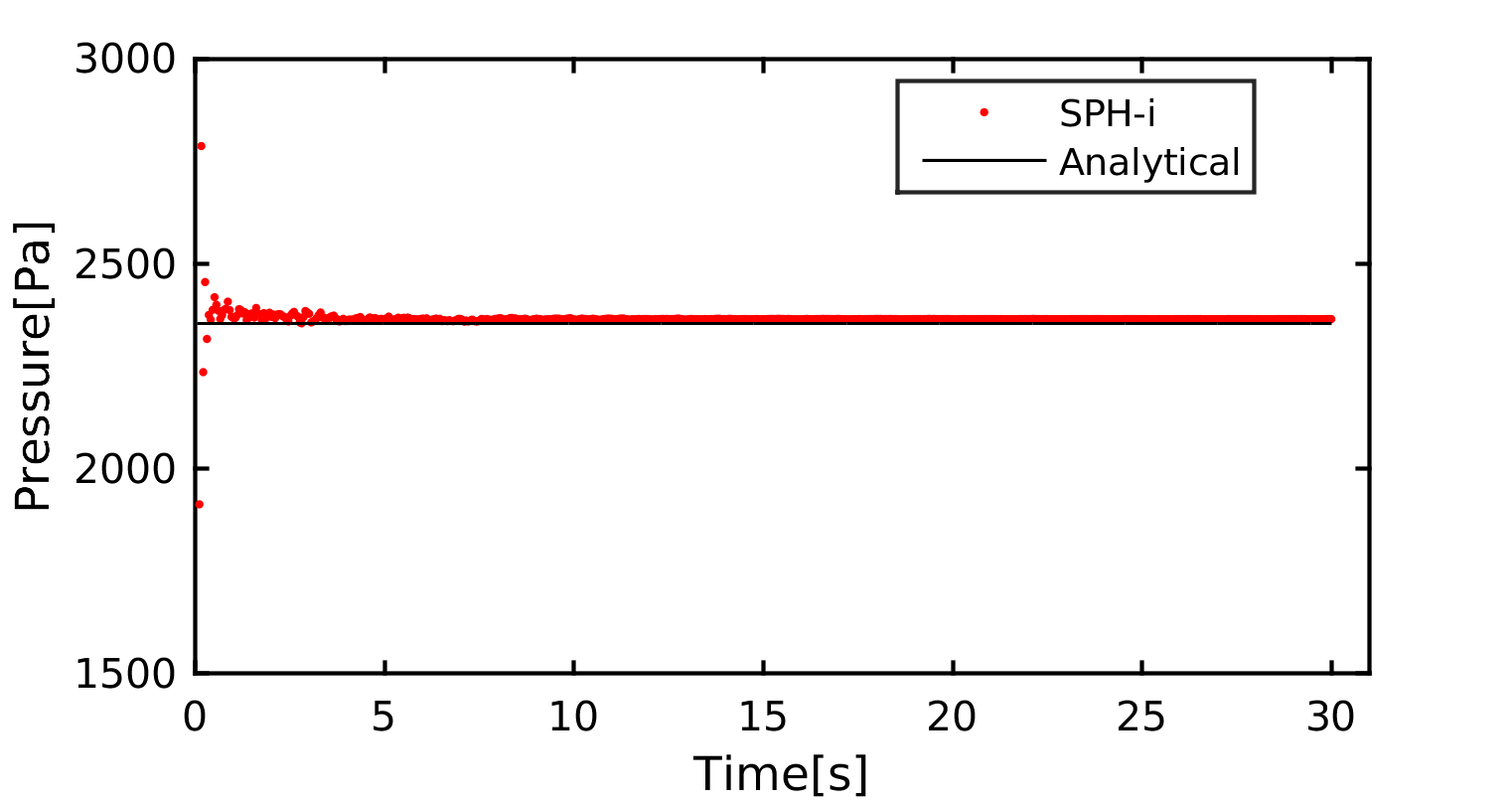}
		\caption[Pressure time history]{Pressure time history}
		\label{fig:PressureHistory2}
	\end{subfigure}%
	\begin{subfigure}[b]{0.4\textwidth}
		\includegraphics[width=0.8\linewidth]{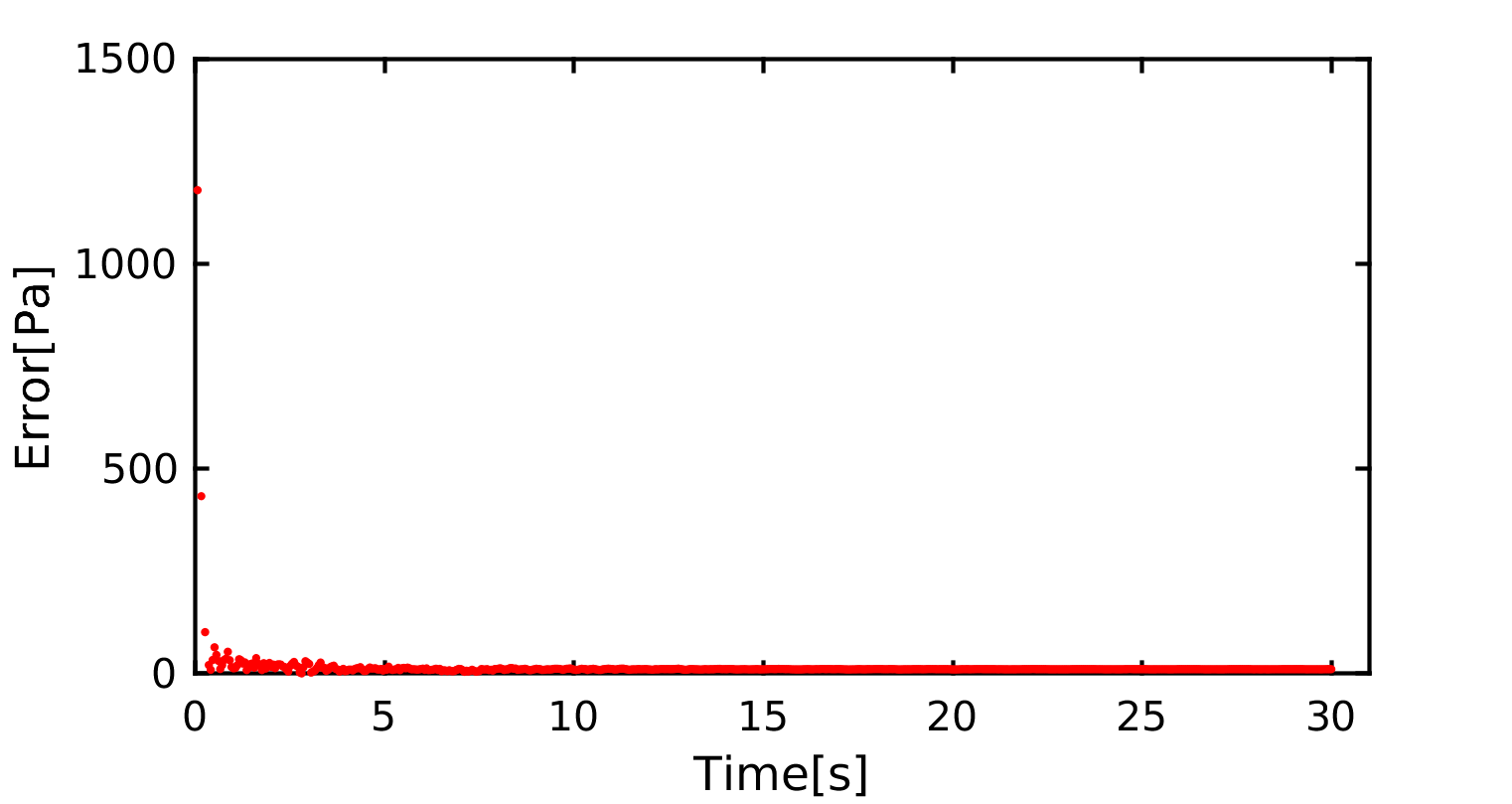}
		\caption[Error]{Error}
		\label{fig:error}
	\end{subfigure}
	\caption[Quasi-hydrostatatic equilibrium:long simulation time]{Time history of the pressure at the probe point P $(y=0.06\textup{m})$ obtained by the SPH$-i$ model in \ref{fig:PressureHistory2}. Here \ref{fig:error} is the residual between the numerical and exact solutions.}
	\label{fig:staticHistory2}
\end{figure}

  
A time history of the pressure at the probe point $P$ is as shown in figure \ref{fig:PressureHistory} and \ref{fig:PressureHistory2}. Initially there is a time lag of about $0.5\textup{s}$ during which time gravity squeezes the fluid down. However, pressure gradients quickly build up, and together with boundary forces counterbalances the gravitational push. Thus the fluid adjusts to a new equilibrium position through an oscillation mode. The SPH$-i$ models clearly preserves hydrostatic pressure equilibrium for long-time simulations as can be seen in figure \ref{fig:PressureHistory2}. This also indicates that the diffusion terms in the pressure equation can effectively smooth out numerical noise in the pressure field. Furthermore, the kinetic energy asymptotically decays to zero as shown in figure \ref{fig:Kinetic}. At finite times, however, the kinetic energy is small but non-zero. Therefore, the fluid particles will tend to oscillate about their mean positions; consistent with a physical system. 

In some SPH models the hydrostatic pressure profile slowly diverges from the numerical solution at long-time simulation \cite{Chen2013}. This is attributed to numerical diffusion in those models. To ascertain the possibility of any instability build-up due to numerical diffusion, the simulation was allowed to progress for a much long time as shown in figure \ref{fig:staticHistory2}.

\subsection{Dam break on a dry bed}
\begin{figure}
	\centering
	\begin{tikzpicture}[scale=1.2]
	\begin{axis}[ticks=none,xmax=20,ymax=10,width=10cm,height=6cm]
	\addplot[draw=none,no marks] coordinates {(0,0) (20,10)};
	\foreach \x in {-1,0,...,19}
	{
		\foreach \y in {0,1,...,9}
		{\x --["\ifthenelse{\NOT\x<10) \OR \NOT\y<5}{}{	\begingroup\edef\temp{\endgroup\noexpand\draw[fill]  (axis cs:\x,\y) circle (2pt);}\temp}"] \y;
		}
	}
	\draw[blue, very thick, {Stealth}-{Stealth}, postaction={decoration={raise=3pt, text along path, text={W = 1.6m},text align=center}, decorate}] (-20,80) -- (200,80);
	\draw[blue, very thick, {Stealth}-{Stealth}, postaction={decoration={raise=3pt, text along path, text={L = 2H},text align=center}, decorate}] (-20,50) -- (100,50);
	\draw[blue, very thick, {Stealth}-{Stealth}, postaction={decoration={raise=2pt, text along path, text={H = 0.3m},text align=center}, decorate}] (110,-10) -- (110,42);
	\node[label={180:{P}},circle,fill,inner sep=2pt] at (axis cs:19.99,0.2) {};
	\end{axis}
	\end{tikzpicture}
	\caption[Schematic of dam break on dry bed]{Initial set up of the numerical dam break problem.}
	\label{fig:damSchematic}
\end{figure}
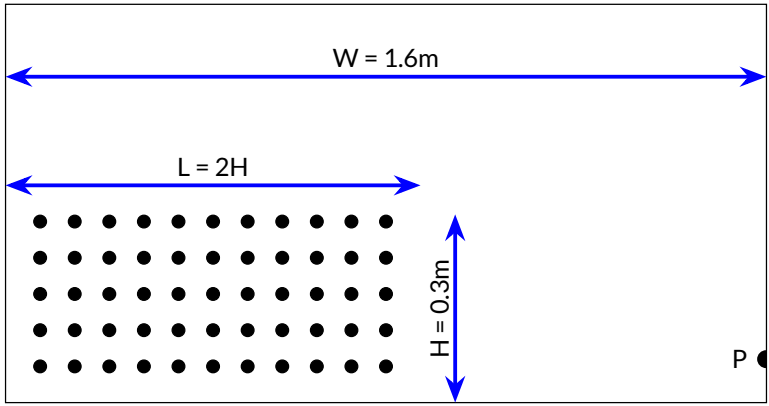

The second validation test case that was considered in this work is the violent shallow water breaking wave process generated by a dam break in a finite domain. It is one of the fundamental benchmark problems in the numerical study of free surface flows. The dam break flow is a highly nonlinear, complex phenomena that is characterized by large free surface deformation, splash up and multiple breaking events. Figure \ref{fig:damSchematic} shows a schematic diagram of the experiment reported in \cite{Lee2002} of which the experimental data was obtained from Chen et.al. \cite{Chen2013}. This experiment has been widely used in literature for numerical validations (refer to \cite{Colagrossi2003,Chen2013,Colagrossi2015}) with comparisons made using  various SPH models with varying degrees of accuracy. As discussed in \cite{Colagrossi2015}, the dam break problem has several characteristic features including;
(i) irrotational fluid deformation
(ii) water impact on vertical wall,
(iii) backward plunging jet formation,
(iv) several splashing cycles,
(v)	final sloshing flow regime, and
(vi) adjustment to an hydrostatic equilibrium. 	

The computational domain is of width $W=1.6\textup{m}$ and the water column is initially set at $L\times W=0.6\textup{m}\times0.3\textup{m}$. The initial spacing between fluid particles is $dr=0.002\textup{m}$ and in total $N=45,000$ particles were used in the simulation. The average number of near neighbors was fixed at 91; chosen so as to minimize numerical dissipation attributable to filtering/de-filtering processes. The compact support radius in units of $h$ is then $\xi=\sqrt{N_{n}/4\pi}=2.69$. The smoothing length is thus $h=\xi\Delta x=0.00538$m. The sound speed was reduced to $c_{s}=10\sqrt{2gH}$, thermal diffusivity $\alpha=0.00015\textup{m}^{2}\textup{s}^{-1}$ and kinematic viscosity $\nu=1.0\times10^{-6}\textup{m}^{2}\textup{s}^{-1}$ were used. To guarantee long term stability of the time integration a time step of $\Delta t=1.0\times10^{-5}\textup{s}$ was chosen. In the experiment a circularly shaped pressure gauge of diameter $9\textup{cm}(\approx0.15H)$ located on the vertical wall with center at $0.267\textup{H}$ above the deck as shown in figure \ref{fig:damSchematic} was used to record the pressure time history at the probe point P.  

For this test case, the Reynolds number, $Re=H\sqrt{gH}/\nu$, was $5.1\times10^{5}$ at a spatial resolution $H/dx=150$. For the initial conditions on the field values, the fluid density was set to its rest value $\rho_{0}=1000\textup{Kgm}^{-3}$ while the pressure and velocity were all initialized to zero.

\begin{figure}
	\centering
	\begin{subfigure}[b]{.4\textwidth}
		\includegraphics[width=\textwidth,height=2cm]{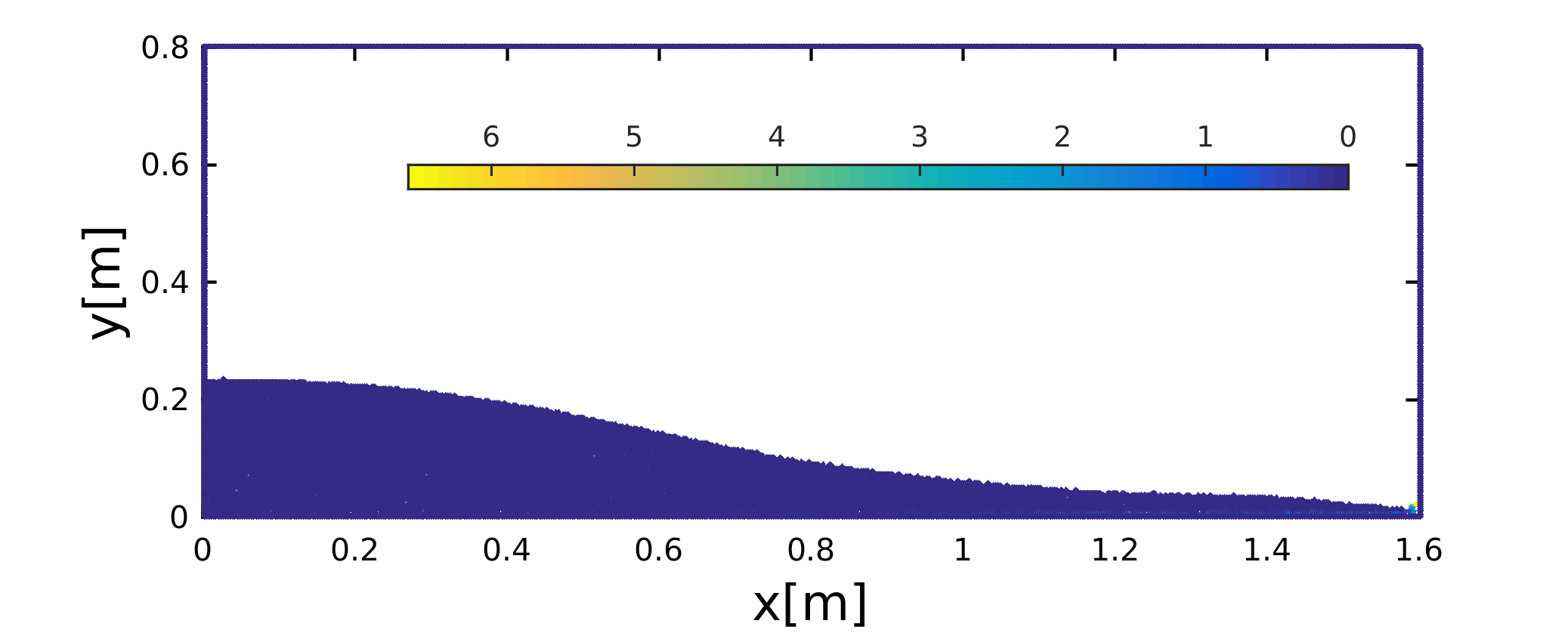}
		\caption{$t=0.425\textup{s}$}
		\label{fig:tke1}
		\vspace{0.1ex}
		
		\includegraphics[width=\textwidth,height=2cm]{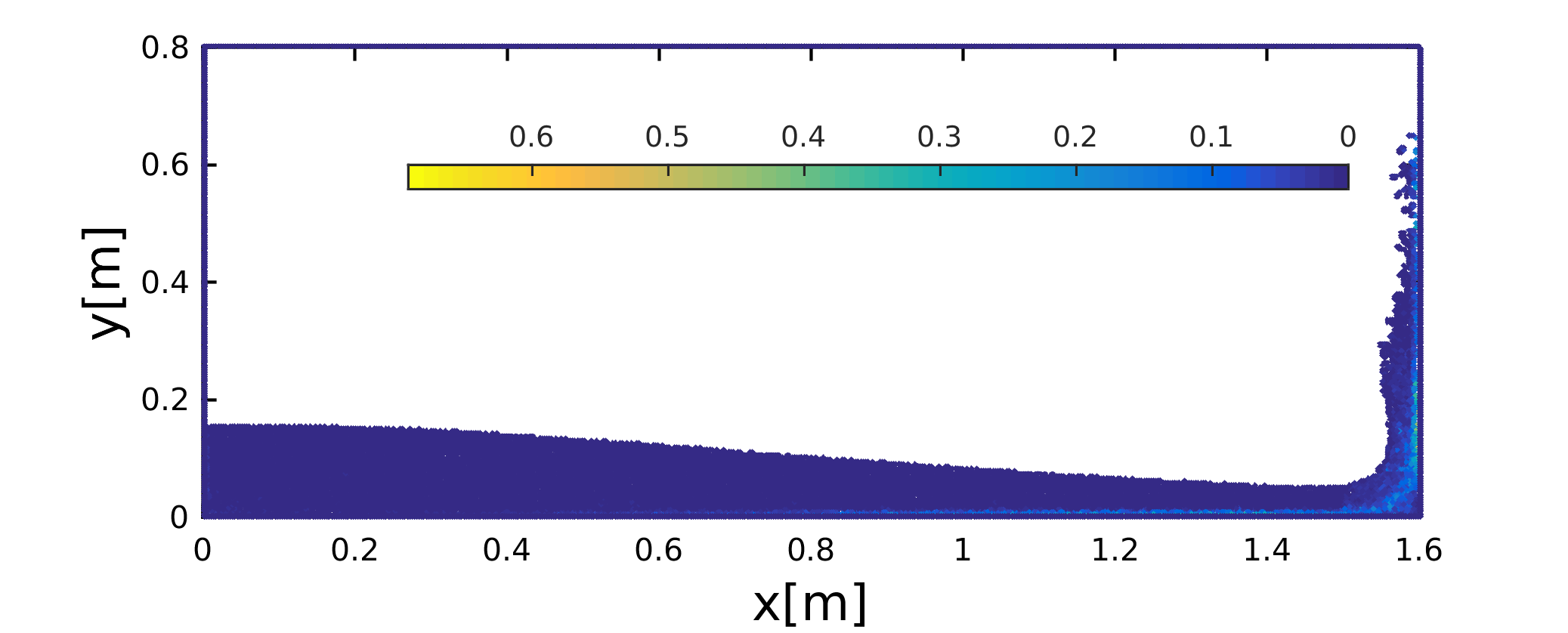}
		\caption{$t=0.700\textup{s}$}
		\label{fig:tke2}
		\vspace{0.1ex}
		
		\includegraphics[width=\textwidth,height=2cm]{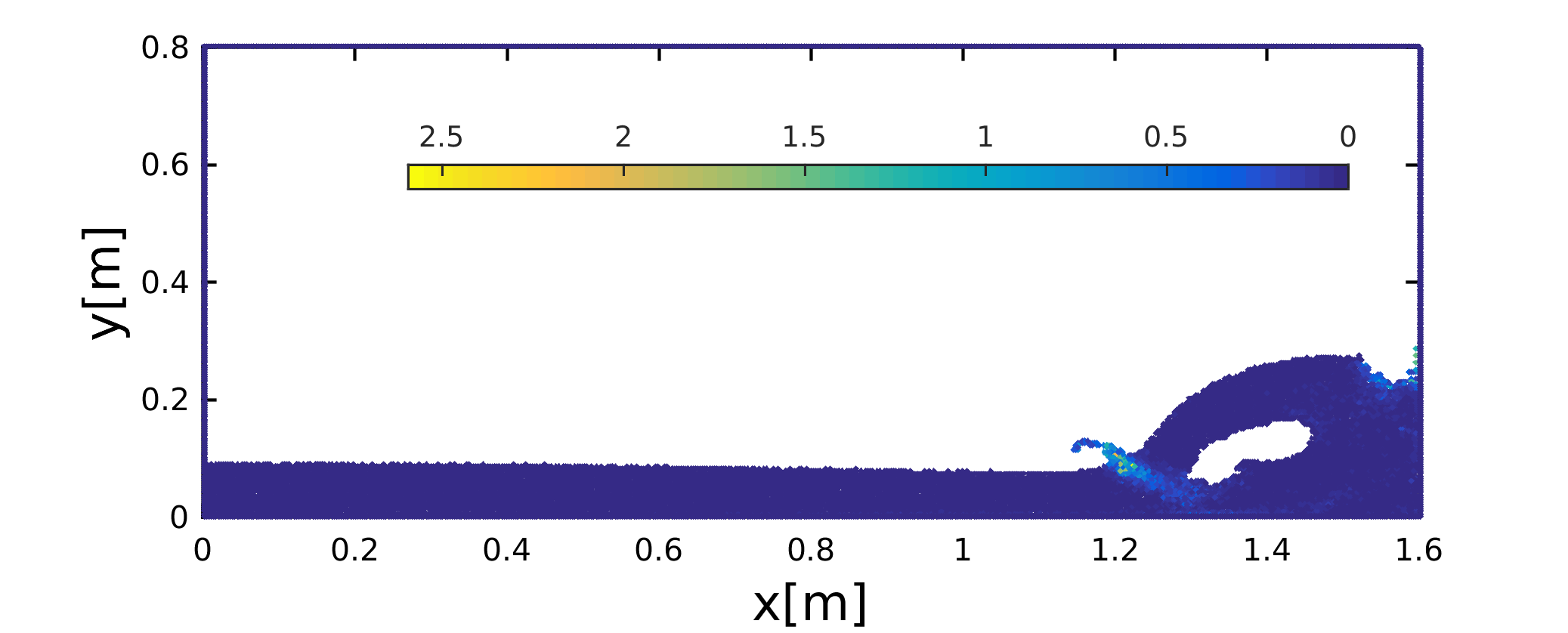}
		\caption{$t=1.100\textup{s}$}
		\label{fig:tke3}
		\vspace{0.1ex}
		
		\includegraphics[width=\textwidth,height=2cm]{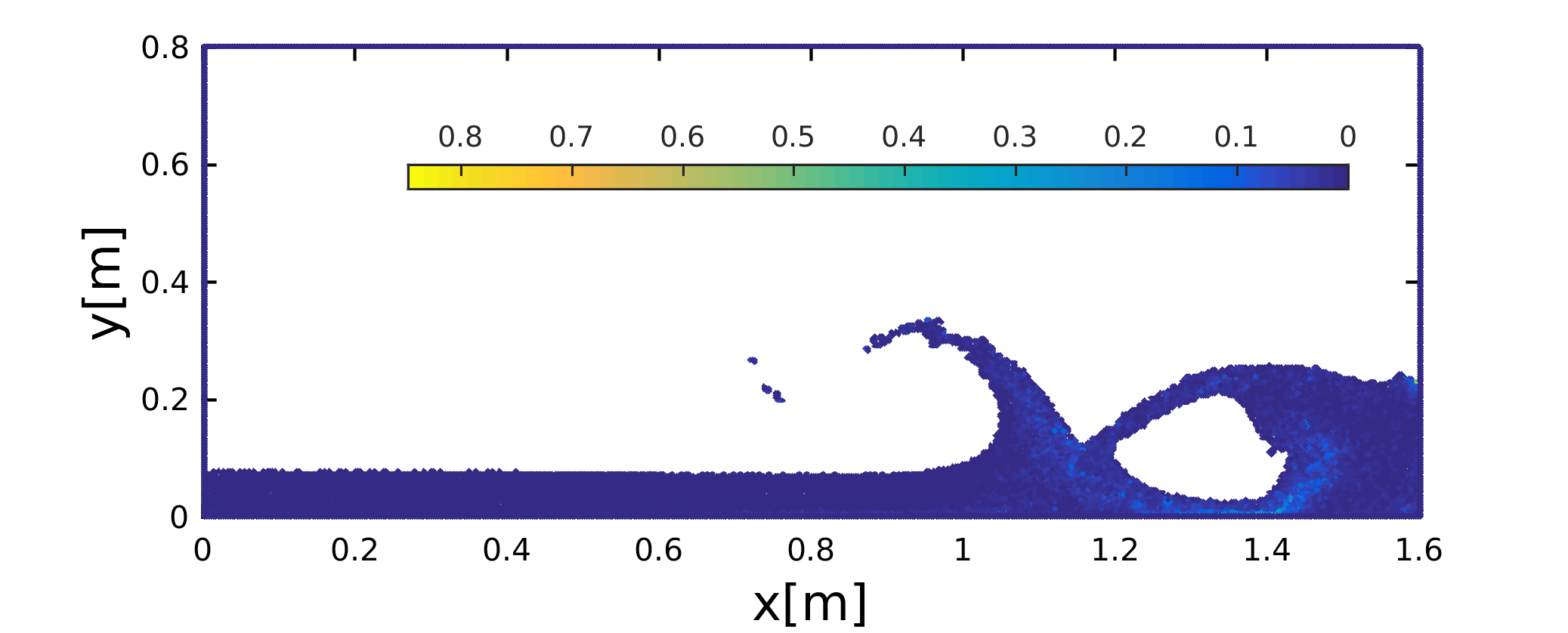}
		\caption{$t=1.250\textup{s}$}
		\label{fig:tke4}
		\vspace{0.1ex}
		
		\includegraphics[width=\textwidth,height=2cm]{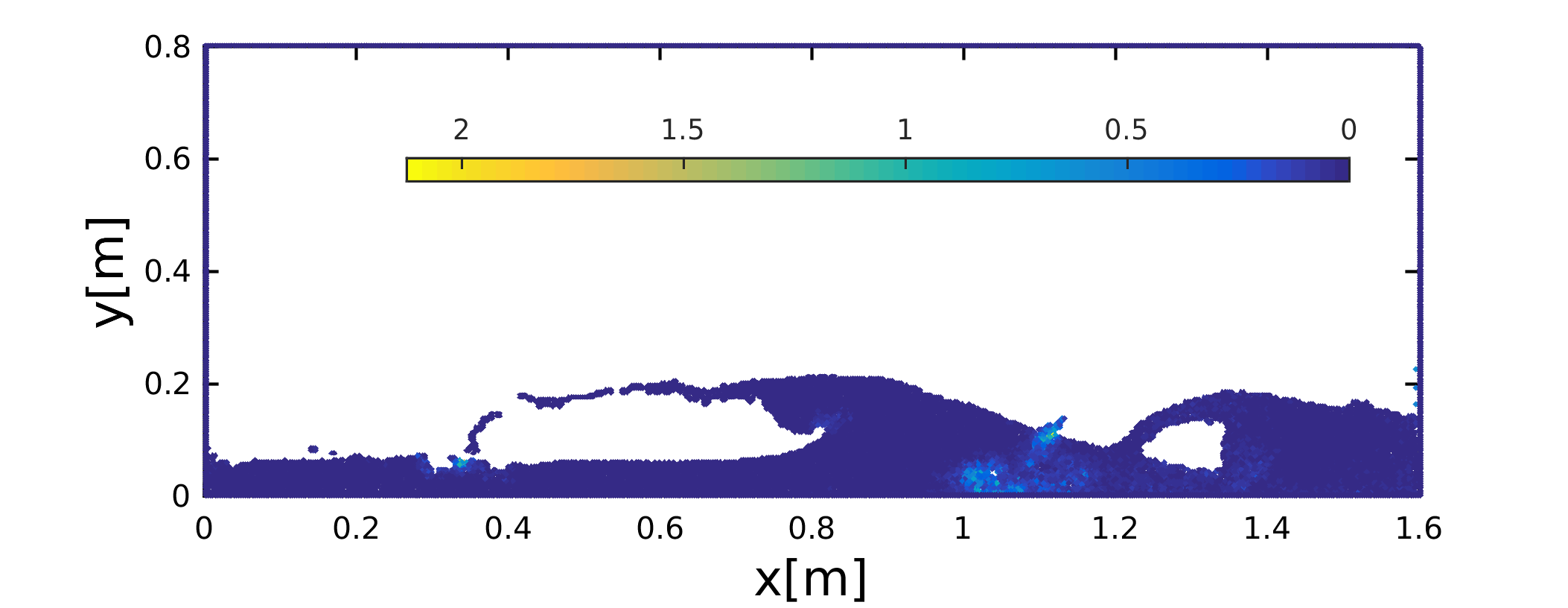}
		\caption{$t=1.550\textup{s}$}
		\label{fig:tke5}
		\vspace{0.1ex}
		
		\includegraphics[width=\textwidth,height=2cm]{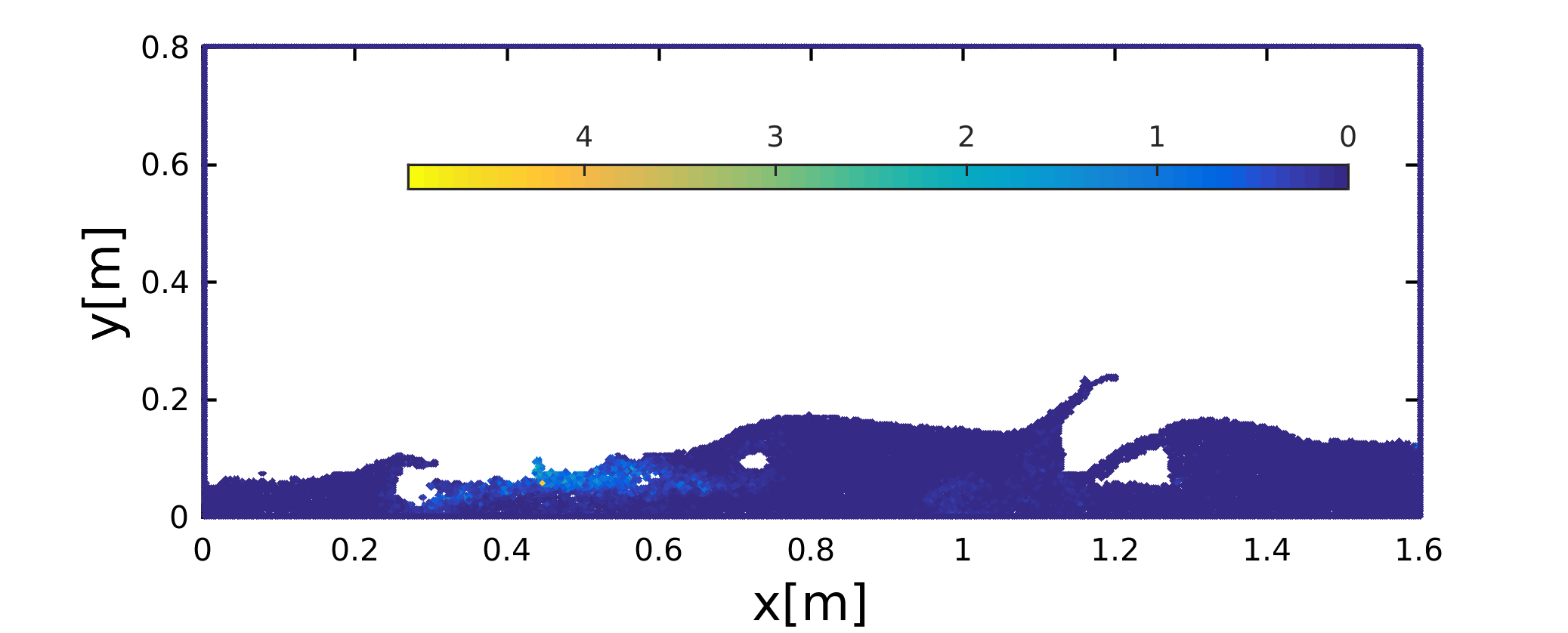}
		\caption{$t=1.625\textup{s}$}
		\label{fig:tke6}
		\vspace{0.1ex}
		
		\includegraphics[width=\textwidth,height=2cm]{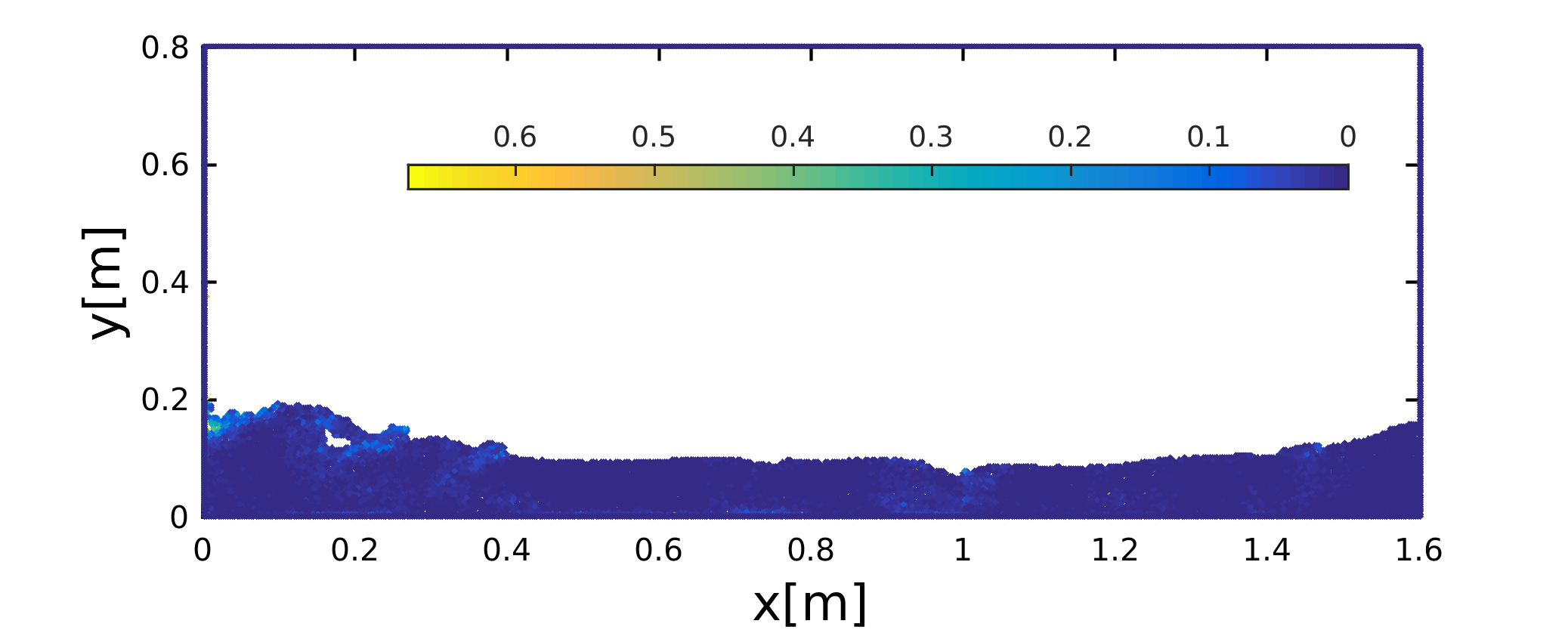}
		\caption{$t=2.275\textup{s}$}
		\label{fig:tke7}
	\end{subfigure}\qquad
	\begin{subfigure}[b]{.4\textwidth}
		\includegraphics[width=\textwidth,height=2cm]{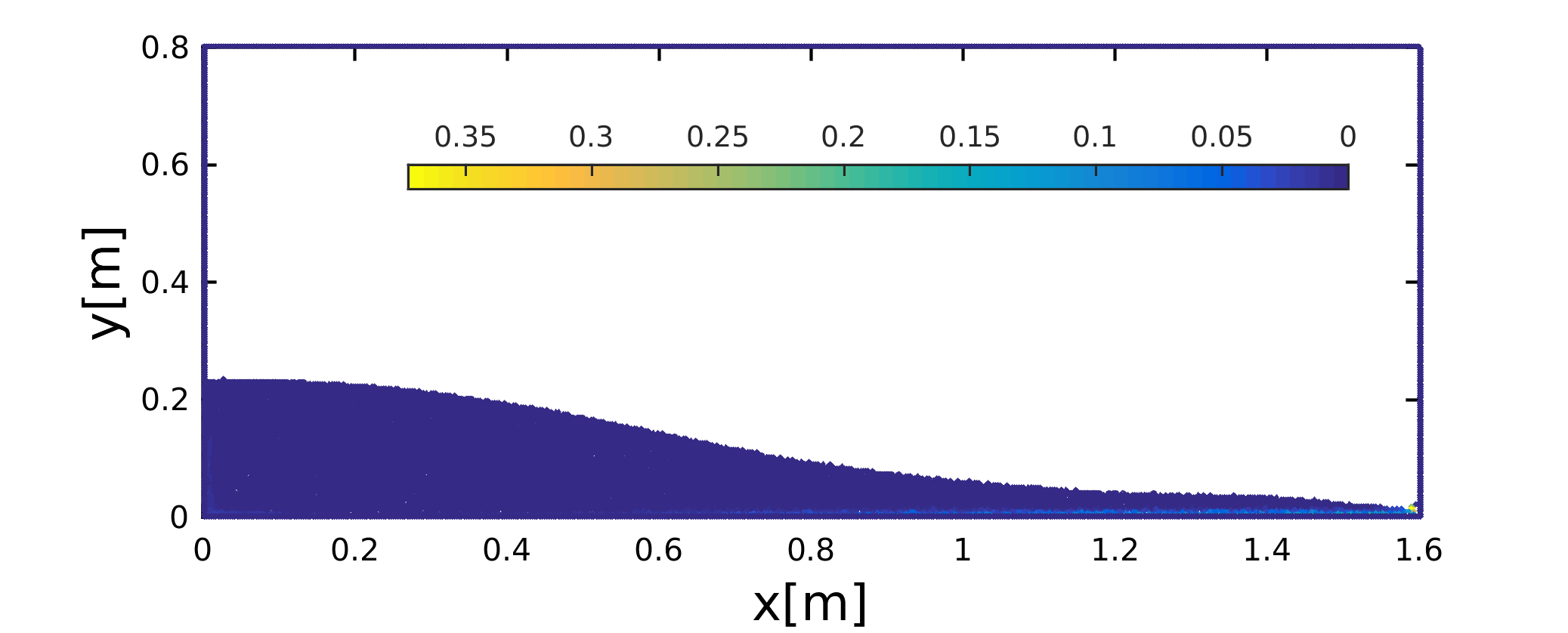}
		\caption{$t=0.425\textup{s}$}
		\label{fig:tdr1}
		\vspace{0.1ex}
		
		\includegraphics[width=\textwidth,height=2cm]{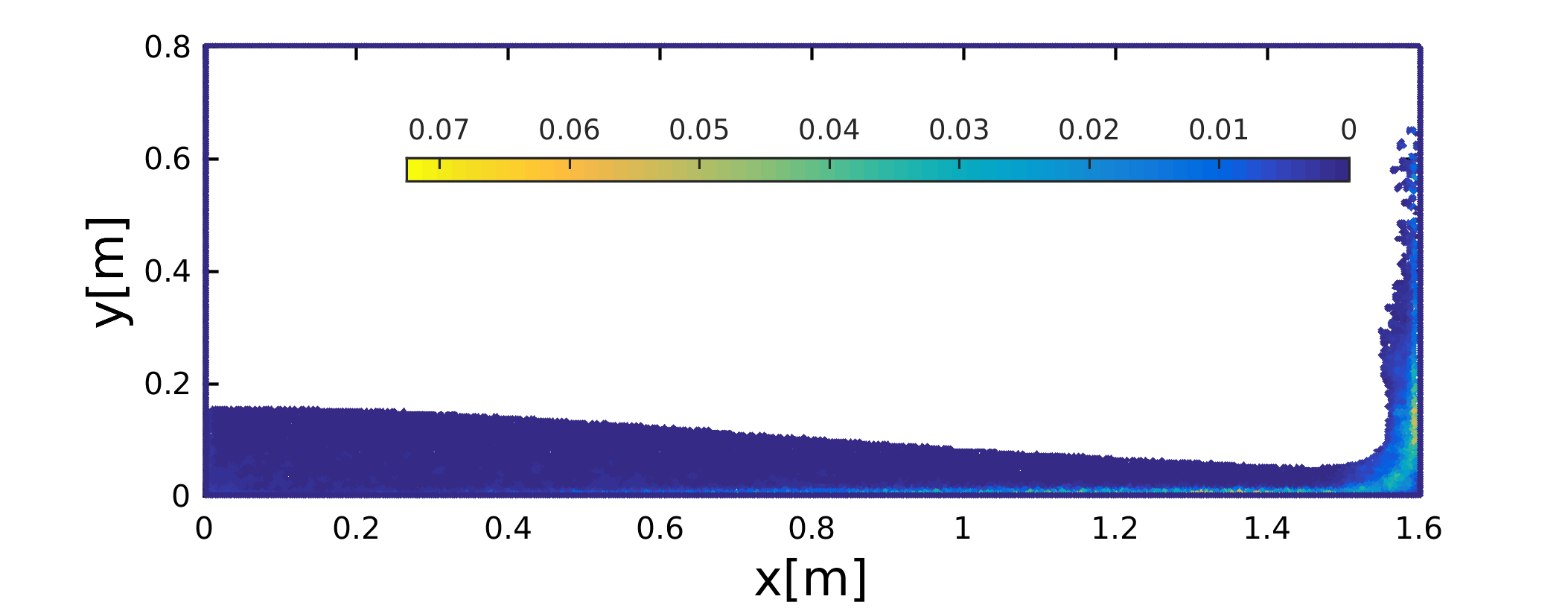}
		\caption{$t=0.700\textup{s}$}
		\label{fig:tdr2}
		\vspace{0.1ex}
		
		\includegraphics[width=\textwidth,height=2cm]{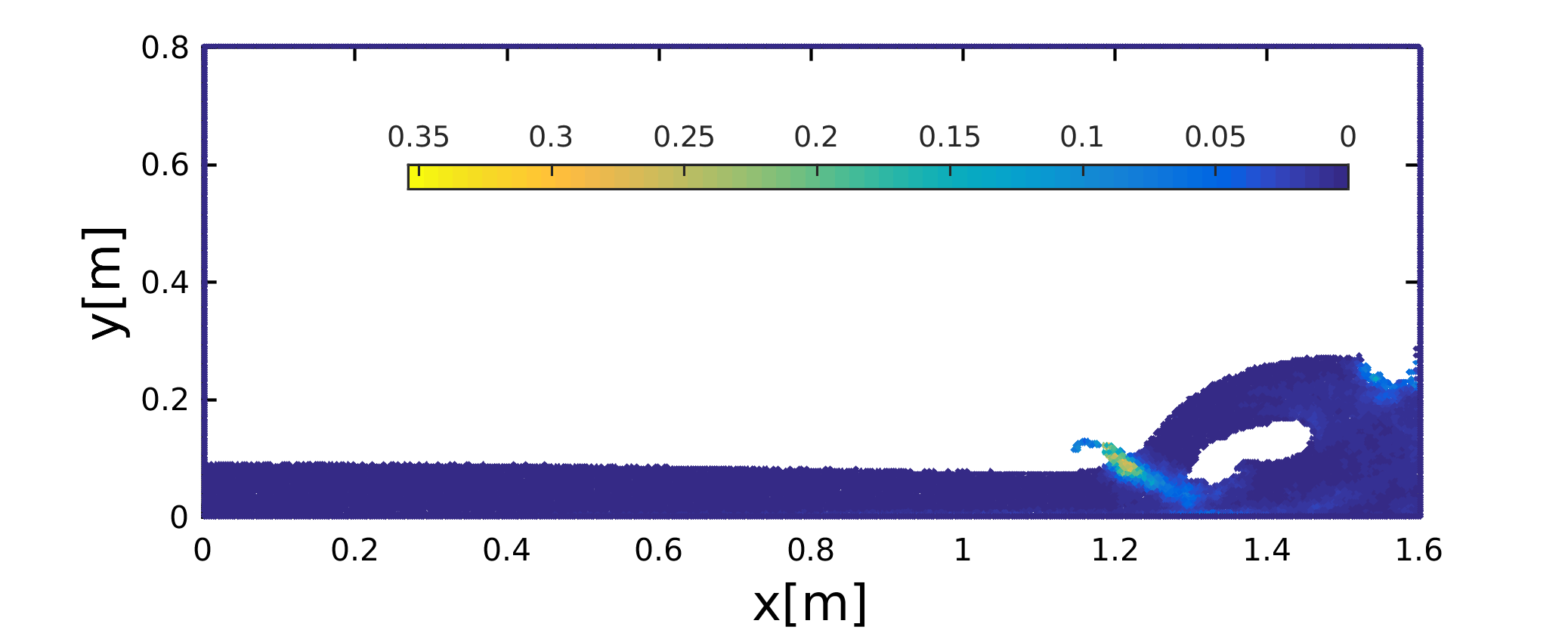}
		\caption{$t=1.100\textup{s}$}
		\label{fig:tdr3}
		\vspace{0.1ex}
		
		\includegraphics[width=\textwidth,height=2cm]{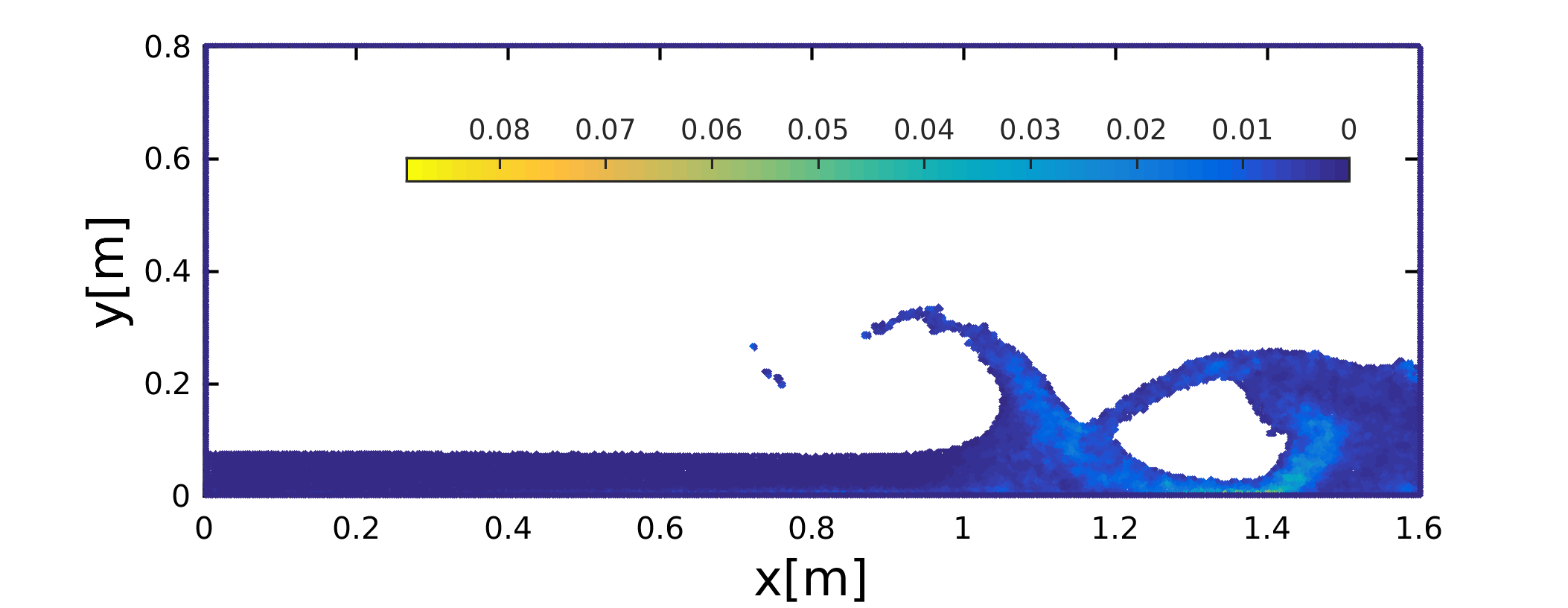}
		\caption{$t=1.250\textup{s}$}
		\label{fig:tdr4}
		\vspace{0.1ex}
		
		\includegraphics[width=\textwidth,height=2cm]{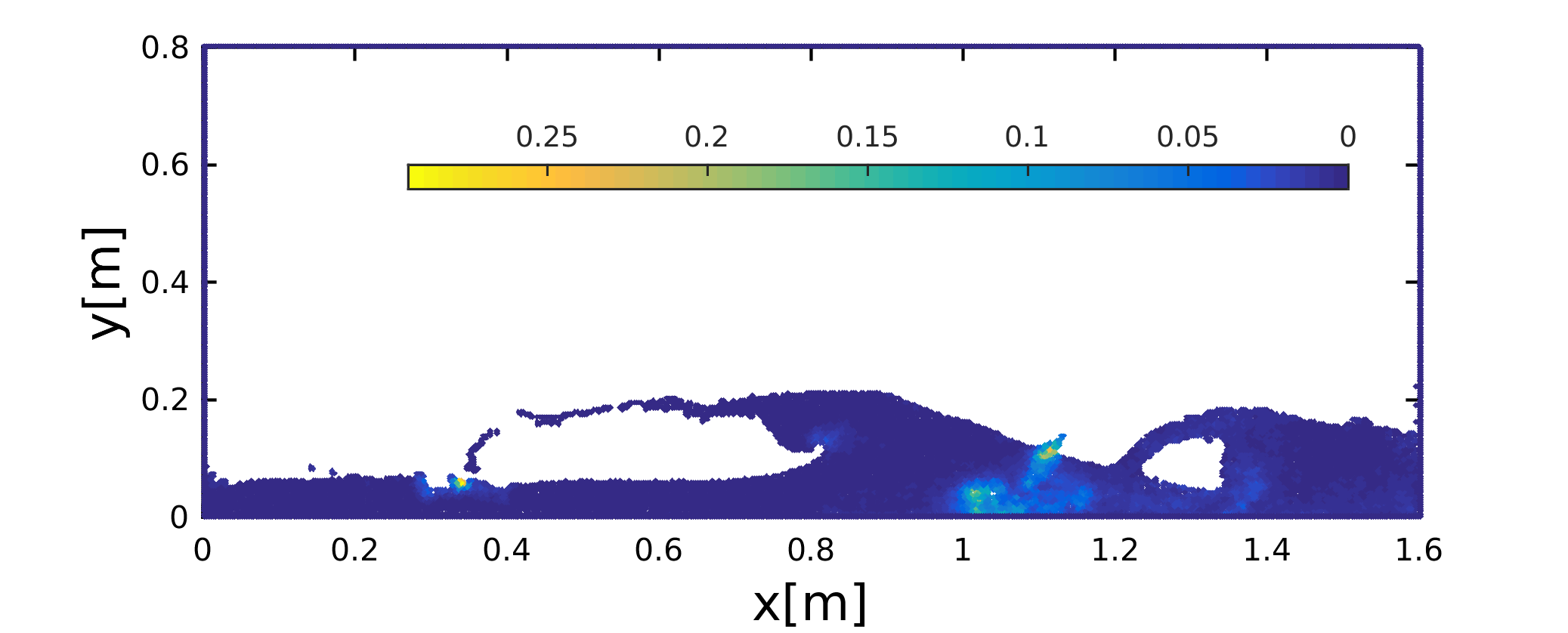}
		\caption{$t=1.550\textup{s}$}
		\label{fig:tdr5}
		\vspace{0.1ex}
		
		\includegraphics[width=\textwidth,height=2cm]{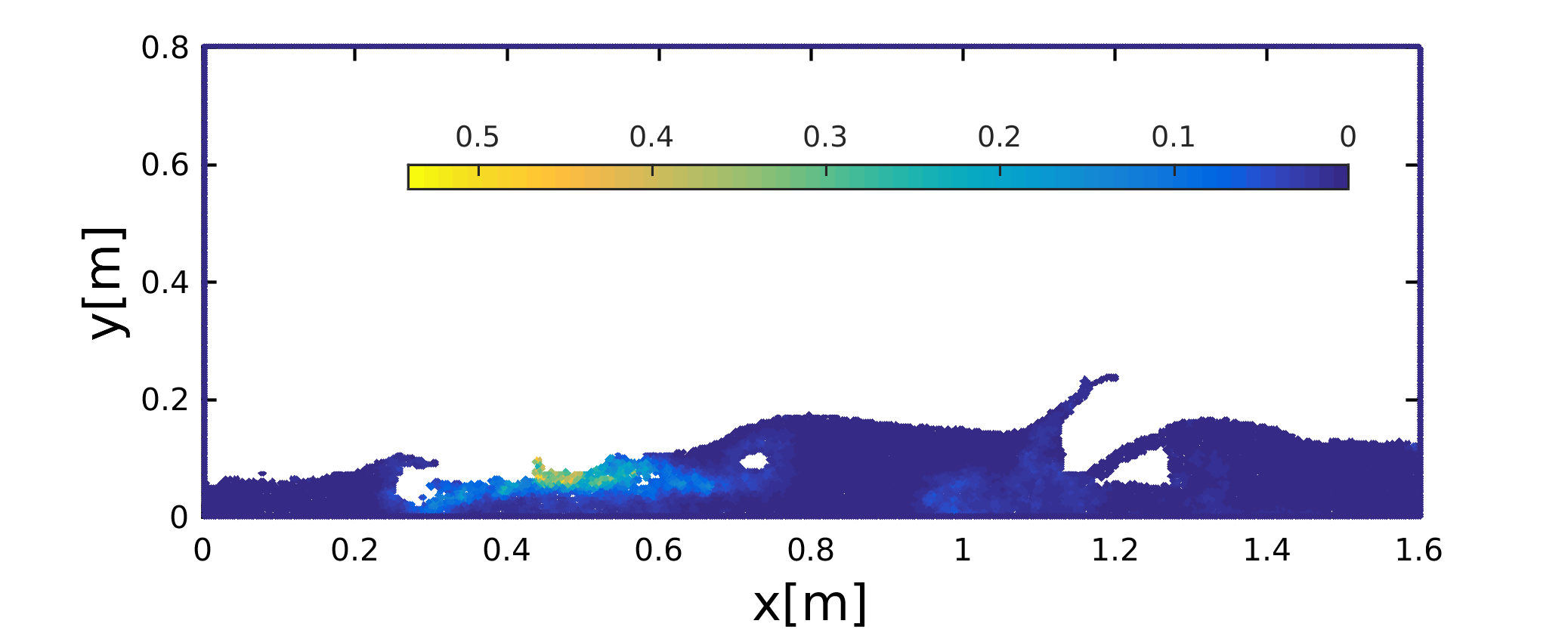}
		\caption{$t=1.625\textup{s}$}
		\label{fig:tdr6}
		\vspace{0.1ex}
		
		\includegraphics[width=\textwidth,height=2cm]{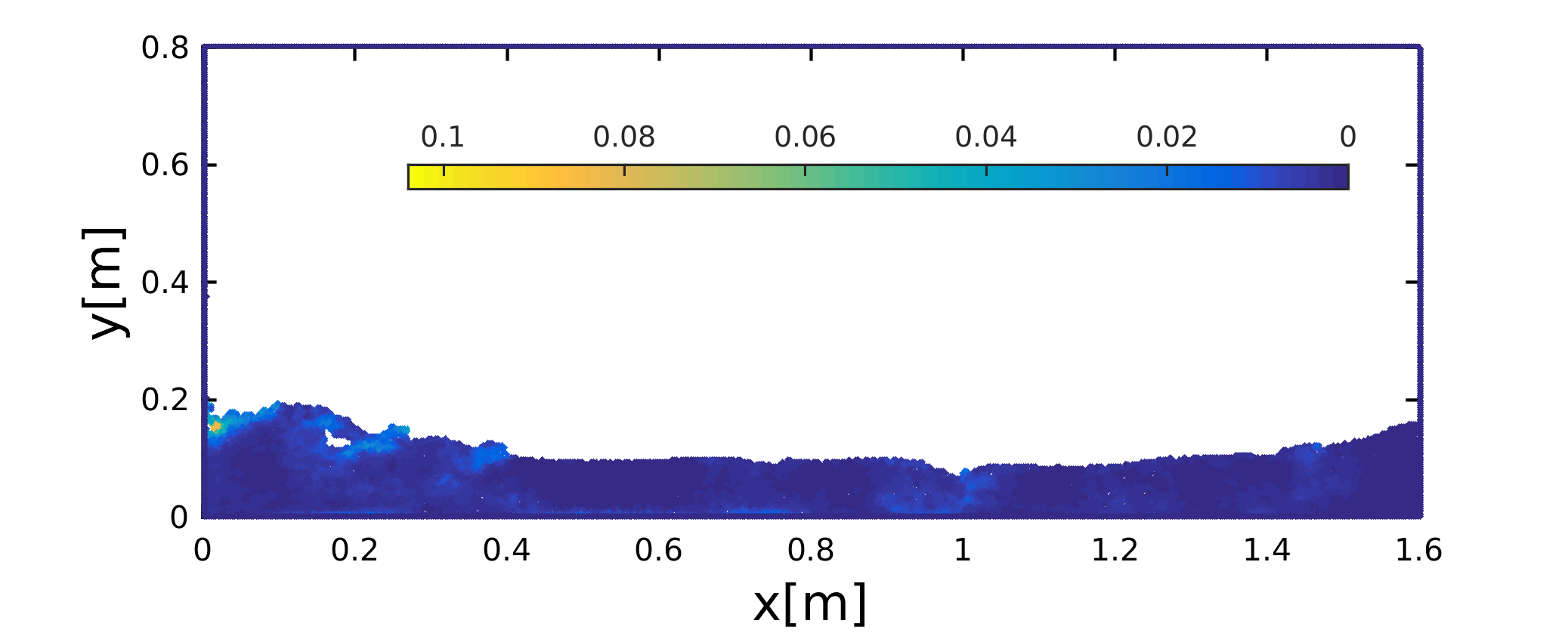}
		\caption{$t=2.275\textup{s}$}
		\label{fig:tdr7}
	\end{subfigure}
	\caption[Dam break on dry bed]{Production and dissipation of turbulence in shallow water breaking waves, $\textup{Re}=5.1\times10^{5}$. Turbulent kinetic energy (left) and turbulent dissipation rate (right) at seven time instants.}
	\label{fig:dryBreak}
\end{figure} 

The time evolution of the dam break flow is depicted in figure \ref{fig:dryBreak} along with the associated turbulent kinetic energy and turbulent dissipation rate.
Initially the flow is practically uniform so that any turbulent fluctuations are negligibly small as discussed in proposition \cite{2018arXiv180711244C}. Therefore, up until $t=0.425\textup{s}$ both turbulent kinetic energy and the associated turbulence dissipation rate are negligibly small except near solid boundaries (see figure \ref{fig:tke1} and \ref{fig:tdr1}). After the moving front hits and momentarily interacts with the right wall, turbulent kinetic energy is gradually produced (see figure \ref{fig:tke2}) and is quickly dissipated (see figures \ref{fig:tdr2}).
After impact with the right wall, a backward plunging jet is formed. As the plunging jet impinges on the free surface, at $t=1.100\textup{s}$, significant turbulent kinetic energy is produced (see figure \ref{fig:tke3}) and is quickly dissipated (see figure \ref{fig:tdr3}).
At times $t=1.550\textup{s}$ and $t=1.625\textup{s}$, due to the collapse of entrapped cavities, more turbulent kinetic energy is produced and quickly dissipated . For $t>1.625\textup{s}$ small secondary splash-ups develop, and finally flow enters a shallow water sloshing regime.   

\begin{figure}
	\centering
	\begin{subfigure}[b]{.4\textwidth}
		\includegraphics[width=\textwidth,height=4cm]{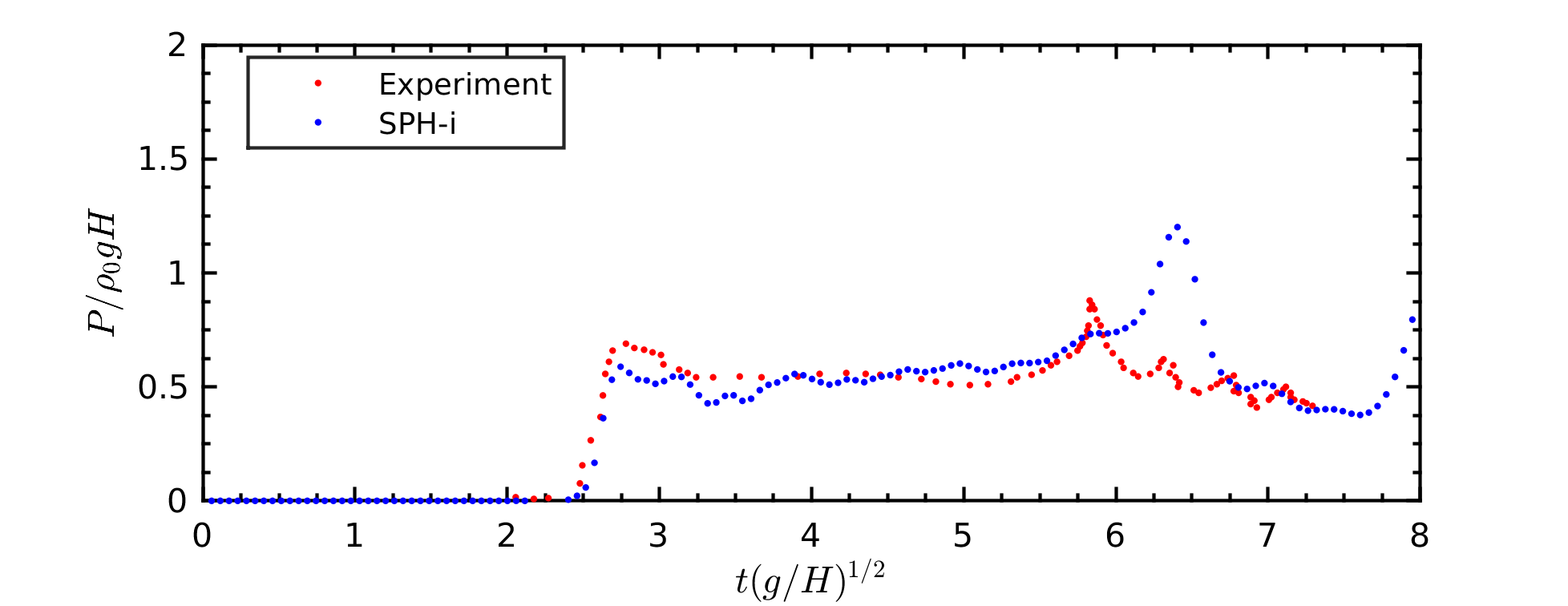}
		\vspace{0.1ex}
		
		\includegraphics[width=\textwidth,height=4cm]{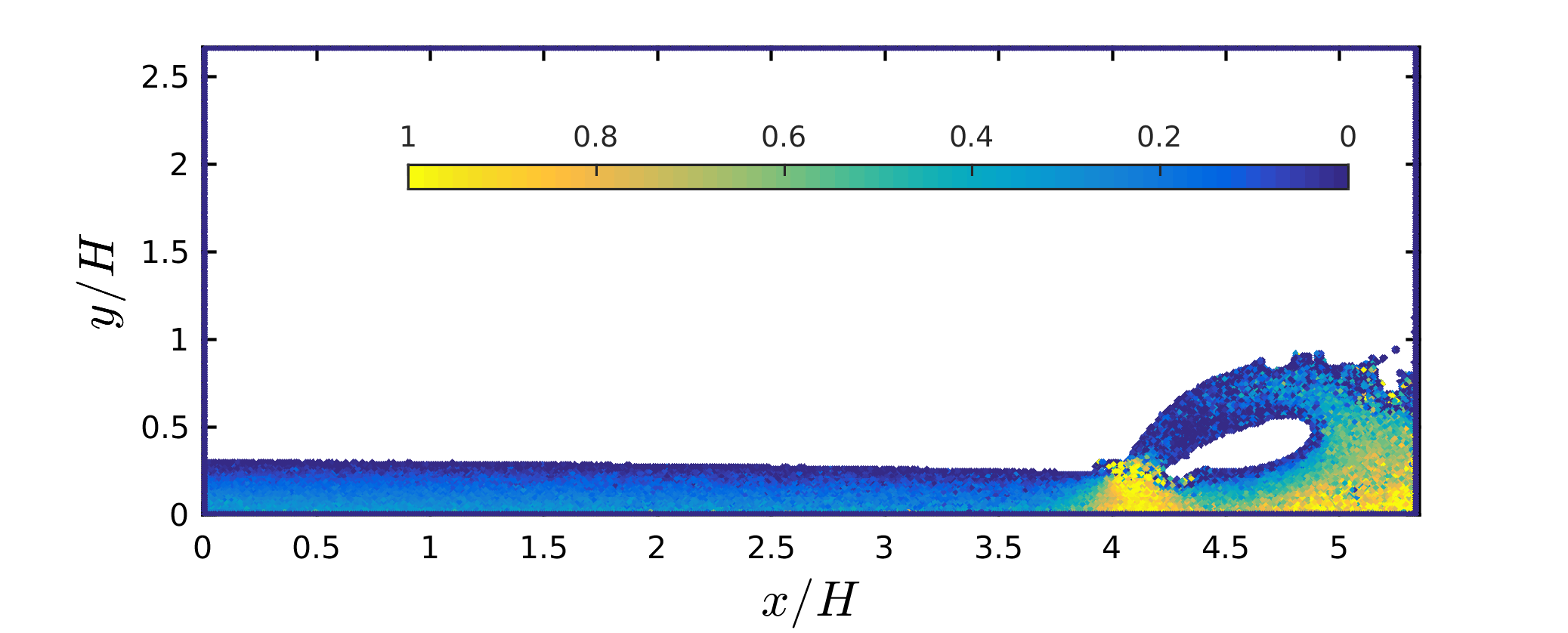}
	\end{subfigure}\qquad
	\begin{subfigure}[b]{.4\textwidth}
		\includegraphics[width=\textwidth,height=4cm]{out110000cp.png}
		\vspace{0.1ex}
		
		\includegraphics[width=\textwidth,height=4cm]{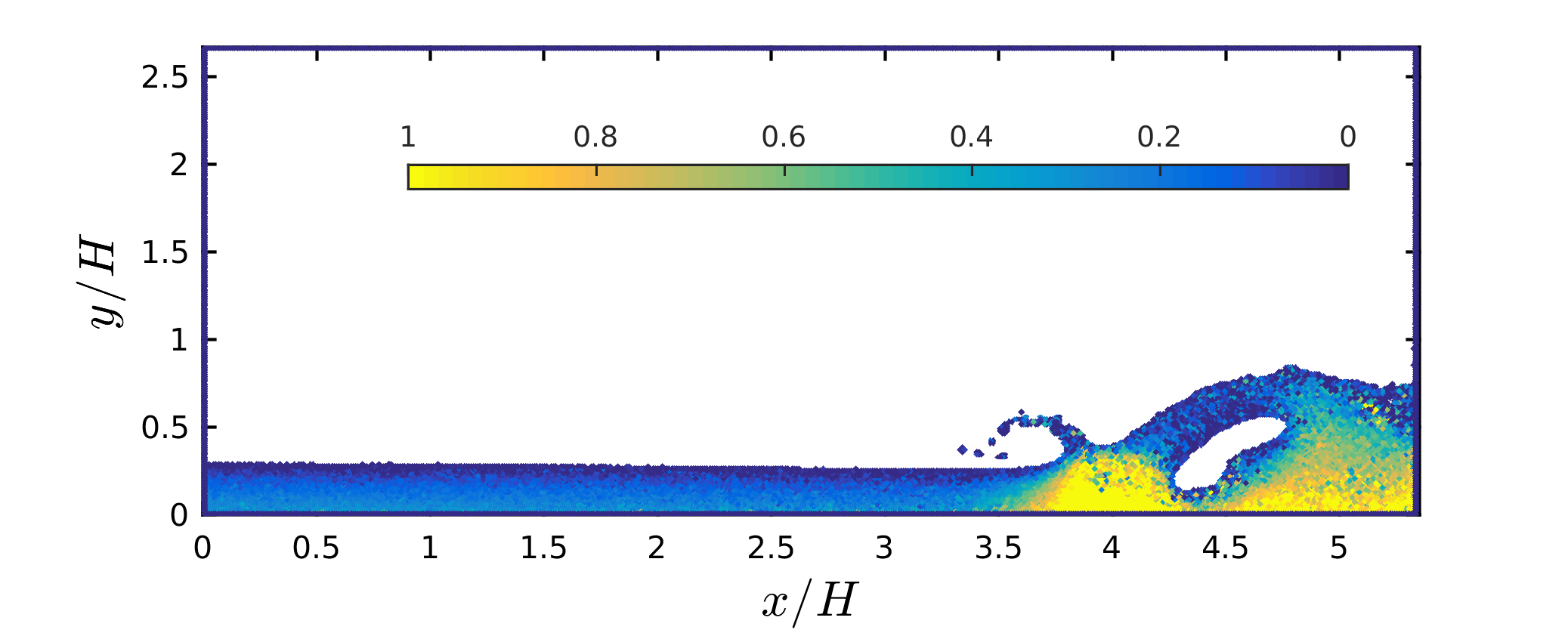}
	\end{subfigure}
	\caption[Dam break flow and impact against a vertical wall]{Dam break flow and impact against a vertical wall.Top left: pressure evolution on the wall; free-surface simulation by SPH$-i$ and experimental data from \cite{Lee2002}. Bottom: free surface-flow configurations corresponding to peaks at $t\sqrt{\mathnormal{g}/H}\simeq6.16$ and $t\sqrt{\mathnormal{g}/H}\simeq6.62$ in the pressure evolution left and right respectively. Color: non-dimensional pressure field $P/\rho\mathnormal{g}H$ at two time instants.}
	\label{fig:DryDam}
\end{figure}

The plots in figure \ref{fig:DryDam} show pressure plots obtained numerically by SPH$-i$ compared with experimental data \cite{Lee2002}. As intuitively expected \cite{Colagrossi2003}, when the moving front hits the vertical wall on the right, an impulse in the pressure is recorded at non-dimensional time $t\sqrt{\mathnormal{g}/H}\simeq2.4$ and the numerical simulation by SPH$-i$ recovers the measurements relatively well.   

\begin{figure}
	\centering
	\begin{subfigure}[b]{.4\textwidth}
		\includegraphics[width=\textwidth,height=4cm]{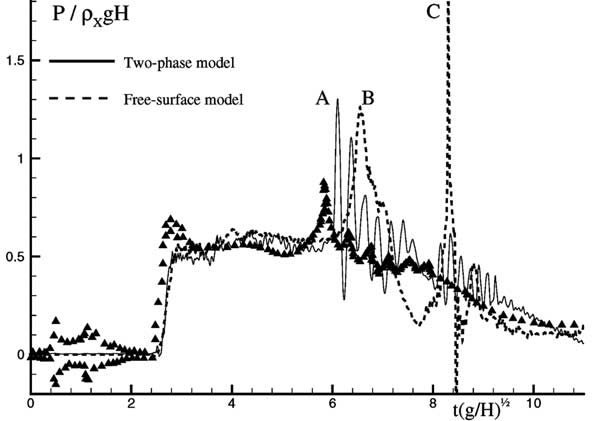}
		\vspace{0.1ex}
		
		\includegraphics[width=\textwidth,height=4cm]{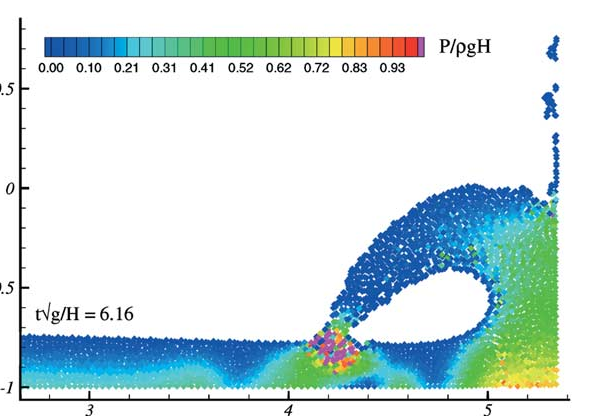}
	\end{subfigure}\qquad
	\begin{subfigure}[b]{.4\textwidth}
		\includegraphics[width=\textwidth,height=4cm]{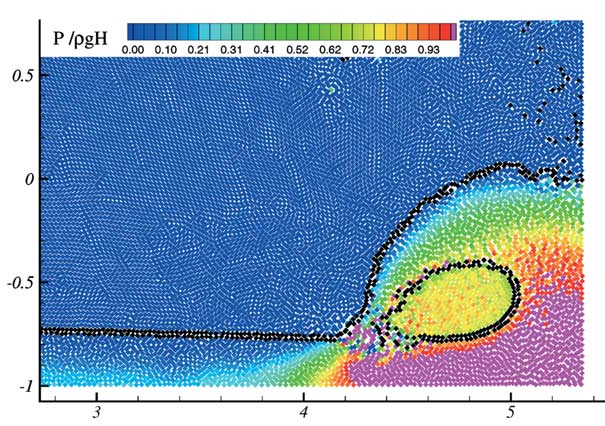}
		\vspace{0.1ex}
		
		\includegraphics[width=\textwidth,height=4cm]{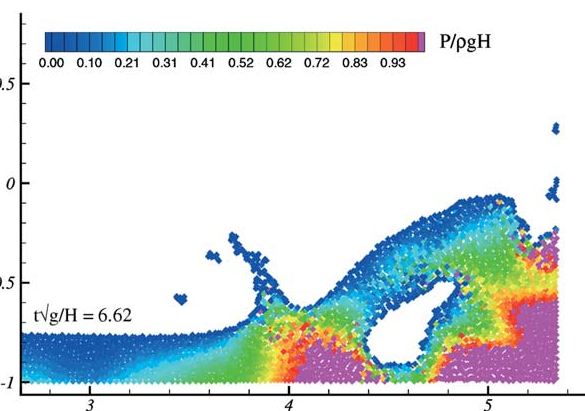}
	\end{subfigure}
	\caption[Dam break flow and impact against a vertical wall]{Dam break flow and impact against a vertical wall.Top left: pressure evolution on the wall; solid line: two-phase simulation; dashed line: free-surface simulation from \cite{Colagrossi2003};($\bullet$) experiments from \cite{Zhou1999}. Top right: air water-flow configuration, $\rho_{Y}/\rho_{X}=0.001$, corresponding to the pressure peak A in the pressure evolution. Bottom: free surface-flow configurations corresponding to peaks B and C in the pressure evolution left and right respectively.}
	\label{fig:colagrossiSimul}
\end{figure}

All free -surface flow simulations with SPH$-i$ were single phase with density ratio $\rho_{Y}/\rho_{X}=0$. The entrapment of air in the developing cavity as the plunging jet impinges on the interface introduces an air-pressure field that differs from the free-surface case. Colagrossi and Landrini \cite{Colagrossi2003} performed numerical simulations with their free surface and two-phase SPH models (see figure \ref{fig:colagrossiSimul}) and the free surface simulation by SPH$-i$ is shown in figure \ref{fig:DryDam}. 

The backward plunging jet induces a second peak and in the experiment this occurs at $t\sqrt{\mathnormal{g}/H}\simeq5.8$. This phenomena corresponds to the formation of a closed cavity and is here discussed by the pressure contours figure \ref{fig:colagrossiSimul} for $\rho_{Y}/\rho_{X}=0.001$ and $0$ (top-right and bottom-left plots, respectively) and in the case of the proposed SPH$-i$ in figure \ref{fig:DryDam} (top-right and bottom-left plots, respectively). In the free surface cases, the air-cushion effect is not observed which may account for the delayed pressure rise in the free surface flow cases since there is a fast circulatory flow around the entrapped cavity shown in the bottom-right plots of figures \ref{fig:colagrossiSimul} and \ref{fig:DryDam}.    

\begin{figure}
	\centering
	\begin{subfigure}[b]{.55\textwidth}
		\includegraphics[width=10cm]{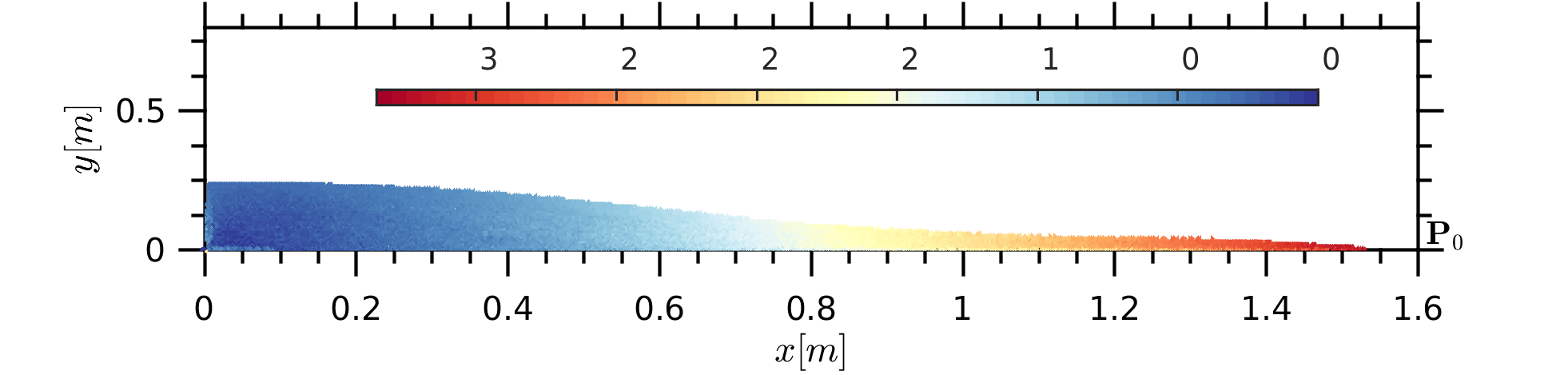}
		\caption{}
		\label{fig:damTongue}
	\end{subfigure}
	\begin{subfigure}[b]{.55\textwidth}
		\includegraphics[width=10cm]{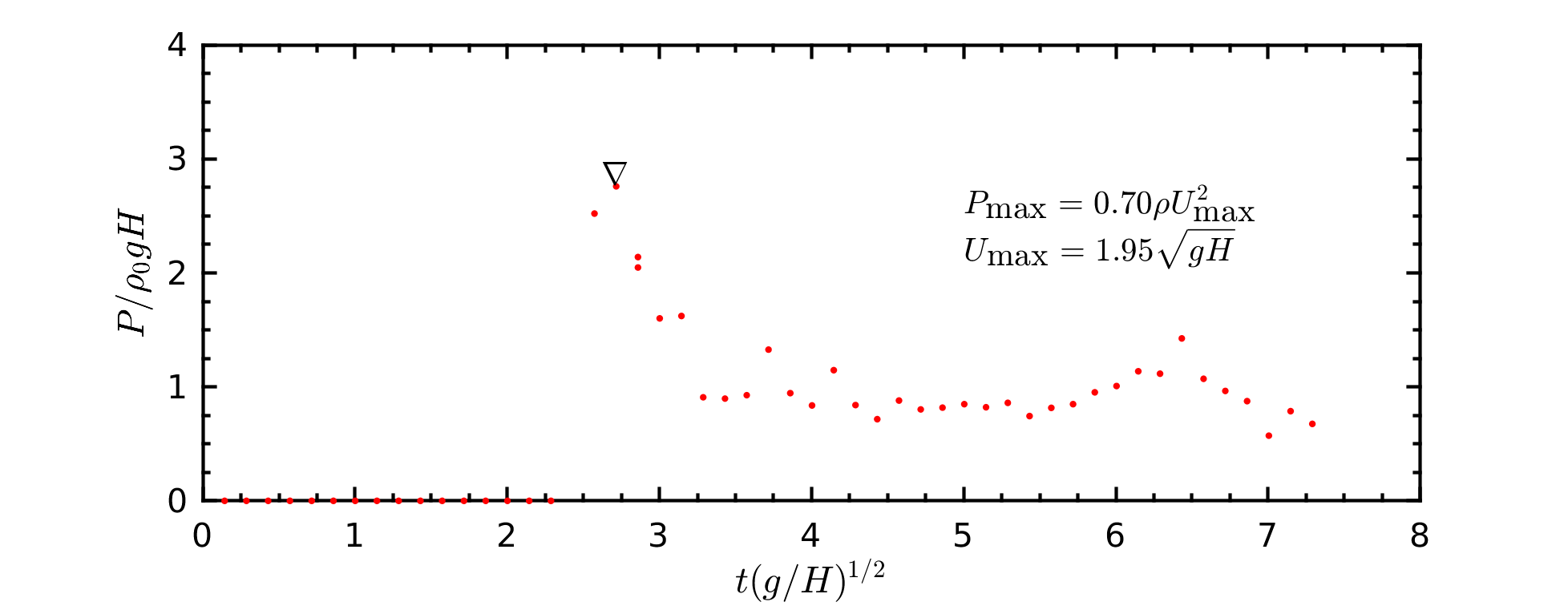}
		\caption{}
		\label{fig:cornerHist}
	\end{subfigure}%
	\caption[Maximum pressure in dam break flow]{(\ref{fig:damTongue}):Fluid flow just before impact against the vertical wall. Particles are colored using the magnitude of the velocity field. (\ref{fig:cornerHist}):shows the time history of SPH$-i$ pressure signals evaluated at the probe point $P_{0}$.}
	\label{fig:cornerPressure}
\end{figure}

Figure \ref{fig:damTongue} shows the wave front just before impact with the vertical wall. The angle between the free surface and the bottom boundary is small $\sim\ang{10}$. For such small angles, an asymptotic solution based on linear wave theory \cite{Marrone2011} can be used for validation. The wave front moves with a velocity of about $U_{\textup{max}}=1.95\sqrt{\mathnormal{g}H}$, the maximum pressure peak predicted by this theory is $P_{\textup{max}}=0.7\rho_{0}U_{\textup{max}}^{2}=2.67\rho_{0}\mathnormal{g}H$. To compare this value with the SPH$-i$, a pressure probe $P_{0}$ was placed at the bottom corner on the right wall. Figure \ref{fig:cornerHist} shows the pressure time history obtained from the SPH$-i$ model. It is clear that the maximum pressure computed from the SPH$-i$ model is very close to the asymptotic solution.       

\section{Model applications}
\subsection{Periodic wave breaking on a plane slope}
Modeling of breaking waves is an important topic in coastal and marine engineering; understanding the energetics of breaking waves is fundamental to predicting the damage caused by breaking waves, tsunamis etc. 

The type of wave profile converter that is proposed in this project is a rotating propeller that converts wave energy into rotational energy. A number of these rotating propellers are placed in the surf zone, near the wave breaking point. The fast forward current flow around the wave crest collides with wave energy converters (WECs), making the propellers to rotate for the duration of wave breaking. Once the wave crest has passed, the propellers stop rotating until the next breaking wave arrives and so the generated electric power is pulsed.
Under the uniform flow assumption, the accessible power of water within the plunging jet can be approximated by using the equation below.
\begin{eqnarray}
P=\frac{1}{2}\rho S\bar{u}^{3}\label{deq:2018k1}
\end{eqnarray}
where $\rho=1000~\textup{Kg}\textup{m}^{-3}$ is water density, $S$ is the surface area swept through by the propellers and $\bar{u}$ is the mean velocity of water. One characteristic feature of equation(\ref{deq:2018k1}) is that wave power is proportional to the cube of fluid velocity in a way similar to wind power. Since water has a high density, high wave power can be extracted especially with the fast flow around the crest.

The generated unregulated AC power is rectified into DC power which is then temporarily stored in a super-capacitor bank. By further converting the DC power into AC power using \textit{power inverters}, the stabilized AC line power is fed into the power grid. The power inverters offer dynamic reactive power control that helps maintain the reliability and integrity of the electric power grid. Regardless of whether there are ocean waves or not the power inverters will provide reactive power continuously. Therefore, installing these inverters on our small scale renewable energy storage system will improve voltage regulation. 
\begin{figure}
	\centering
	\includegraphics[width=10cm]{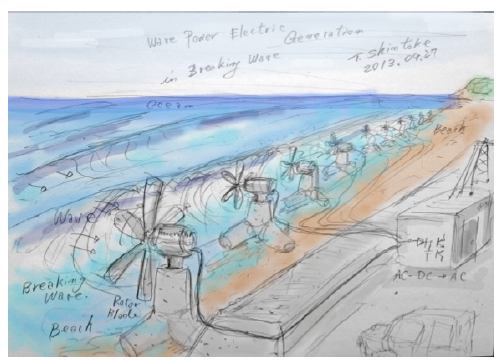}
	\caption[Wave Power Plant]{Wave Power Plant. Credit to my supervisor prof.Shintake for the schematic.}
	\label{wec:types}
\end{figure} 

Standard SPH is conceptually based on the same principle as explicit LES and therefore turbulence modeling is required. Turbulence modeling is of particular concern in modeling breaking wave phenomena. The first turbulence model for SPH was proposed by Gotoh \cite{Gotoh2001}. Lo and Shao \cite{Lo2002} first developed an ISPH-LES model while Dalrymple and Rogers\cite{Dalrymple2006}\cite{Rogers2004} further employed the SPH for breaking waves on a beach. A comprehensive review of the various turbulent models developed for SPH was conducted by Issa and Violeau \cite{Violeau2007}. They found the results to be generally satisfactory and that improvements are necessary by investigating the free-surface influence and wall conditions. Shao and Changming \cite{Songdong2006} devised a 2D SPH-LES model to investigate plunging waves. With their model they found the computations to be in good agreement with documented data. The computed turbulence quantities under breaking waves agreed better with experiments when compared with $k-\epsilon$ models. An important observation arising from their work is that both the turbulence model and the spatial resolution play a fundamental role in the model predictions; with sub-particle effects becoming less significant with particle refinement. 

To study the wave breaking phenomena, the SPH-$i$ model developed in this work will be used. Comparisons with experimental data will be made. 

\begin{figure}
	\centering
	\includegraphics[width=10cm]{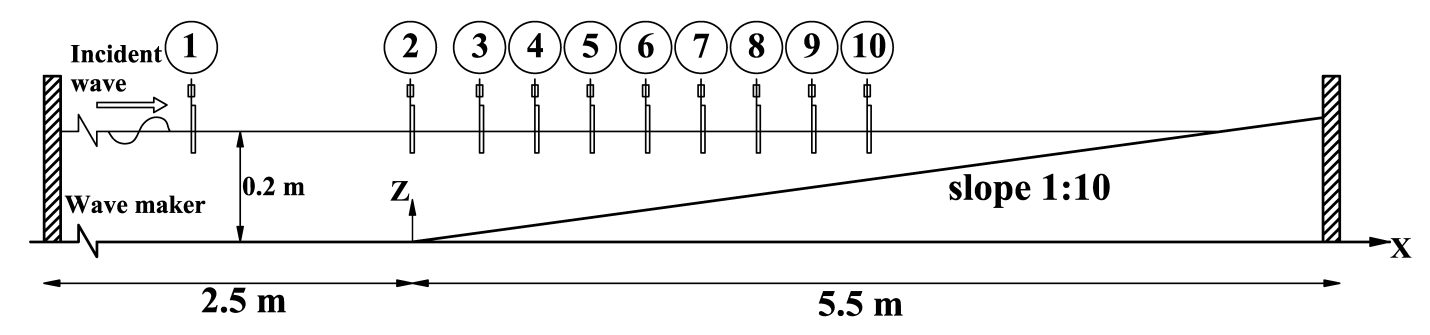}
	\caption[Breaking wave schematic diagram of experimental set-up]{Schematic diagram of experimental set-up courtesy of Mahmoudi et.al \cite{Mahmoudi2016}}
	\label{fig:waveExpt}
\end{figure}

When waves propagate in shallow water, they are influenced by shoaling effects due to the increase in wave height as water depth decreases. To investigate the breaking process, the experimental data from \cite{Mahmoudi2016} will be used for benchmarking. In their experiment the wave propagation breaking process was recorded using a high speed camera placed normal to the glass walls of the wave flume. A schematic of the experiment set-up is shown figure \ref{fig:waveExpt}.
where $H_{0}$ is the wave height in deep water, $H_{b}$ is the wave height at the break point i.e. edge of the surf zone and $(\xi_{0},\xi_{b})$ is the iribarren number. 

There are three types of breaking waves; spilling, plunging and surging. The iribarren number, defined below, is used for classifying the breaking waves. These are summarized in table \ref{tab:iribarren}. In this research plunging and spilling cases of periodic waves breaking on a plane slope were simulated. The initial particle water depth $h_{0}=0.2$m was used in all cases.  The initial particle spacing was $dx=dy=0.005$m leading to a total number of fluid particles $N=27,160$. The average number of near neighbors was fixed at 91; chosen so as to minimize numerical dissipation attributable to filtering/de-filtering processes. The compact support radius in units of $h$ is then $\xi=\sqrt{N_{n}/4\pi}=2.69$. The smoothing length is thus $h=\xi\Delta x=0.00538$m.The stiffness and damping constants were fixed at $k^{s}=0.00589\textup{Nm}^{-1}$ and $k^{d}=00002912\textup{Nsm}^{-1}$.  

\begin{table}
	\caption{Wave breaker classification} 
	\label{tab:iribarren}
	\small 
	\centering 
	\begin{tabular}{lcr} 
		\toprule[\heavyrulewidth]\toprule[\heavyrulewidth]
		\textbf{Breaker type} & $\xi_{0}$-\textbf{Range} & $\xi_{b}$-\textbf{Range} \\ 
		\midrule
		collapsing & $\xi_{0}>3.3$ & $\xi_{b}>2.0$\\
		plunging & $0.5<\xi_{0}<3.3$ & $0.4<\xi_{b}<2.0$\\
		spilling & $\xi_{0}<0.5$ & $\xi_{b}<0.4$\\
		\bottomrule[\heavyrulewidth] 
	\end{tabular}
\end{table}

\begin{table}
	\caption[parameters for breaking wave test cases]{Parameter set-up for breaking wave test cases} 
	\label{tab:wave-param}
	\small 
	\centering 
	\begin{tabular}{lcccr} 
		\toprule[\heavyrulewidth]\toprule[\heavyrulewidth]
		\textbf{Simulation case} & \textbf{Wave height} & \textbf{Wave period}& \textbf{Stroke} & \textbf{Average Power}\\ 
		\midrule
		case 1 & $0.0664$ & $1.8$& $0.127$& $6.81$\\
		case 2 & $0.0758$ & $2.7$& $0.224$& $9.52$\\
		case 3 & $0.07$ & $1.14$& $0.08$& $6.25$\\
		\bottomrule[\heavyrulewidth] 
	\end{tabular}
\end{table}

\begin{align}
\xi_{0}&=\frac{\tan{\alpha}}{\sqrt{H_{0}/L_{0}}}\qquad\text{or}\qquad\xi_{b}=\frac{\tan{\alpha}}{\sqrt{H_{b}/L_{0}}}
\end{align}  

In SPH, there are two types of wavemakers are generally used; piston-type and flap-type. The piston wavemaker is used in this thesis. It is represented by a vertical column of boundary particle whose horizontal displacement, according to linear wave theory, is given by
\begin{align}
x(t)&=\frac{S}{2}\sin(\omega t)\label{deq:2018w1}
\end{align}
where $S$ is called the stroke i.e. the amplitude  of the paddle oscillation. The oscillation frequency of the piston should be identical to the frequency of the generated waves. This type of wavemaker will generate a wave of height $H$ provided that the stroke and wave height satisfy the following relation
\begin{align}
\frac{H}{S}&=2\frac{\cosh(2kh_{0})-1}{\sinh(2kh_{0})+2kh_{0}}\label{deq:2018w3}
\end{align} 
which is the transfer function of the wave paddle. Here $h_{0}$ is the local fluid depth at the wavemaker or deep water depth.
Fenton and McKee \cite{Fenton1990} derived an approximate equation for computing the wavelength of the generated waves
\begin{align}
L=L_{0}\bigg(\tanh\left(\frac{2\pi h_{0}}{L_{0}}\right)^{\frac{3}{4}}\bigg)^{\frac{2}{3}}\qquad L_{0}=\frac{\mathnormal{g}T^{2}}{2\pi}\label{deq:2018w4}
\end{align} 
in which $L_{0}$ is the deep water wavelength and $L$ is the shallow water wavelength. 
From extended practical experience with this project, it was found that the paddle motion given by (\ref{deq:2018w1}) could not lead to very stable simulations results. Therefore, for all test cases presented in this thesis the paddle motion was enforced according to the following 
\begin{align}
x(t)&=\frac{S}{2}\bigg(1-\cos(\omega t)\bigg)\label{deq:2018w2}
\end{align}

By prescribing $\{\textup{T}, \textup{h}_{0}, \textup{H}\}$, the wavemaker can be setup by solving equations (\ref{deq:2018w4}), (\ref{deq:2018w3}) and (\ref{deq:2018w2}). 
The speed of sound for determining the incompressibility modulus is obtained by $c_{0}=10\sqrt{\mathnormal{g}h_{0}}$ and the kinematic viscosity is taken as $\nu=1.0\times 10^{-6}\textup{m}^{2}/\textup{s}$, thermal diffusivity $\nu=0.015\textup{m}^{2}/\textup{s}$ and adiabatic index $\gamma=7$.
\begin{figure}
	\centering
	\begin{subfigure}[b]{0.5\textwidth}
		\includegraphics[width=0.75\linewidth]{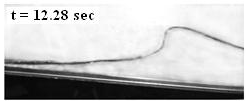}
		\caption[Breaking wave simulation]{}
		\label{fig:ewaves-1}
	\end{subfigure}%
	\begin{subfigure}[b]{0.5\textwidth}
		\includegraphics[width=0.85\linewidth]{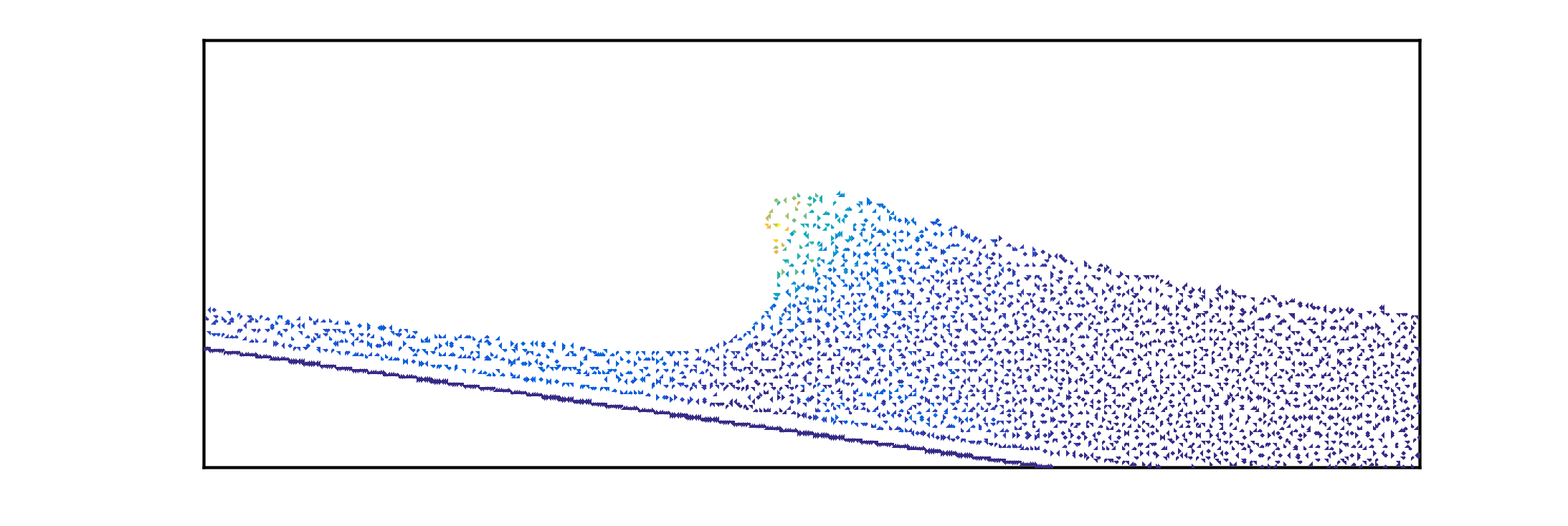}
		\caption[Breaking wave simulation]{$t=12.28\textup{s}$}
		\label{fig:waves-1}
	\end{subfigure}
	
	\begin{subfigure}[b]{0.5\textwidth}
		\includegraphics[width=0.85\linewidth]{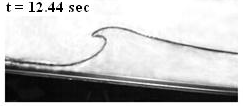}
		\caption[Breaking wave simulation.]{}
		\label{fig:ewaves-2}
	\end{subfigure}%
	\begin{subfigure}[b]{0.5\textwidth}
		\includegraphics[width=0.85\linewidth]{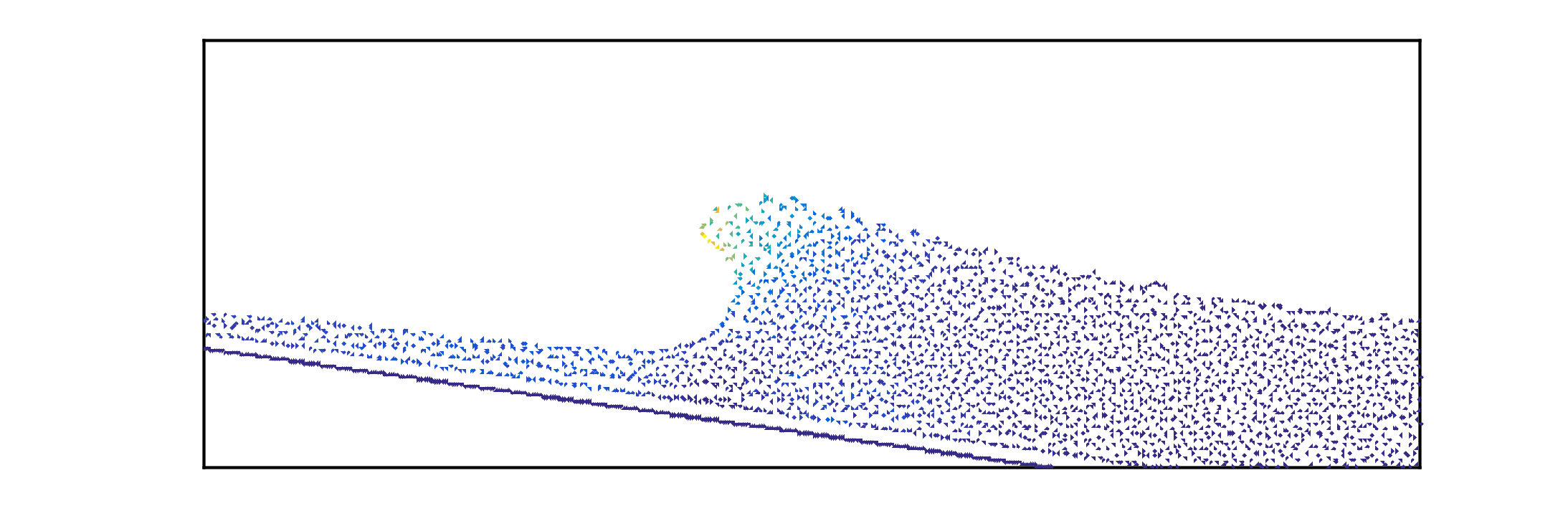}
		\caption[Breaking wave simulation.]{$t=12.44\textup{s}$}
		\label{fig:waves-2}
	\end{subfigure}
	
	\begin{subfigure}[b]{0.5\linewidth}
		\includegraphics[width=0.85\textwidth]{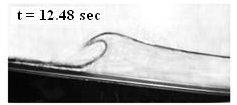}
		\caption[Breaking wave simulation.]{}
		\label{fig:ewaves-3}
	\end{subfigure}%
	\begin{subfigure}[b]{0.5\linewidth}
		\includegraphics[width=0.85\textwidth]{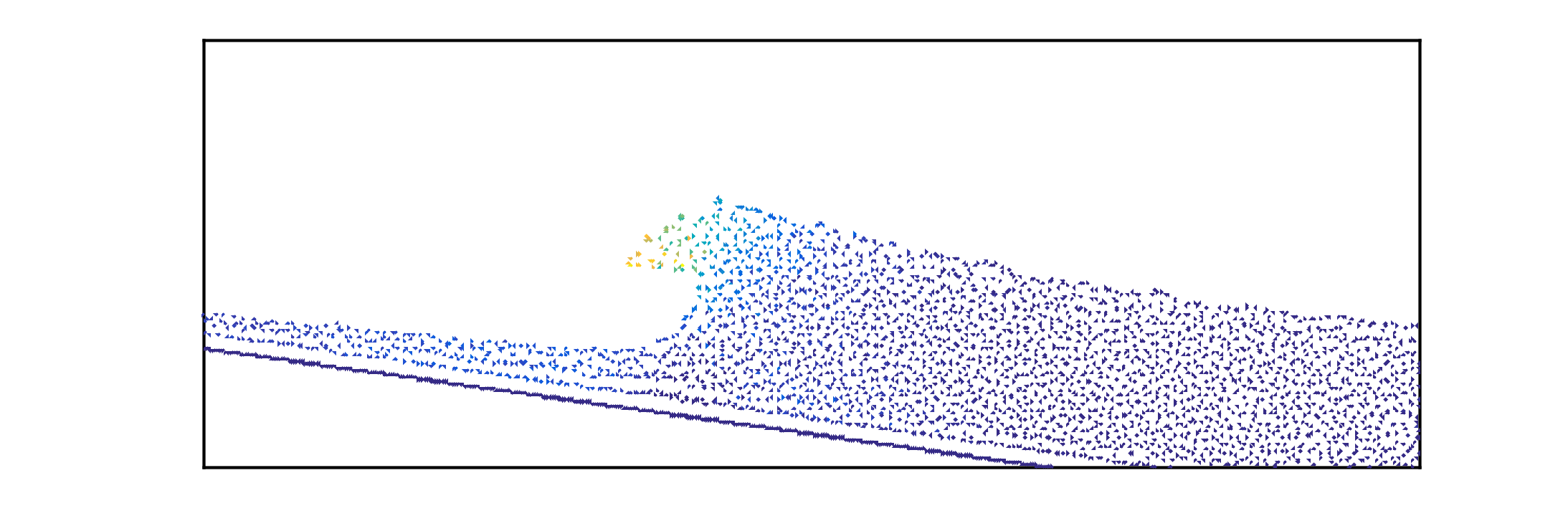}
		\caption[Breaking wave simulation.]{$t=12.48\textup{s}$}
		\label{fig:waves-3}
	\end{subfigure}

	\begin{subfigure}[b]{0.5\linewidth}
		\includegraphics[width=0.85\textwidth]{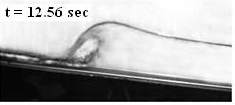}
		\caption[Breaking wave simulation.]{}
		\label{fig:ewaves-4}
	\end{subfigure}%
	\begin{subfigure}[b]{0.5\linewidth}
		\includegraphics[width=0.85\textwidth]{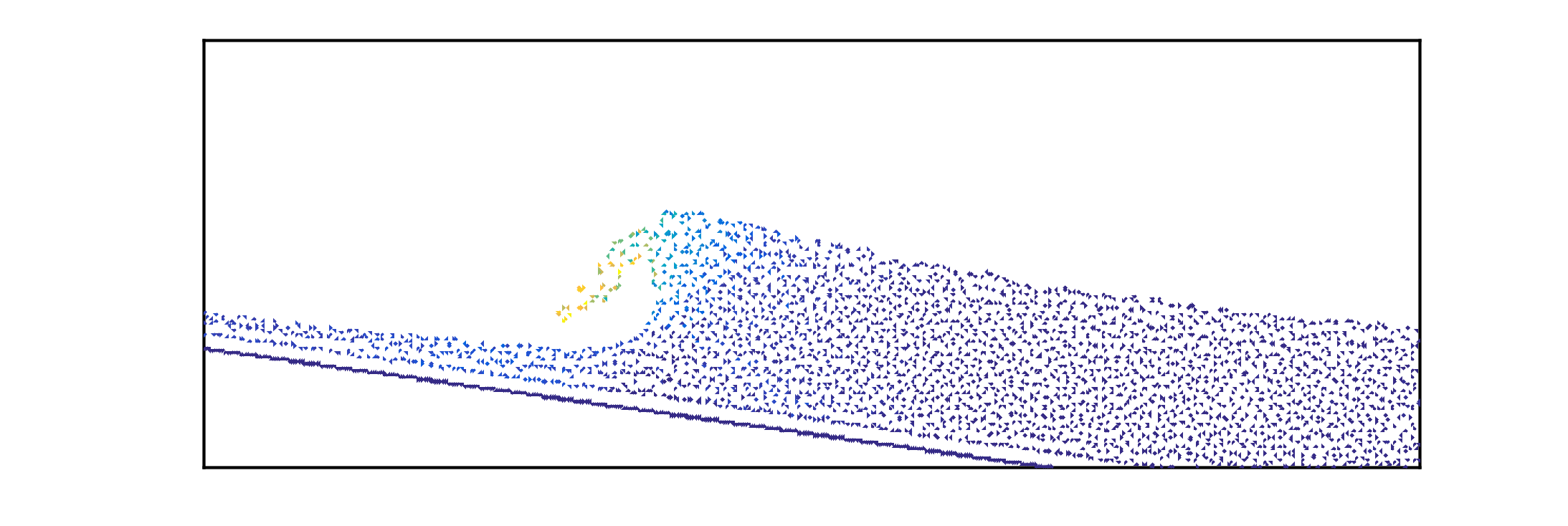}
		\caption[Breaking wave simulation.]{$t=12.56\textup{s}$}
		\label{fig:waves-4}
	\end{subfigure}
	
	\begin{subfigure}[b]{0.5\linewidth}
		\includegraphics[width=0.85\textwidth]{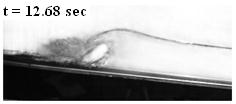}
		\caption[Breaking wave simulation.]{}
		\label{fig:ewaves-5}
	\end{subfigure}%
	\begin{subfigure}[b]{0.5\linewidth}
		\includegraphics[width=0.85\textwidth]{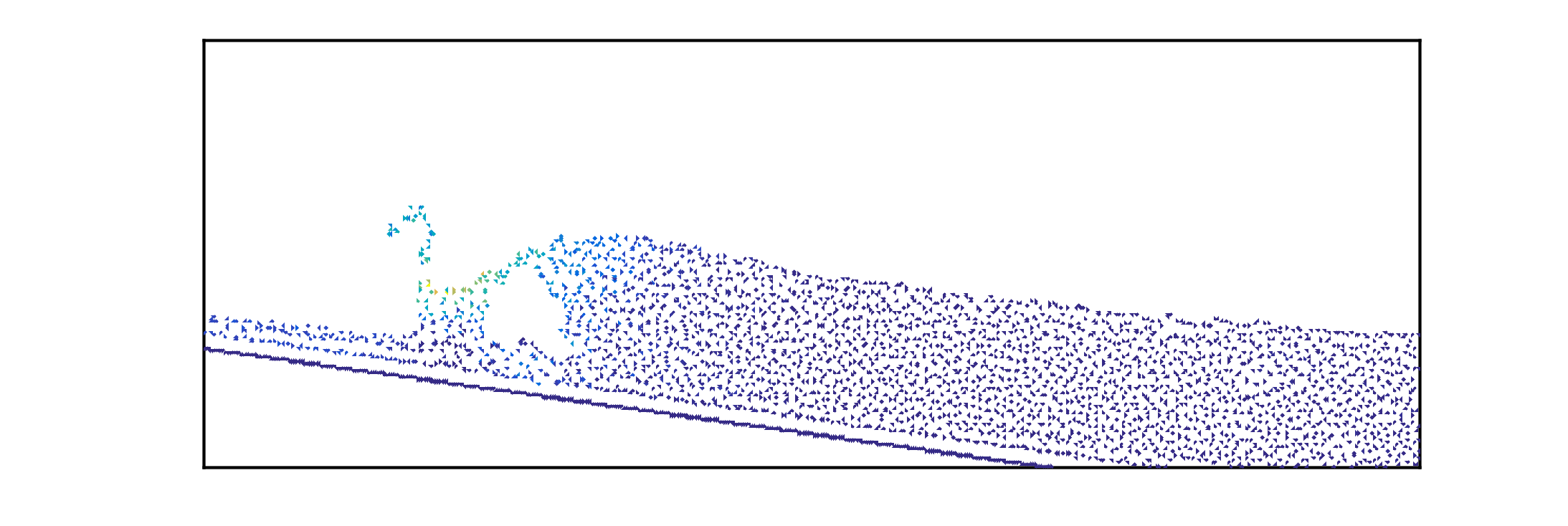}
		\caption[Breaking wave simulation.]{$t=12.68\textup{s}$}
		\label{fig:waves-5}
	\end{subfigure}
	
	\begin{subfigure}[b]{0.5\linewidth}
		\includegraphics[width=0.85\textwidth]{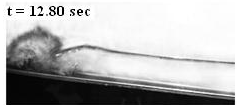}
		\caption[Breaking wave simulation.]{}
		\label{fig:ewaves-6}
	\end{subfigure}%
	\begin{subfigure}[b]{0.5\linewidth}
		\includegraphics[width=0.85\textwidth]{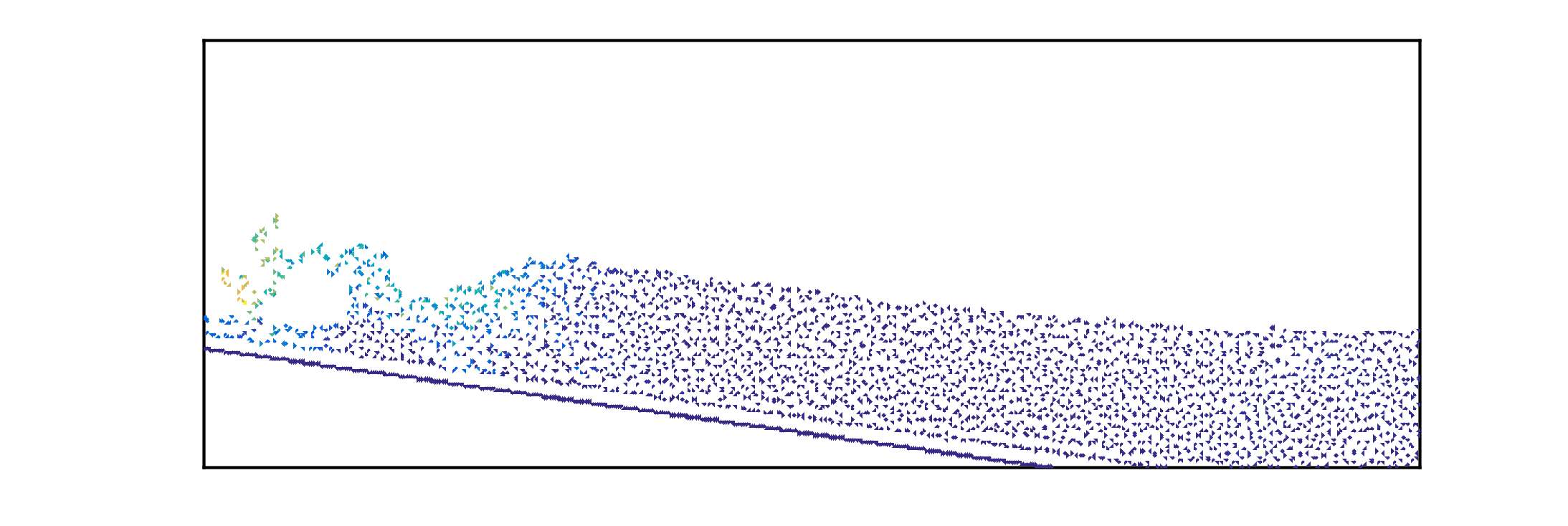}
		\caption[Breaking wave simulation.]{$t=12.80\textup{s}$}
		\label{fig:waves-6}
	\end{subfigure}
	
	\caption[Breaking wave dissipation: case 1]{Dissipation for a shallow water breaking wave. Kinetic energy density at eight time instants for case 1.}
	\label{fig:breakWave}
\end{figure}

\begin{figure}
	\centering
	\begin{subfigure}[b]{0.5\textwidth}
		\includegraphics[width=0.75\linewidth]{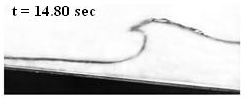}
		\caption[Breaking wave simulation]{}
		\label{fig:ewaves2-1}
	\end{subfigure}%
	\begin{subfigure}[b]{0.5\textwidth}
		\includegraphics[height=2.5cm,width=0.85\linewidth]{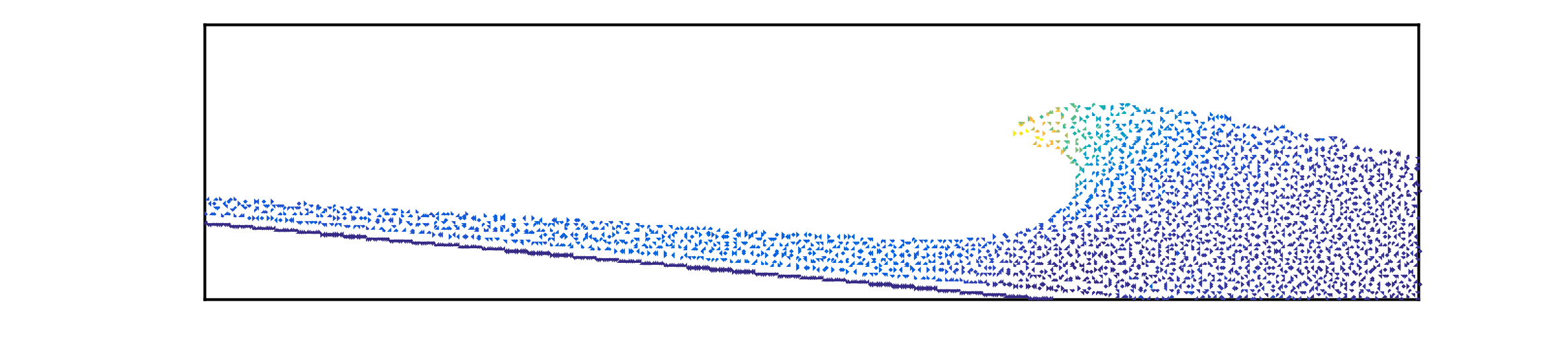}
		\caption[Breaking wave simulation]{$t=14.80\textup{s}$}
		\label{fig:waves2-1}
	\end{subfigure}
	
	\begin{subfigure}[b]{0.5\textwidth}
		\includegraphics[width=0.85\linewidth]{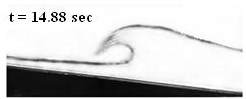}
		\caption[Breaking wave simulation.]{}
		\label{fig:ewaves2-2}
	\end{subfigure}%
	\begin{subfigure}[b]{0.5\textwidth}
		\includegraphics[height=2.5cm,width=0.85\linewidth]{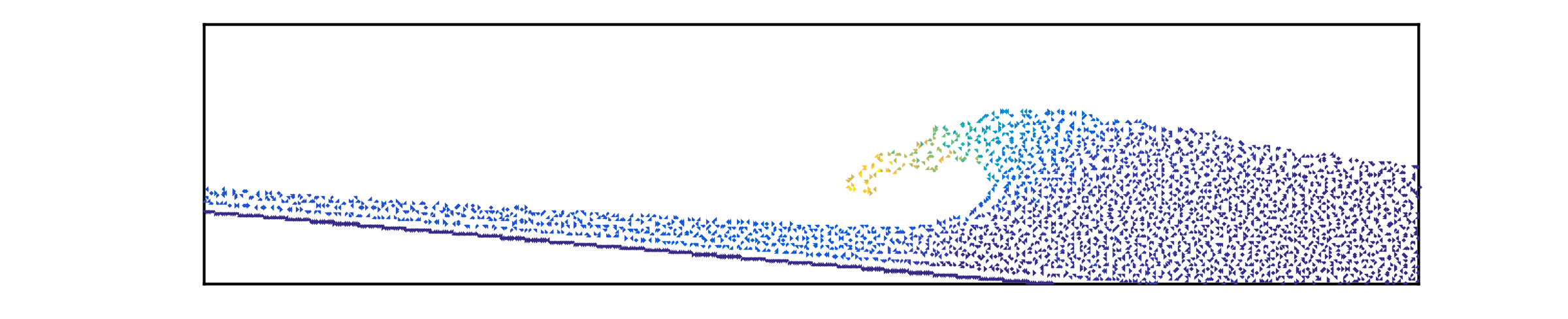}
		\caption[Breaking wave simulation.]{$t=14.88\textup{s}$}
		\label{fig:waves2-2}
	\end{subfigure}
	
	\begin{subfigure}[b]{0.5\linewidth}
		\includegraphics[width=0.85\textwidth]{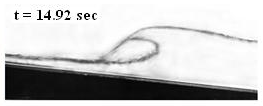}
		\caption[Breaking wave simulation.]{}
		\label{fig:ewaves2-3}
	\end{subfigure}%
	\begin{subfigure}[b]{0.5\linewidth}
		\includegraphics[height=2.5cm,width=0.85\textwidth]{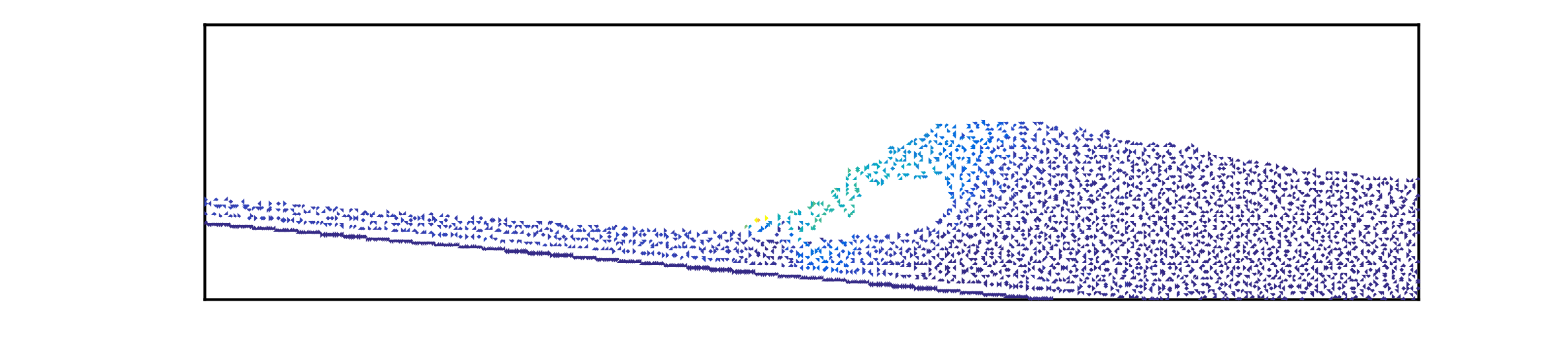}
		\caption[Breaking wave simulation.]{$t=14.92\textup{s}$}
		\label{fig:waves2-3}
	\end{subfigure}
	
	\begin{subfigure}[b]{0.5\linewidth}
		\includegraphics[width=0.85\textwidth]{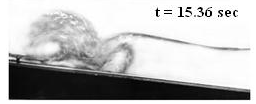}
		\caption[Breaking wave simulation.]{}
		\label{fig:ewaves2-4}
	\end{subfigure}%
	\begin{subfigure}[b]{0.5\linewidth}
		\includegraphics[height=2.5cm,width=0.85\textwidth]{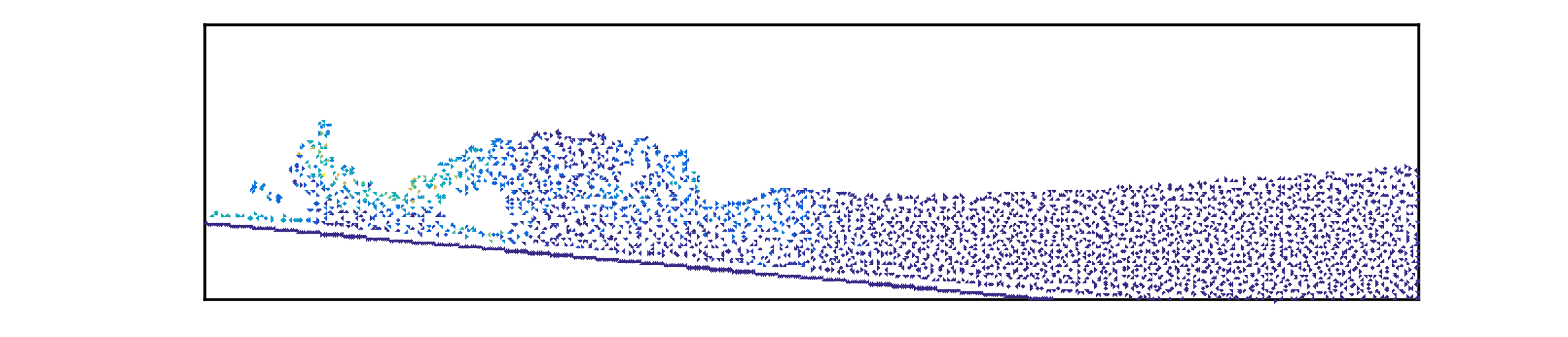}
		\caption[Breaking wave simulation.]{$t=15.36\textup{s}$}
		\label{fig:waves2-4}
	\end{subfigure}
	
	\caption[Breaking wave dissipation: case 2]{Dissipation for a shallow water breaking wave. Kinetic energy density at four time instants for case 2.}
	\label{fig:breakWave2}
\end{figure}

Figures \ref{fig:breakWave} and \ref{fig:breakWave} illustrates the plunging and splash-up processes of a periodic breaking wave whose physical parameters are respectively specified by case 1 and case 2 given in table \ref{tab:wave-param}. On the right-hand side are the snapshots taken using a high speed camera during laboratory experiments conducted by Mahmoudi et. al \cite{Mahmoudi2016} while the SPH-$i$ results are shown on the left hand side. 

When waves propagate on a slope, wave shoaling occurs due to decreased water depth. The wave front continuously steepens until breaking occurs. As the wave propagates with phase velocity $v_{p}=\sqrt{\mathnormal{g}H}$, until breaking when when the fluid velocity $\vert\vert\mathbf{u}\vert\vert$ exceeds the phase velocity i.e. $\vert\vert\mathbf{u}\vert\vert\geq v_{p}$.

The comparison between experimental and SPH-$i$ results shows that the SPH-$i$ model was able to successfully simulate the plunging breaking wave. In the initial set up of the breaker figures (\ref{fig:waves-1}),(\ref{fig:waves-2}) and (\ref{fig:waves-3}) is the on-set of turbulence. As the plunging jet hits the water surface, a splash-up process is generated; this leads to the formation of a turbulent bore that propagates towards the shore. This process is associated with high shear stress generated around the point of impact hence high turbulent kinetic energy is generated. Furthermore, as the wave undergoes breaking the kinetic energy of the fluid increases (see figure \ref{fig:breakWave}). It is this increase in kinetic energy that can be used to harness energy from the breaking wave.  

\begin{figure}[ht!]
	\centering
	\includegraphics[height=5cm,width=10cm]{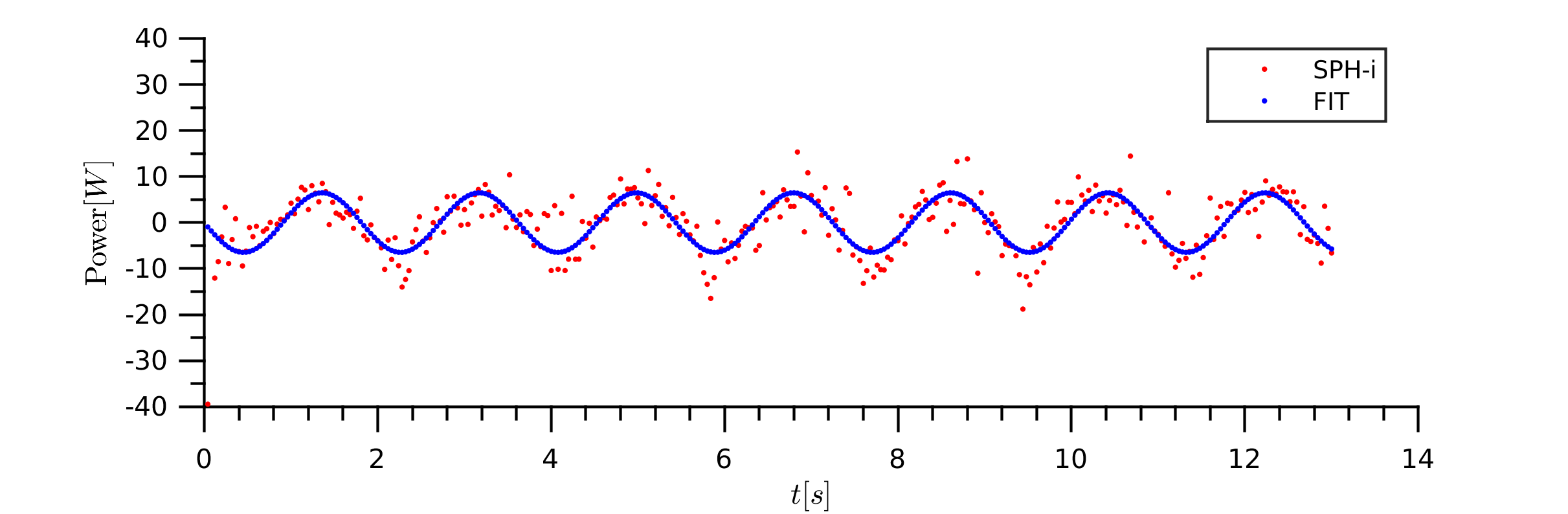}
	\caption{Instantaneous power delivered to fluid body $\Omega$}
	\label{fig:wavePower1}
\end{figure}

The average power required to generate the waves for case 1, as shown in table \ref{tab:wave-param} is $6.81\textup{W}$. Using equation (\ref{eqn:2018d3}), the instantaneous power delivered to the fluid body was computed and is shown in figure \ref{fig:wavePower1}. 

The instantaneous power is a sinusoidal signal with period of $T=1.8\textup{s}$ and amplitude of about $P_{m}=10\textup{W}$. Furthermore the average power over each half cycle can be computed as
\begin{align}
\Pb^{i}&=\frac{2}{T}\int_{0}^{\frac{T}{2}}P_{m}\sin\left(\frac{2\pi}{T}t\right)dt\\
&\approx 6.37\textup{W}\nonumber
\end{align}
which compared well with the value given in table \ref{tab:wave-param}

\subsubsection{Turbulent production and dissipation of breaking waves}
During the wave breaking process, wave power is dissipated due to various mechanisms including 
(i) turbulent dissipation
(ii) viscous dissipation and
(iii) boundary dissipation.
The density weighted turbulent kinetic energy $\widetilde{k}_{h}$ is defined to be half the trace of the SGS tensor (\ref{deq:2018d11}) and is thus
\begin{align}
\langle\rho_{h}(\mathbf{r})\rangle\widetilde{k}_{h}(\mathbf{r})&:=\frac{1}{2}\tr\bigg(\langle\underline{\underline{\mathcal{H}}}_{h}(\mathbf{r})\rangle\bigg)\nonumber\\
&=\frac{1}{2}\int_{\Omega(\mathbf{r})}\rho(\mathbf{r}^{\prime})\vert\vert\widetilde{\mathbf{u}}_{h}(\mathbf{r})-\mathbf{u}(\mathbf{r}^{\prime})\vert\vert^{2}w_{h}d\Omega(\mathbf{r}^{\prime})\label{deq:2018d11a}
\end{align}
It is the locally averaged kinetic energy per unit mass of the fluctuating velocity field.

By direct filtering of the energy conservation law (\ref{eq2016:2aa}), a sub-grid term called density weighted turbulent dissipation rate appears in the filtered energy balance equation. It is the rate at which turbulent kinetic energy is converted to thermal internal energy of the system. This is defined as
\begin{align}
\langle\rho_{h}(\mathbf{r})\rangle\widetilde{\varepsilon}_{h}(\mathbf{r})&=\int_{\Omega(\mathbf{r})}\rho(\mathbf{r}^{\prime})\nu\underline{\underline{\hat{\sigma}}}_{h}(\mathbf{r}^{\prime}):\nabla^{\prime}\hat{\mathbf{u}}_{h}(\mathbf{r}^{\prime}) d\Omega(\mathbf{r}^{\prime})\label{deq:2018m1}
\end{align}
where $\nu^{\textup{eff}}$ is the effective kinematic viscosity and the small-scale dissipation function is given as
\begin{align}
\nu\underline{\underline{\hat{\sigma}}}_{h}(\mathbf{r}):\nabla\hat{\mathbf{u}}_{h}(\mathbf{r})&=-\int_{\Omega(\mathbf{r})}\nu^{\textup{eff}}\bigg(\frac{\langle\rho_{h}(\mathbf{r}^{\prime})\rangle}{\rho(\mathbf{r})}+\frac{\langle\rho_{h}(\mathbf{r})\rangle}{\rho(\mathbf{r}^{\prime})}\bigg)\vert\vert\widetilde{\mathbf{u}}_{h}(\mathbf{r})-\mathbf{u}(\mathbf{r}^{\prime})\vert\vert^{2}\frac{(\mathbf{r}-\mathbf{r}^{\prime})\cdot\nabla\varphi_{h}}{\vert\vert\mathbf{r}-\mathbf{r}^{\prime}\vert\vert^{2}}d\Omega(\mathbf{r}^{\prime})\label{deq:2018m2}
\end{align}

It is important to note that the turbulent dissipation function  satisfies the physical requirement that it be negative definite i.e. $\widetilde{\varepsilon}_{h}\leq0$. 

\begin{figure}
	\centering
	\includegraphics[width=14cm]{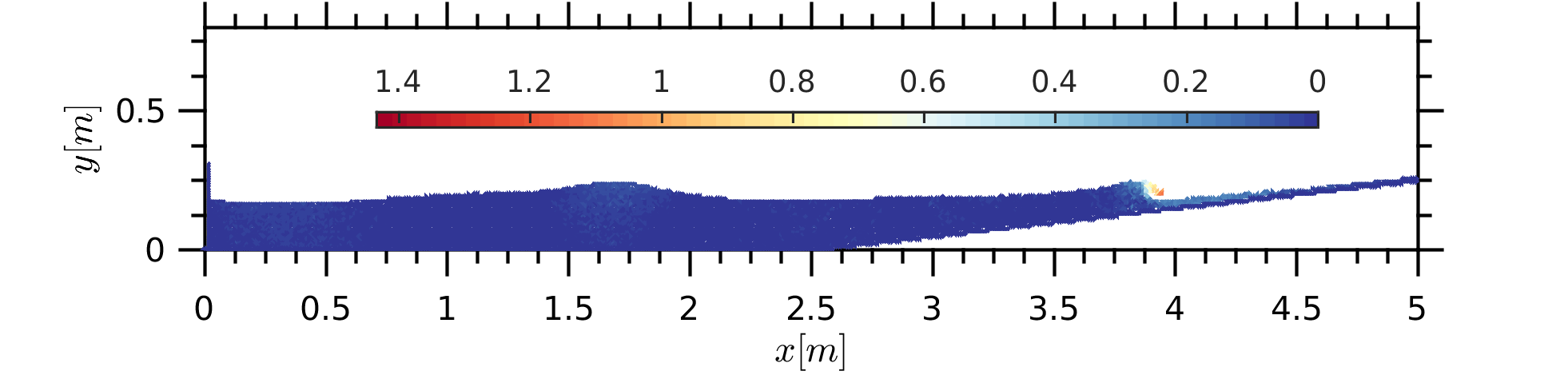}
	\caption[Kinetic Energy of Waves]{ Kinetic energy of breaking waves. Color: Specific kinetic energy of each particle $\frac{1}{2}\vert\vert\mathbf{u}_{i}\vert\vert^{2}\stackrel{\textup{D}}{\sim}\textup{m}^{2}\textup{s}^{-2}$ at time $t=5.68\textup{s}$.}
	\label{fig:SpecificKineticEnergy}
\end{figure}

\begin{figure}
	\centering
	\includegraphics[width=14cm]{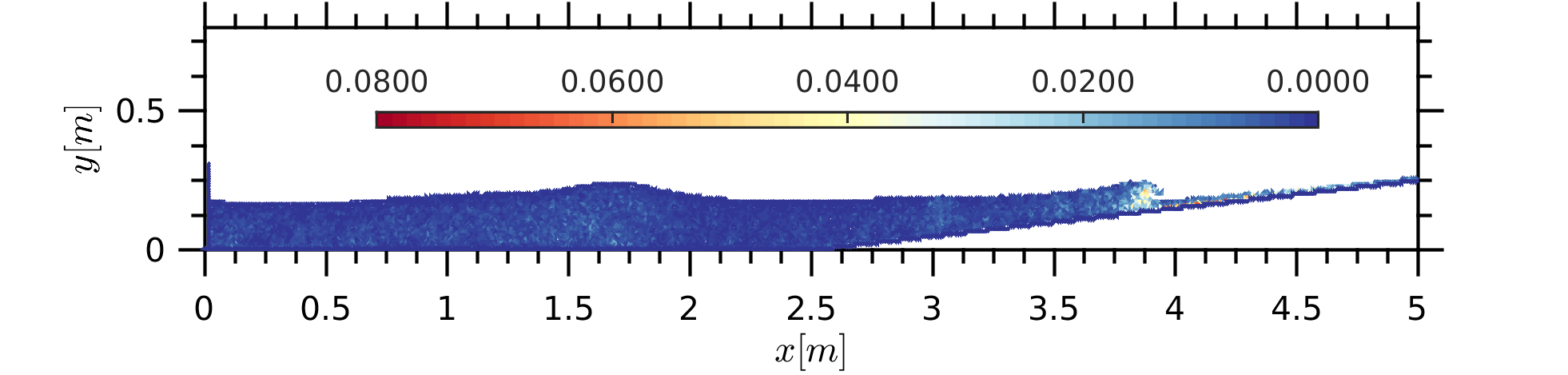}
	\caption[Turbulent kinetic energy]{ Turbulent kinetic energy production of breaking waves. Color: density weighted turbulent kinetic energy $\widetilde{k}_{h}\stackrel{\textup{D}}{\sim}\textup{m}^{2}\textup{s}^{-2}$ at time $t=5.68\textup{s}$.}
	\label{fig:TKE}
\end{figure}

\begin{figure}
	\centering
	\includegraphics[width=14cm]{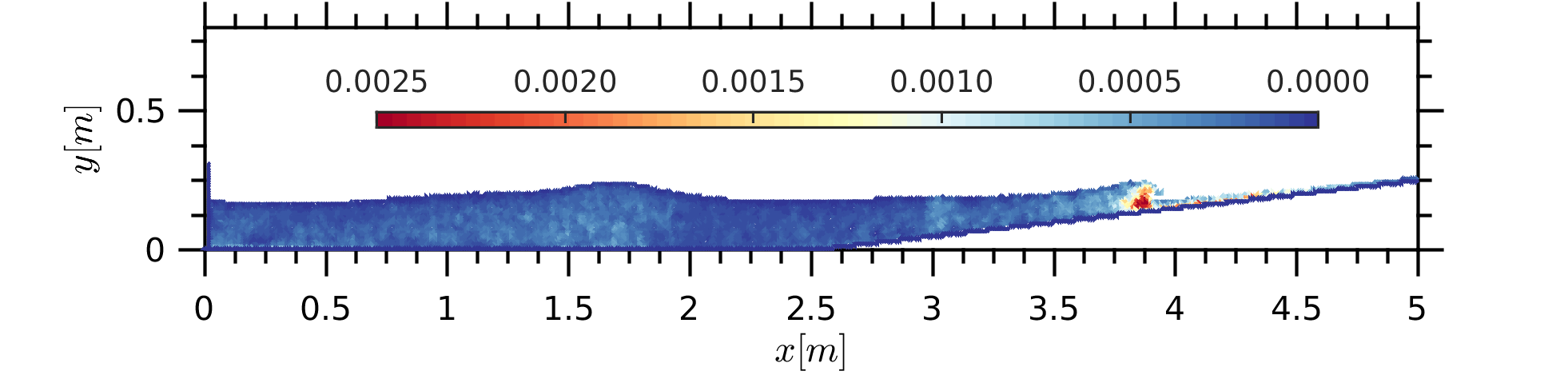}
	\caption[Turbulent dissipation]{ Turbulent dissipation dissipation of breaking waves. Color: density weighted local turbulent dissipation rate average $\widetilde{\varepsilon}_{h}\stackrel{\textup{D}}{\sim}\textup{m}^{2}\textup{s}^{-1}$ at time $t=5.68\textup{s}$.}
	\label{fig:TDR}
\end{figure}

Figure \ref{fig:SpecificKineticEnergy} depicts periodic waves undergoing the wave breaking process on a plane slope and the associated specific kinetic energy $\frac{1}{2}\vert\vert\mathbf{u}_{i}\vert\vert^{2}$ of each fluid particle at time $t=5.68\textup{s}$. As the wave approaches the breaking point, fluid velocity $\mathbf{u}_{i}$ approaches and exceeds the phase velocity $c=\sqrt{\mathnormal{g}H}$. During this phase, the  kinetic energy of the fluid increases, particularly around plunging jet. During this energy transformation process, an increase in kinetic energy  is accompanied by an in increase in turbulent kinetic average $\widetilde{k}_{h}$ given by equation (\ref{deq:2018d11a}), as shown in figure \ref{fig:TKE}. However, as soon as the turbulent kinetic energy is produced, it is quickly dissipated. Figure (\ref{fig:TDR}) shows the turbulent dissipation rate, computed from equation (\ref{deq:2018m1}). As the wave approaches the breaking point, some of the wave energy is transformed into turbulent kinetic energy.


\subsubsection{Viscous dissipation of breaking waves}
Another important dissipation mechanism for a shallow water breaking wave is viscous dissipation.  
\begin{figure}
	\centering
	\includegraphics[width=14cm]{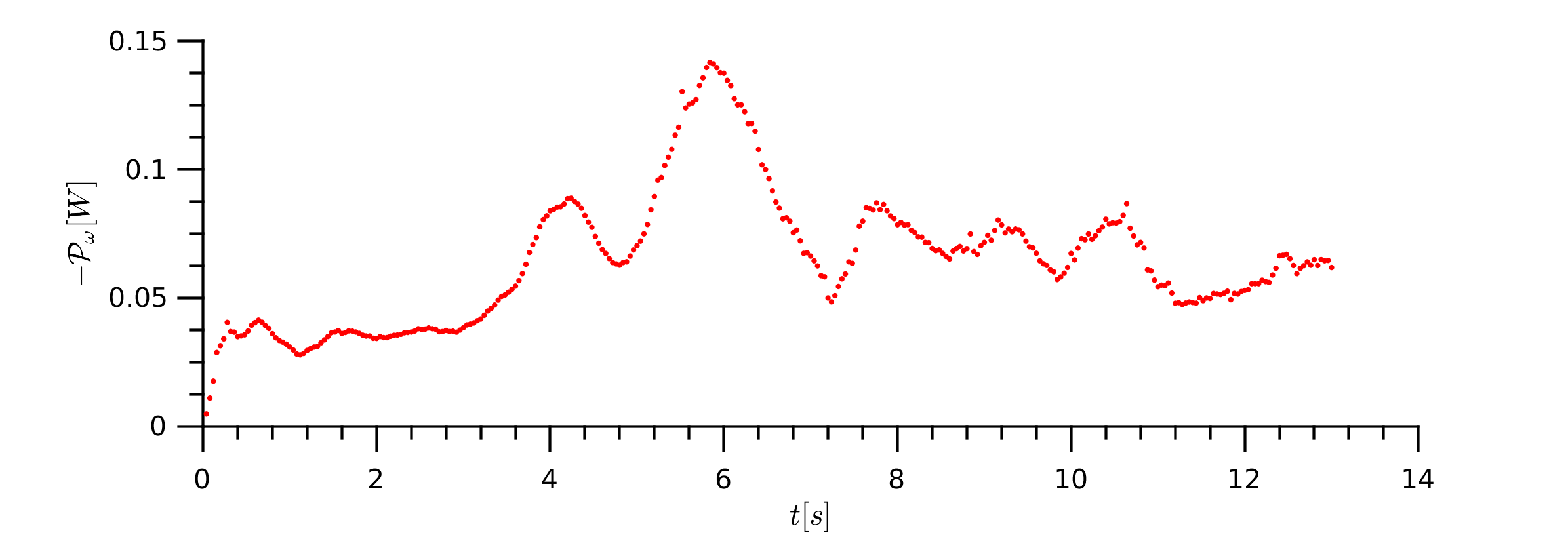}
	\caption[Viscous dissipation]{Viscous  dissipation of shallow breaking waves. Time history of the power $\Pb_{\omega}$.}
	\label{fig:viscDissp}
\end{figure} 
Figure \ref{fig:viscDissp} shows the time history of the power due to enstrophy $\Pb_{\omega}$ and is associated to the vorticity. Until time $t=2.8\textup{s}$, the power $\Pb_{\omega}$ is negligible. At the plunging jet closure $t=5.8\textup{s}$, as the intensity of the vorticity field increases, the power term $\Pb_{\omega}$ is no longer negligible. The power term $\Pb_{\omega}$ attains its maximum value at the instant the plunging impinges on the free surface, as the cavity collapses. The effect of air entrapment in these cavities would have an effect on the intensity of the power dissipation and thus a two phase model must be considered for higher fidelity of the numerical solution.

\subsection{Mechanical power delivered to the surf zone}
In order to develop suitable engineering device for harnessing energy from breaking waves, it is important to study and quantify the amount of power available. 
For an arbitrary fluid body $\Omega$, the total power delivered to it can be computed as
\begin{align}
\Pb_{in}&=\int_{\Omega}\rho\mathbf{b}\cdot\mathbf{u}d\Omega+\int_{\partial\Omega}\underline{\underline{\tau}}:\mathbf{u}\otimes\hat{\mathbf{n}}dS-\int_{\partial\Omega}\mathbf{q}\cdot\hat{\mathbf{n}}dS\nonumber\\
&=\int_{\Omega}\rho\mathbf{b}\cdot\mathbf{u}d\Omega+\int_{\Omega}\nabla\cdot\left(\underline{\underline{\tau}}\cdot\mathbf{u}\right)d\Omega+\int_{\Omega}\nabla\cdot\left(k\nabla T\right)d\Omega\label{eqn:2018d3}
\end{align}
assuming Fourier's heat conduction law $\mathbf{q}:=-k\bm{\nabla} T$ holds. A derivation of this follows from the laws of conservation of energy and momentum. 

\begin{figure}
	\centering
	\begin{subfigure}[b]{.85\textwidth}
		\includegraphics[width=12cm]{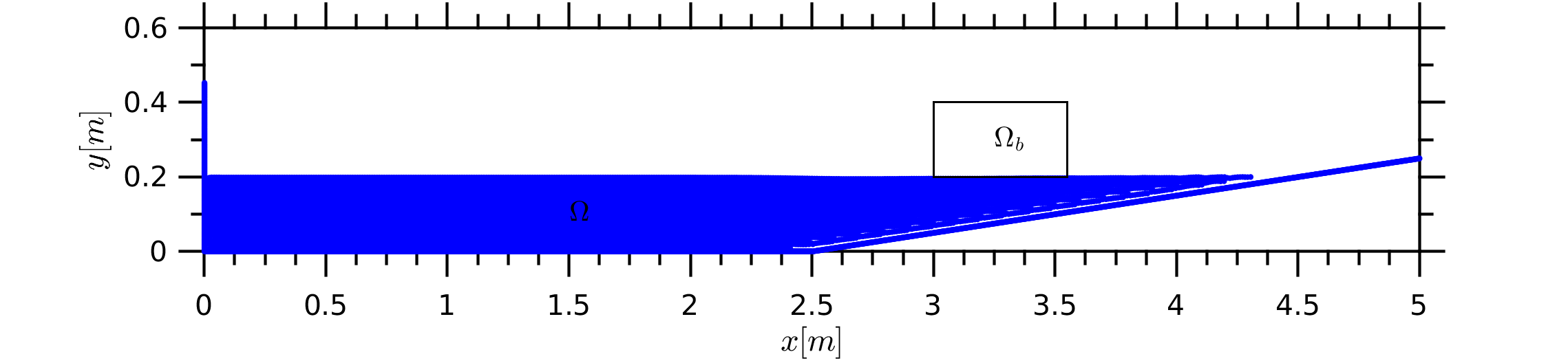}
		\caption{}
		\label{fig:surfCV}
	\end{subfigure}
	\begin{subfigure}[b]{.85\textwidth}
		\includegraphics[width=12cm]{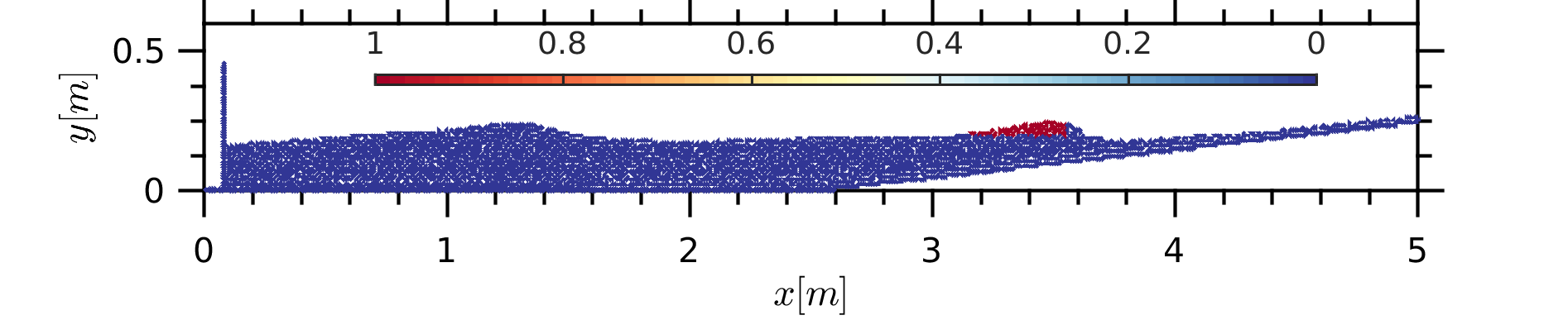}
		\caption{}
		\label{fig:surfWave}
	\end{subfigure}%
	\caption[Breaking wave Power transport]{(\ref{fig:surfCV}):Numerical simulation set up for case 1 with fluid domain $\Omega$ and surf zone control volume $\Omega_{b}$.  (\ref{fig:cornerHist}):shows the breaking wave at time $t=4.88\textup{s}$. Particles are colored using a tag $\psi=1$ if a fluid particle is in $\Omega_{b}$ and a tag of $\psi=0$ otherwise.}
	\label{fig:transportPower}
\end{figure}

For free surface flows, the power due to viscous forces i.e. the second term in (\ref{eqn:2018d3}) is dominantly expended in the formation of vortex motion and by carefully decomposing this term, it can be shown to be given by
\begin{align}
\Pb_{\omega}&=-2\mu\int_{\Omega}\frac{1}{2}\vert\vert\bm{\omega}\vert\vert^{2}dV\label{deq:2018f3}
\end{align} 
where $\omega:=\bm{\nabla\times\mathbf{u}}$ is the vorticity vector.
For applications in shallow water, wave energy can be extracted from breaking waves in the surf zone. Figure \ref{fig:surfCV} shows a schematic diagram of the numerical wave flume for case 1. A much lower resolution of $dx=dy=0.01\textup{m}$ was used as the initial particle spacing of fluid particles on a rectangular grid. In the figure, $\Omega$ is the fluid domain whereas $\Omega_{b}$ is a control volume in the surf zone. The goal is to compute the instantaneous power $\Pb(\Omega)$ and $\Pb(\Omega_{b})$ delivered to the fluid body and the control volume, respectively.     
Figure \ref{fig:surfWave} shows a snapshot of the wave prior to breaking. To compute the power transfer to the control volume, a particle identifier $\psi$ was defined: if a fluid particles enters $\Omega_{b}$ then it attains a marker value $\psi=1$ otherwise it will retain the initial value of $\psi=0$. In figure \ref{fig:surfWave} the fluid particles are colored using the marker $\psi$.  

\begin{figure}
	\centering
	\begin{subfigure}[b]{0.85\textwidth}
		\includegraphics[width=12cm]{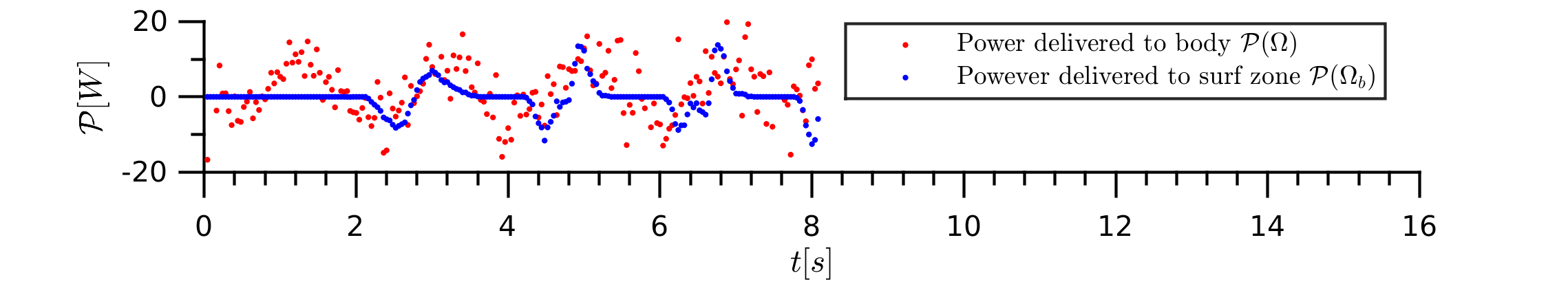}
		\caption[Breaking wave simulation]{}
		\label{fig:totalPower}
	\end{subfigure}
	\begin{subfigure}[b]{0.85\textwidth}
		\includegraphics[width=12cm]{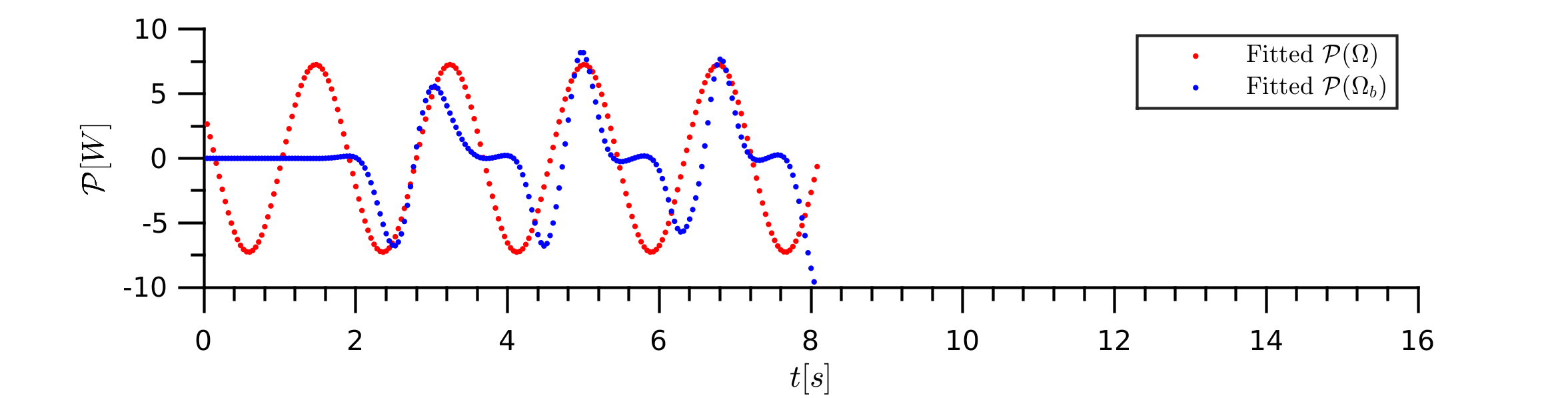}
		\caption[Breaking wave simulation.]{}
		\label{fig:TotalPowerFit}
	\end{subfigure}
	\caption[Instantaneous wave power ]{Instantaneous power for the breaking wave of case 1 at time $t=4.88\textup{s}$ where \ref{fig:totalPower} is SPH$-i$ instantaneous power and \ref{fig:TotalPowerFit} is the model fit using Matlab's fitting tools.}
	\label{fig:breakPowerInst}
\end{figure}

Figure \ref{fig:breakPowerInst} shows a plot of the instantaneous power terms $\Pb(\Omega)$ and $\Pb(\Omega_{b})$ delivered to the fluid body $\Omega$ and the control volume $\Omega_{b}$, respectively. Since the waves are generated at a wave period of $T=1.8\textup{s}$, the power $\Pb(\Omega)$ will be also be delivered to the fluid body $\Omega$ at this same period. Furthermore, power pulses will be registered in the control volume $\Omega_{b}$ whenever a wave is incident on  $\Omega_{b}$. Hence, the control volume power pulse $\Pb(\Omega_{b})$ will also have a period of $T=1.8\textup{s}$ as can be seen in figure \ref{fig:breakPowerInst}.     

Prior to breaking, dissipative mechanisms do not significantly dissipate wave power. However, during and after the breaking process much of the wave power will be dissipated via several mechanisms. 
\begin{figure}
	\centering
	\includegraphics[width=14cm]{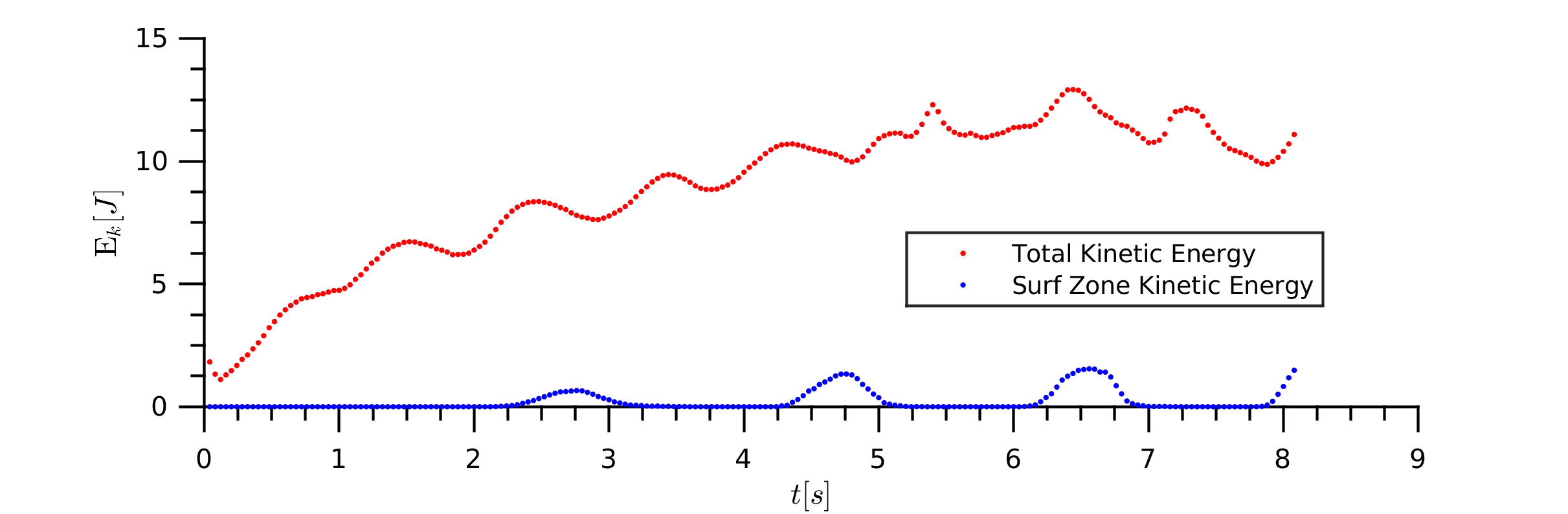}
	\caption[Kinetic energy of beaking waves]{Time history of the total kinetic energy $E(\Omega)$ and the kinetic energy $E(\Omega_{b})$ of waves in the control volume $\Omega_{b}$.}
	\label{fig:kinEnergy}
\end{figure}
Similarly, the pulses of the mechanical energy are registered is the control volume $\Omega_{b}$ at the same period as the incoming waves $T=1.8\textup{s}$. As the waves approach the break point, kinetic energy increases and there is a fast forward current around the plunging jet. Therefore, by placing a suitable wave energy converter (WEC) under a breaking wave, the wave power $\Pb(\Omega_{b})$ can be harnessed into electrical energy; directly in the case a WEC based on high efficiency blade technology. 

\subsection{Effect of artificial wave breaker}
\begin{figure}
	\centering
	\includegraphics[width=16cm]{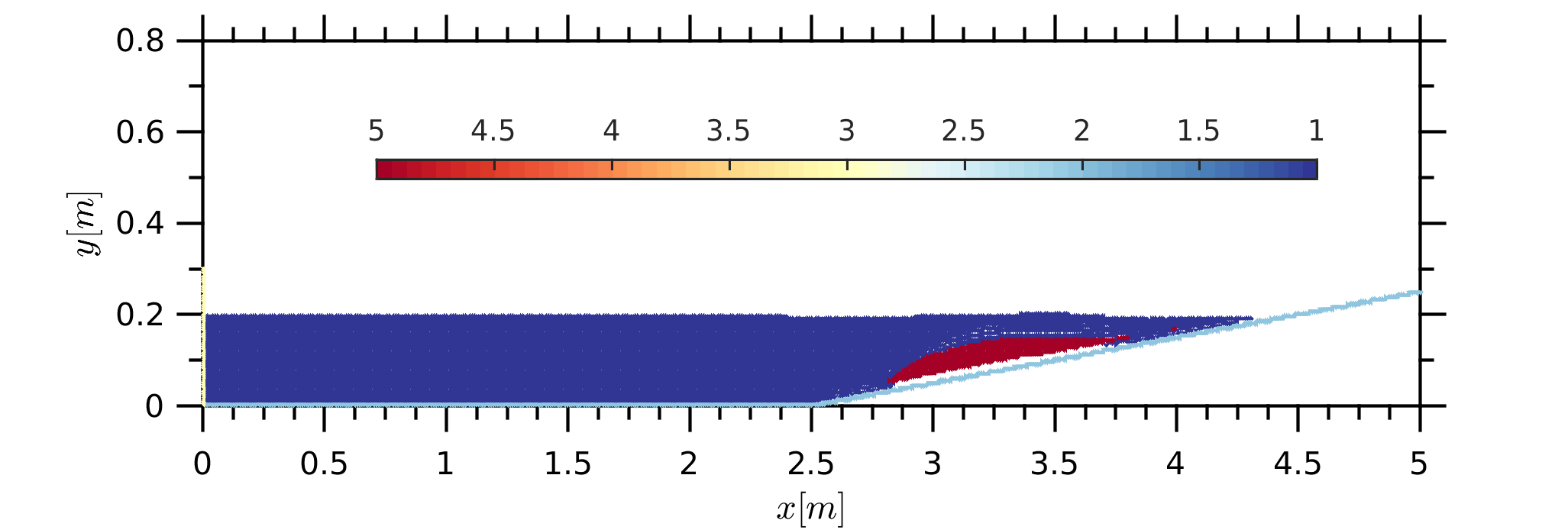}
	\caption[Initial particle setup for case 1 with smooth articfial breaker]{Simulation of case 1 with a smooth artificial breaker inserted in the topography. Particle color based on particle type: 1 for fluid particle, 2 for fixed boundary particle, 3 for moving boundary particle, 4 for ghost particle and 5 for solid particles.}
	\label{fig:SmoothBreaker}
\end{figure}

\begin{figure}
	\centering
	\begin{subfigure}[b]{0.5\textwidth}
		\includegraphics[height=3.0cm,width=0.85\linewidth]{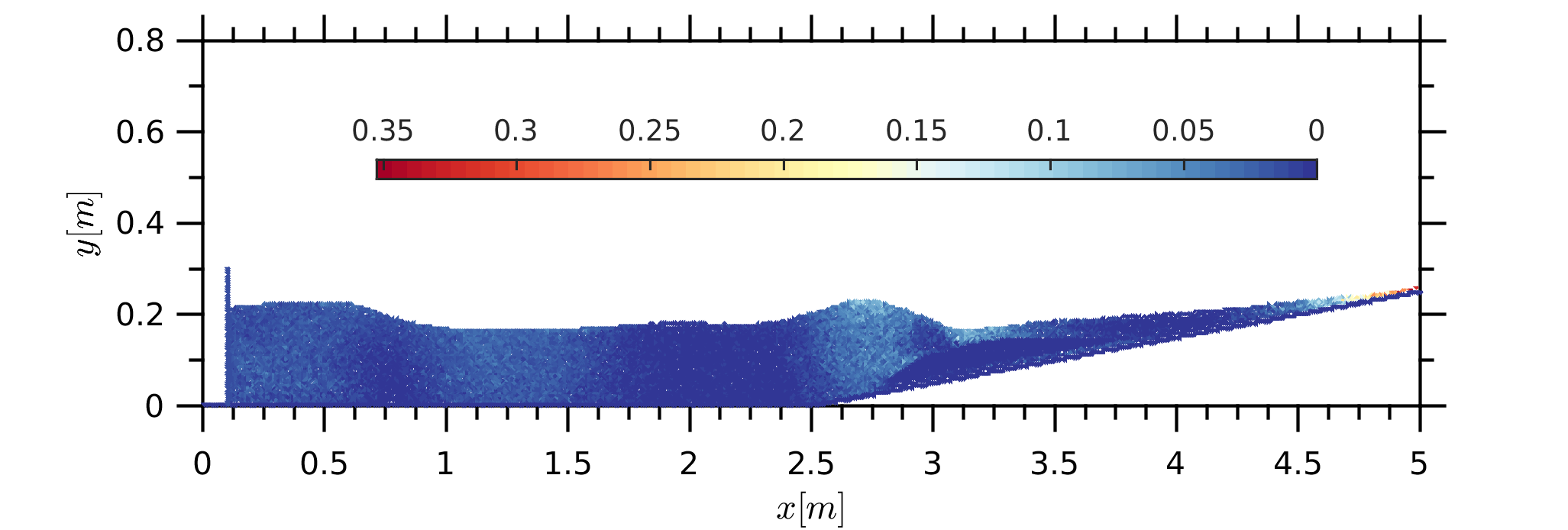}
		\caption[Breaking wave simulation]{$t=4.25\textup{2}$}
		\label{fig:surf1}
	\end{subfigure}%
	\begin{subfigure}[b]{0.5\textwidth}
		\includegraphics[height=3.0cm,width=0.85\linewidth]{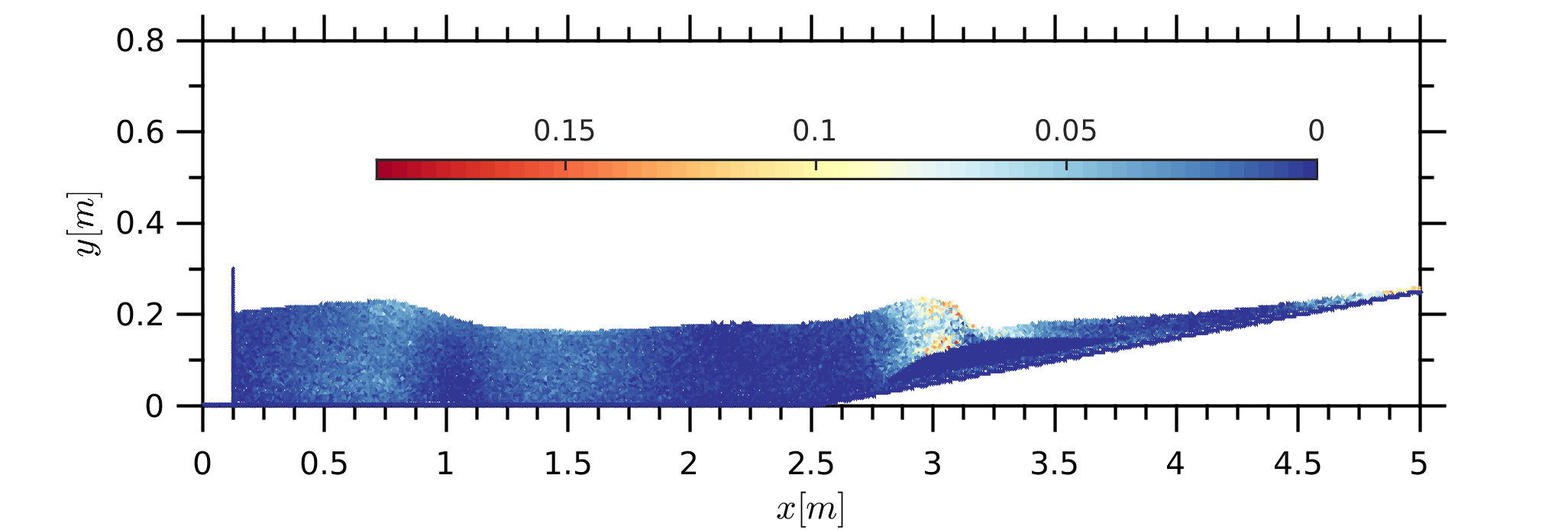}
		\caption[Breaking wave simulation]{$t=4.45\textup{2}$}
		\label{fig:surf2}
	\end{subfigure}
	
	\begin{subfigure}[b]{0.5\textwidth}
		\includegraphics[height=3.0cm,width=0.85\linewidth]{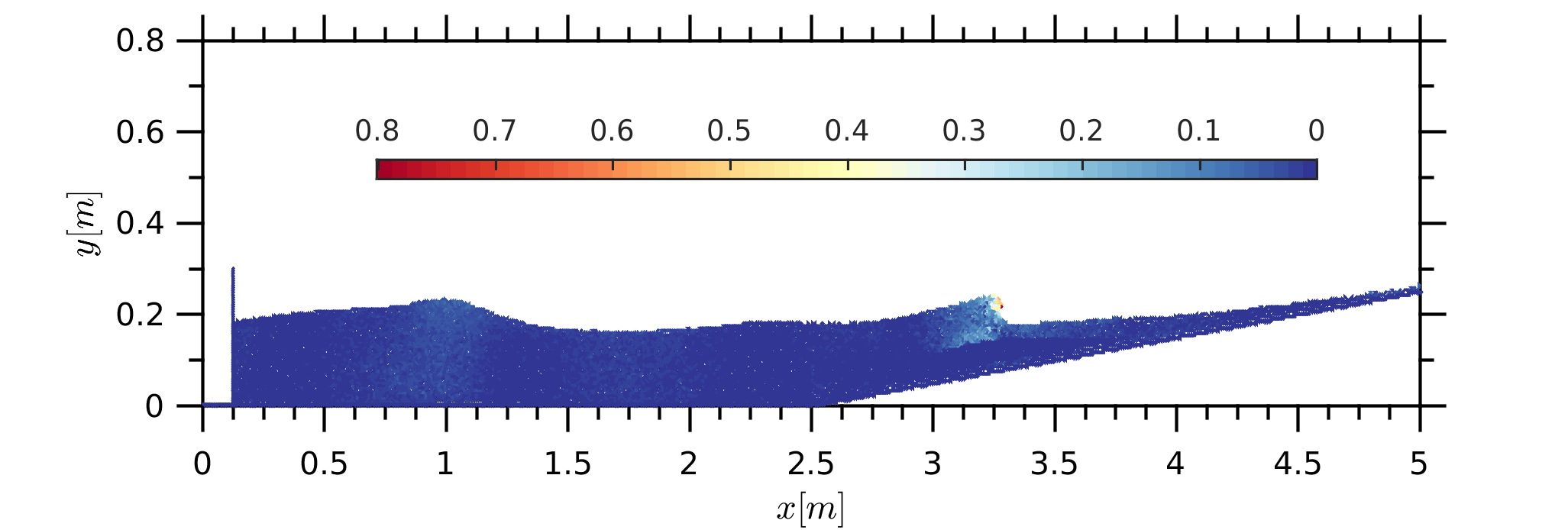}
		\caption[Breaking wave simulation.]{$t=4.65\textup{2}$}
		\label{fig:surf3}
	\end{subfigure}%
	\begin{subfigure}[b]{0.5\textwidth}
		\includegraphics[height=3.0cm,,width=0.85\linewidth]{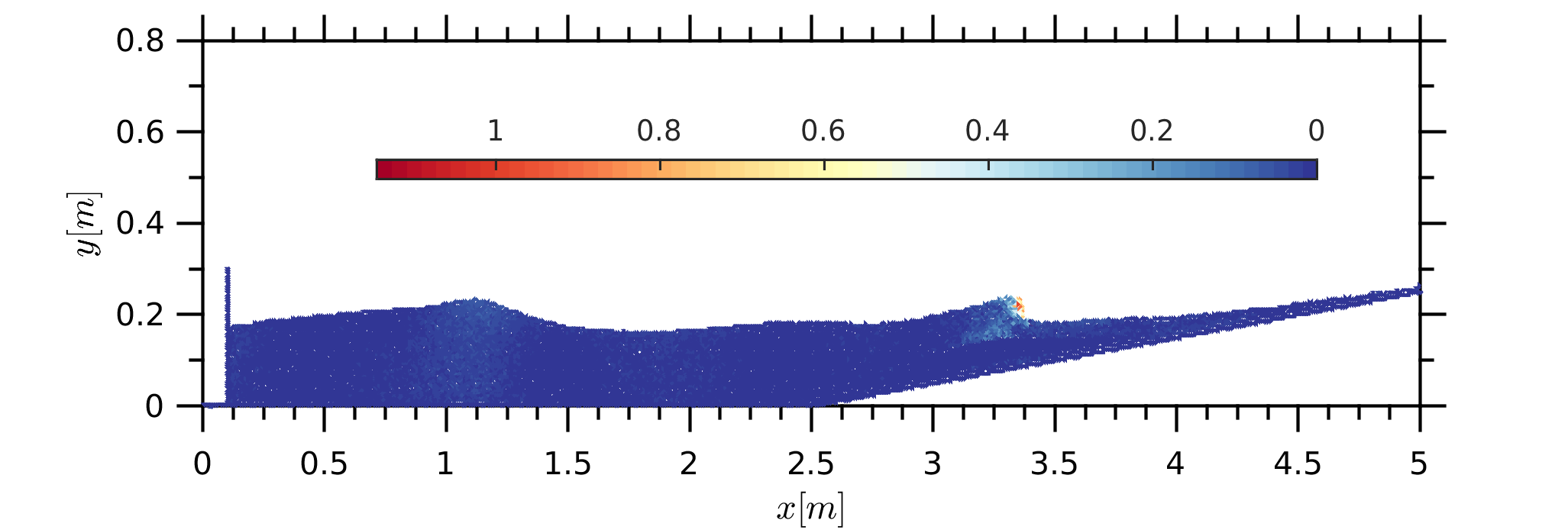}
		\caption[Breaking wave simulation.]{$t=4.75\textup{2}$}
		\label{fig:surf4}
	\end{subfigure}
	
	\begin{subfigure}[b]{0.5\linewidth}
		\includegraphics[width=0.85\textwidth]{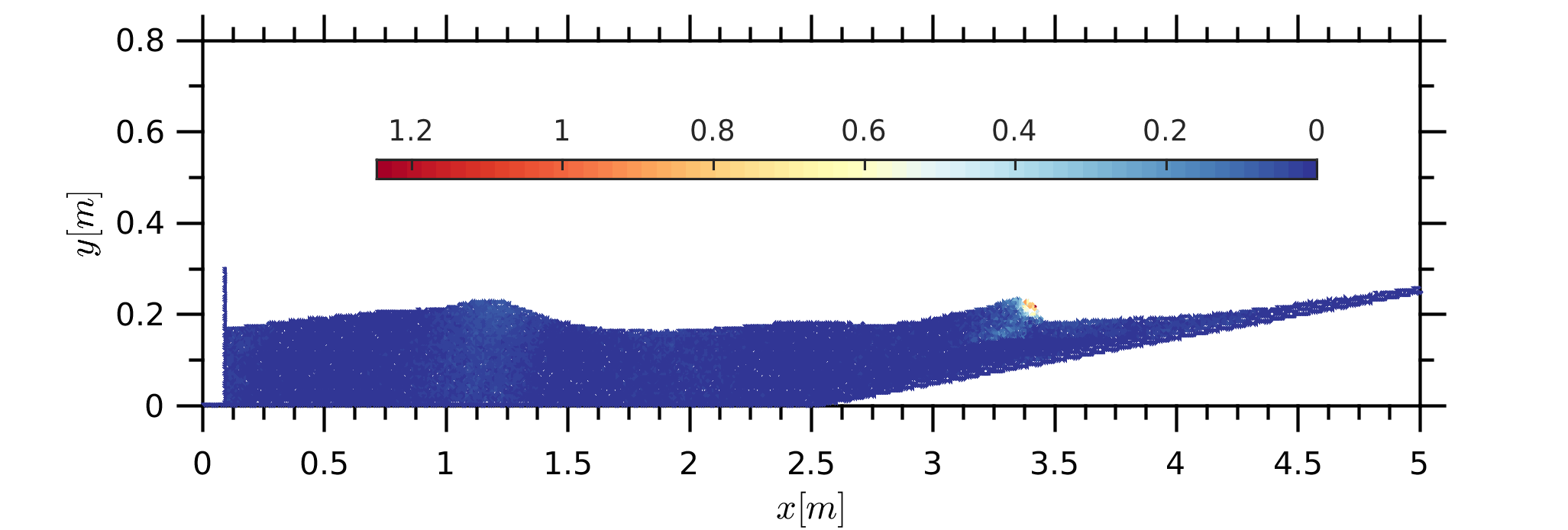}
		\caption[Breaking wave simulation.]{$t=4.80\textup{2}$}
		\label{fig:surf5}
	\end{subfigure}%
	\begin{subfigure}[b]{0.5\linewidth}
		\includegraphics[height=2.5cm,width=0.85\textwidth]{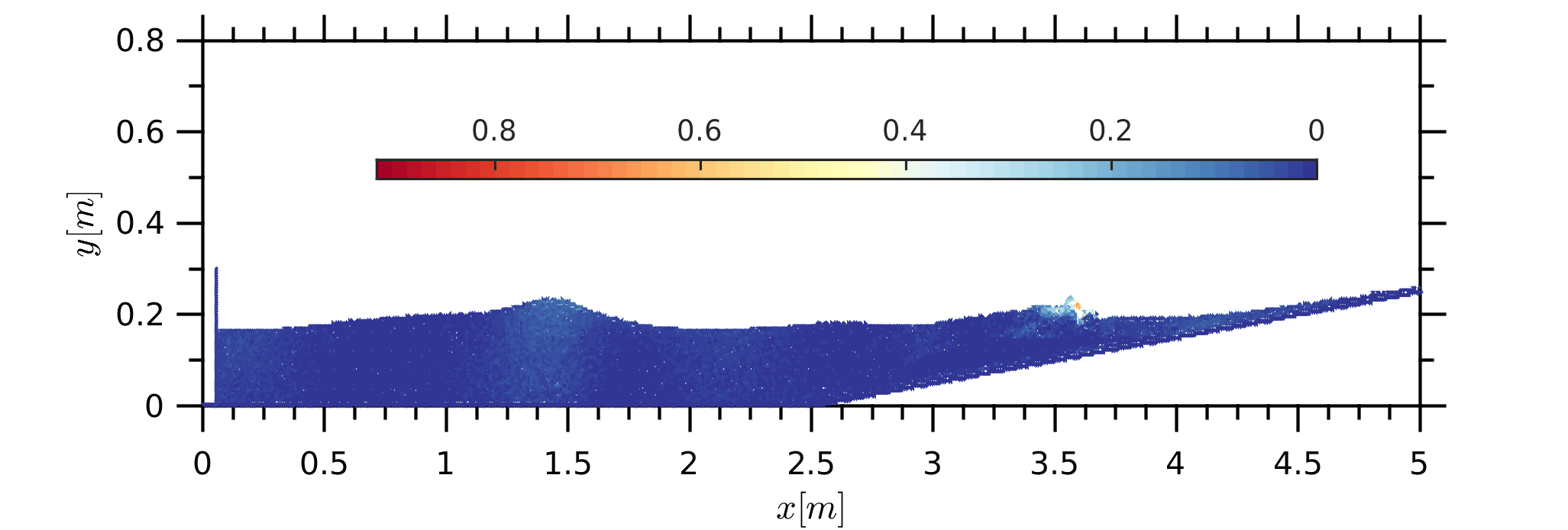}
		\caption[Breaking wave simulation.]{$t=5.0\textup{2}$}
		\label{fig:surf6}
	\end{subfigure}
	
	\caption[Breaking wave over artificial breaker]{Effect of an artificial smooth breaker. Color: kinetic energy density at six time instants for case 1.}
	\label{fig:SmoothbreakWave}
\end{figure}

Figure \ref{fig:SmoothBreaker} shows the breaking wave test case 1 with a smooth breaker inserted in the topography as shown. The effect of the breaker on the wave breaking process is depicted in figure \ref{fig:SmoothbreakWave}. Compared to the case with no breaker in the topography (see figures \ref{fig:TKE}, \ref{fig:TDR} and \ref{fig:viscDissp}) the waves feel the effect of the bottom topography at a much earlier time. Thus, the breaker tends to shift the breaking point away from the shore, as would be expected. The size of the breaker obviously matters. 

\subsection{Mixing process in near-field dam-break flows}
The next validation test case to be considered is the dissipation for a shallow water breaking wave due to a dam break process. This free surface flow problem is a standard benchmark test for numerical methods in CFD. It has significant impact on ecosystems downstream and may cause serious environmental damage due to the associated production of (energetic) breaking waves and flooding \cite{Janosi2004}. The goal of this section is to study the mixing process set up by the collapse of a dam onto a wet bed downstream. The proposed SPH$-i$ model is applied with special attention paid to the mixing process during the onset of a breaking wave. 

\begin{figure}
	\centering
	\includegraphics[width=10cm]{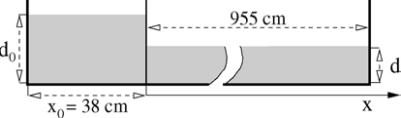}
	\caption{Schematic diagram of the experimental setup in \cite{Janosi2004}}
	\label{fig:gatedDam}
\end{figure}

Janosi \cite{Janosi2004} et.al. (2004) conducted experiments to study the interaction between two fluid bodies in a dam-break process. The experimental setup comprises a long wave flume with a dam upstream and a wet bed downstream as shown in figure \ref{fig:gatedDam}. They also demonstrated that the flow is essentially two-dimensional and hence the effects of the side walls are negligibly small. 

To simulate this problem the above setup would be computationally expensive as the number of fluid particles would be very large. As the maximum recorded time during the actual experiment was $0.6\textup{s}$, it was suggested in \cite{Jian2015} to use a shorter downstream channel of $2.5\textup{m}$ and this would not have any adverse effects on the simulation results. 

The SPH particles were placed on a rectangular grid with initial particle spacing $\Delta r=0.002\textup{m}$ and initial density $\rho_{0}=1000\textup{Kgm}^{-3}$ so that the mass of each fluid particle was $m_{0}=2\textup{Kgm}^{-1}$. The average number of near neighbors was fixed at 91; chosen so as to minimize numerical dissipation attributable to filtering/de-filtering processes. The compact support radius in units of $h$ is then $\xi=\sqrt{N_{n}/4\pi}=2.69$. The smoothing length is thus $h=\xi\Delta x=0.00538$m. The thermal time step dominated the choice of time step and was set at $\Delta t = 2.0\times 10^{-6}\textup{s}$ to guarantee stability of the numerical solution. The pressure and velocity were all initialized to zero at the start of the simulation. 

The gate separating the two water bodies was modeled by a set of boundary particles using the viscously damped boundary force model. The stiffness and damping coefficients were respectively set as $k^{s}=0.000589\textup{Nm}^{-1}$ and $k^{d}=0.002\textup{Nsm}^{-1}$ for each particle. The gate was moved at a constant speed to mimic the removal procedure employed in the physical experiment. Furthermore, the fluid bodies were distinguished by a assigning a flag of $-1$ to the downstream particles and $+1$ for the upstream particles. This enables a clear detection of the mixing interface which manifests as the separation between the flagged particles. 

\begin{figure}
	\centering
	\begin{subfigure}[b]{.4\textwidth}
		\includegraphics[width=\textwidth,height=18cm]{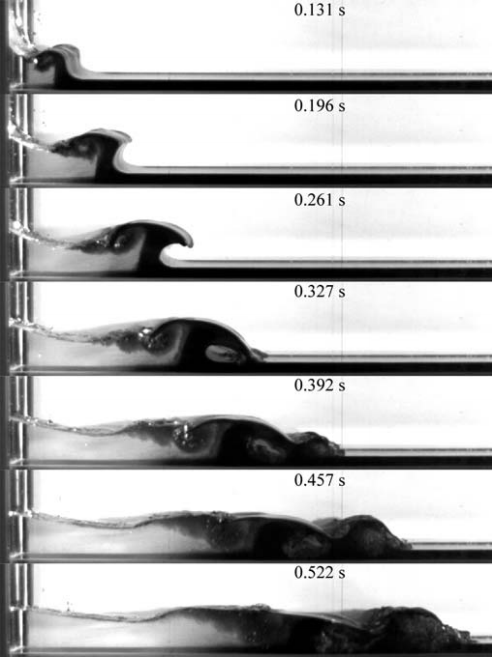}
	\end{subfigure}\qquad
	\begin{subfigure}[b]{.4\textwidth}
		\includegraphics[width=\textwidth,height=2cm]{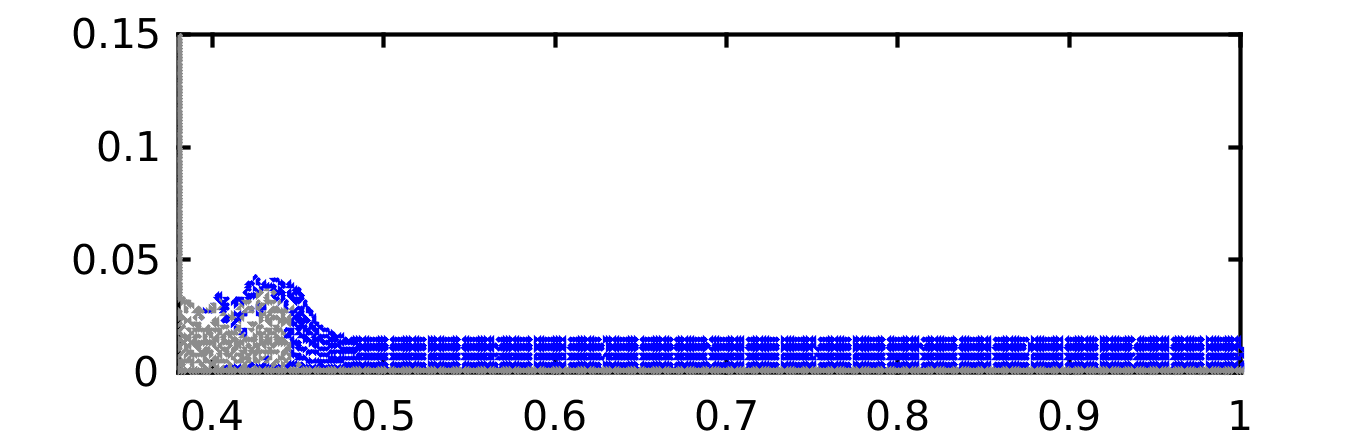}
		\vspace{0.1ex}
		
		\includegraphics[width=\textwidth,height=2cm]{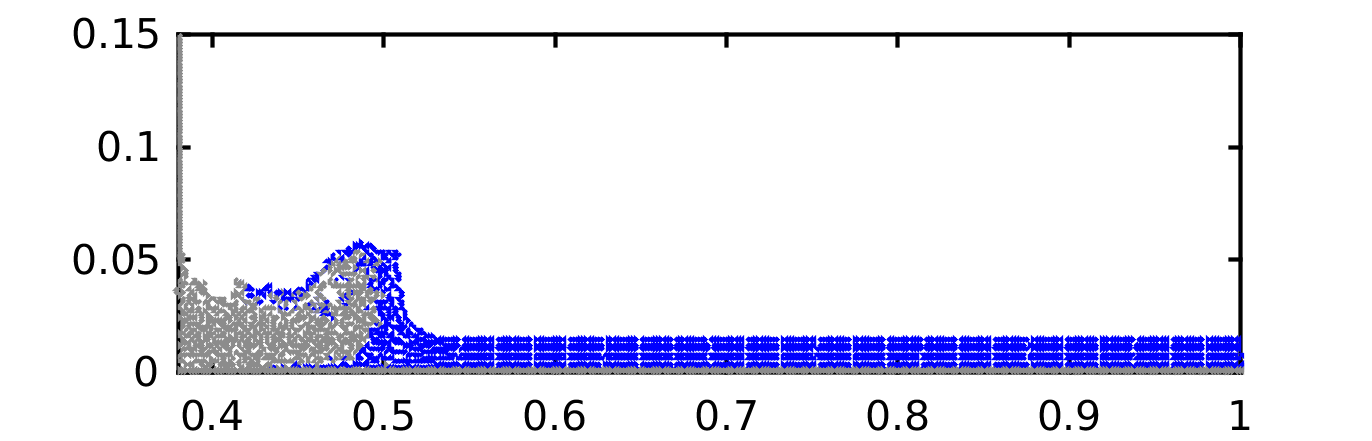}
		\vspace{0.1ex}
		
		\includegraphics[width=\textwidth,height=2cm]{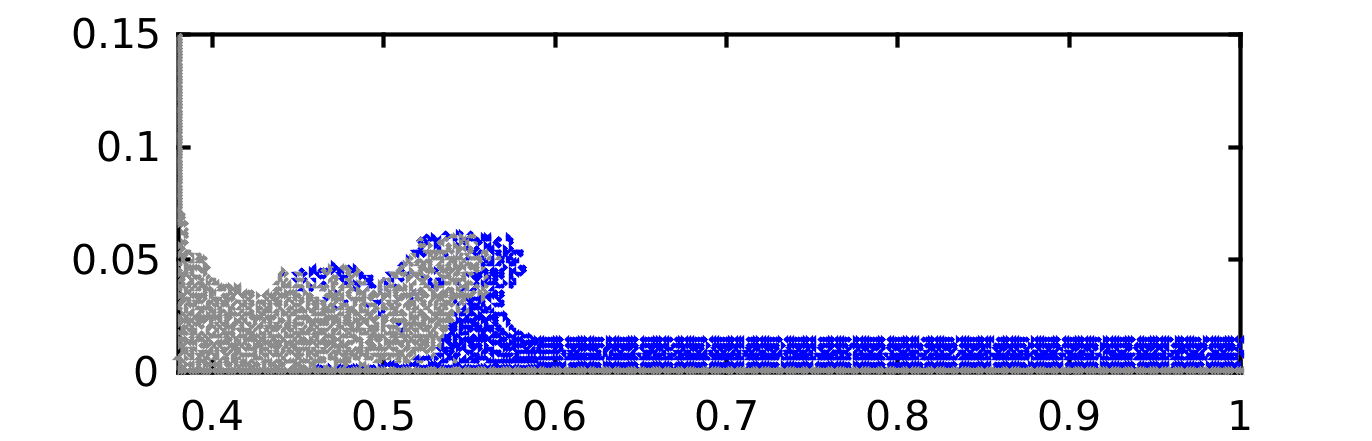}
		
		\vspace{0.1ex}
		
		\includegraphics[width=\textwidth,height=2cm]{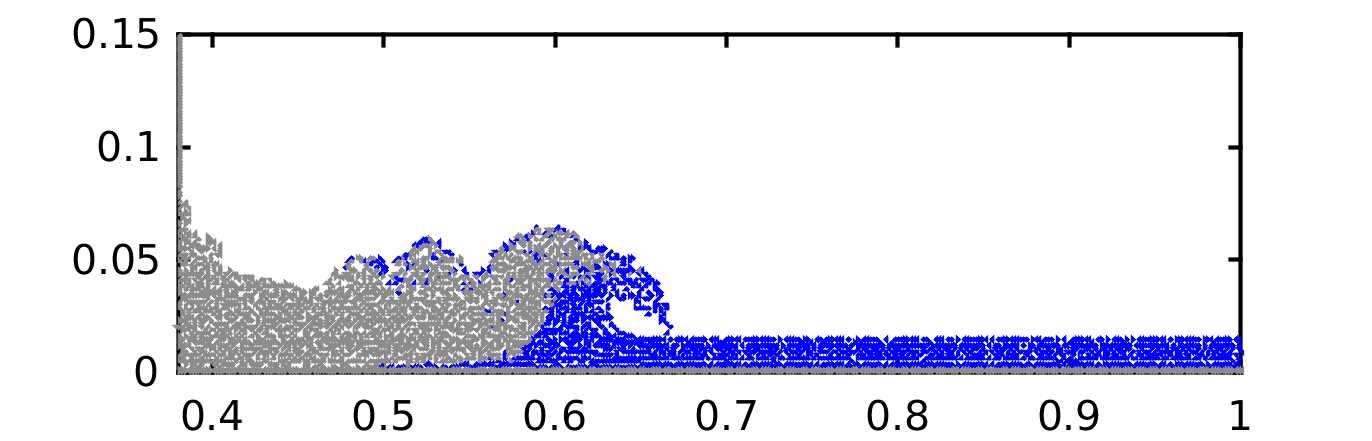}
		\vspace{0.1ex}
		
		\includegraphics[width=\textwidth,height=2cm]{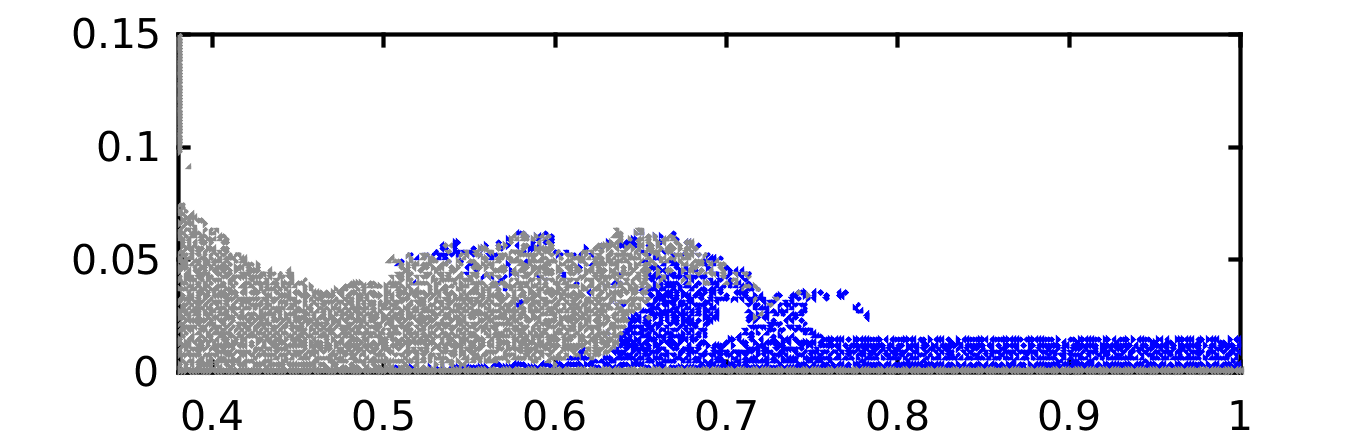}
		\vspace{0.1ex}
		
		\includegraphics[width=\textwidth,height=2cm]{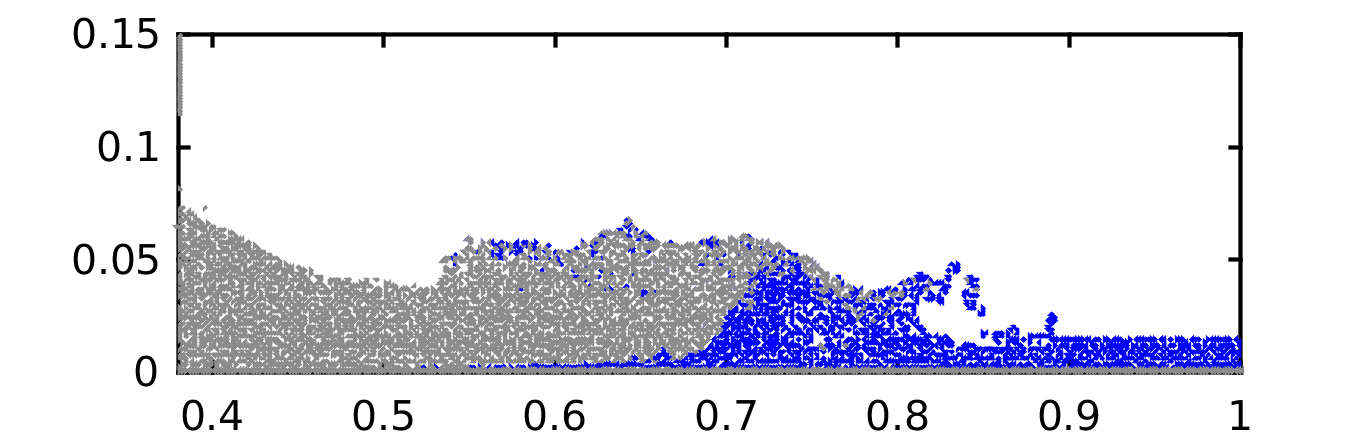}
		\vspace{0.1ex}
		
		\includegraphics[width=\textwidth,height=2cm]{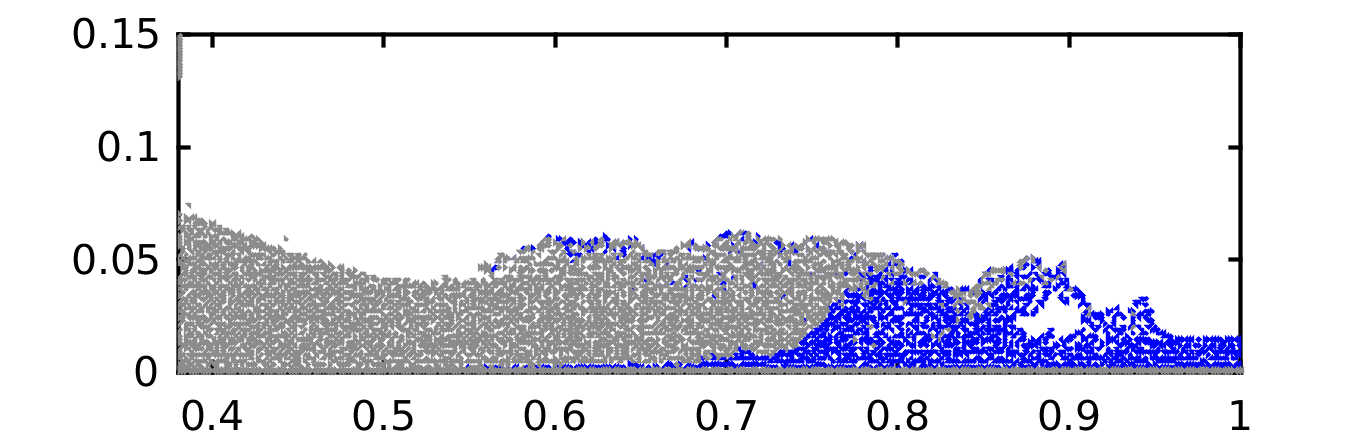}
	\end{subfigure}
	\caption[Mixing in shallow ambient layer]{Comparison of laboratory photographs(left, Janosi et.al. \cite{Janosi2004}) with simulated mixing patterns obtained using the proposed SPH$-i$ model at times $t=0.1962\textup{s},0.2616\textup{s},0.3270\textup{s}$ and $0.3920\textup{s}$.}
	\label{fig:mixing1}
\end{figure}

In the first experiment the wet bed height is set at $d=0.015\textup{m}$. Figure \ref{fig:mixing1} shows that the proposed SPH$-i$ model is able to successfully reproduce the mixing patterns observed in the breaking wave propagation as the gate was gradually moved up. As the gate starts to gradually move up, due to the higher hydrostatic pressure at the bottom, the upstream fluid particles in the bottom are ejected towards the downstream. On the other hand the downstream fluid is still at rest, hence blocking the approaching upstream fluid. Collision of the moving front with the resting fluid in the ambient layer creates an upthrust in the form of a breaking wave, with free surface breaking occurring in both the forward and reverse directions. The formation and evolution of the plunging jet is captured with reasonable accuracy when compared with snapshots from the experiment. At $t=0.1962\textup{s}$ the mixing interface is relatively vertical but it slowly tilts towards the downstream direction as the breaking wave propagates downstream. The formation of the plunging jet was first reported by Stansby et.al. \cite{Stansby1998}.  

\begin{figure}
	\centering
	\begin{subfigure}[b]{.4\textwidth}
		\includegraphics[width=\textwidth,height=13cm]{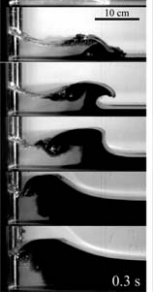}
	\end{subfigure}\qquad
	\begin{subfigure}[b]{.4\textwidth}
		\includegraphics[width=\textwidth,height=2cm]{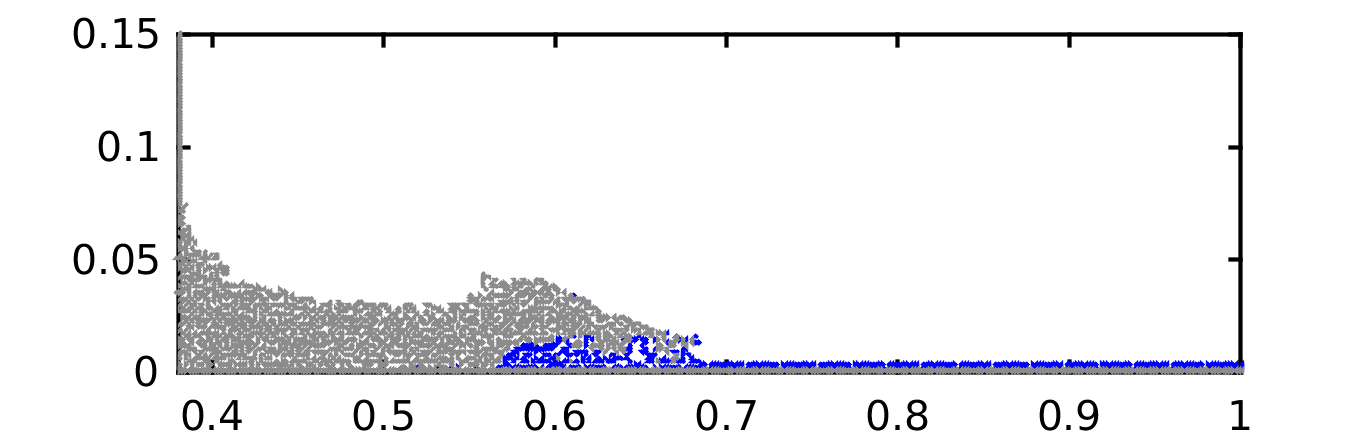}
		\vspace{0.1ex}
		
		\includegraphics[width=\textwidth,height=2cm]{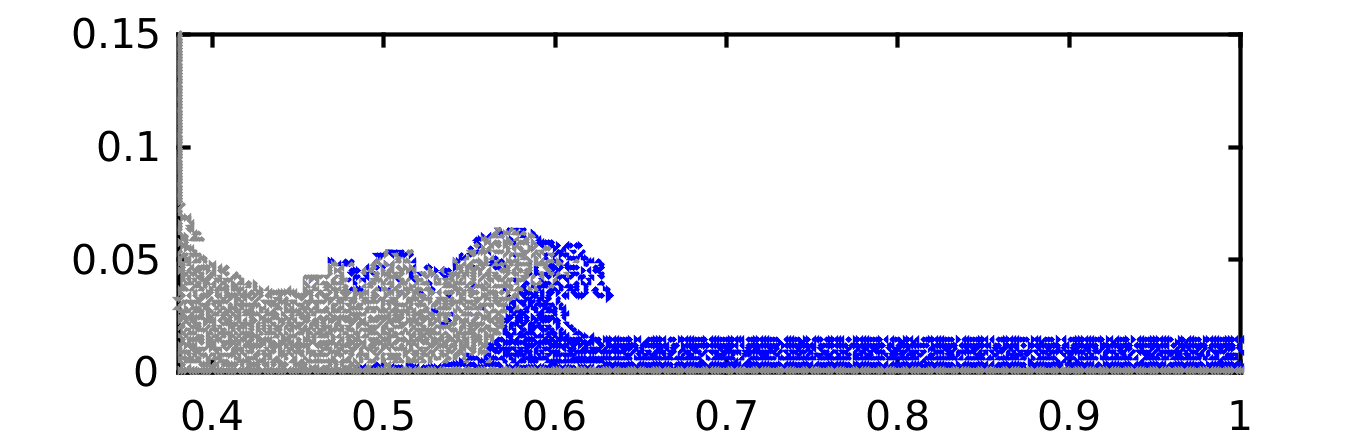}
		\vspace{0.1ex}
		
		\includegraphics[width=\textwidth,height=2cm]{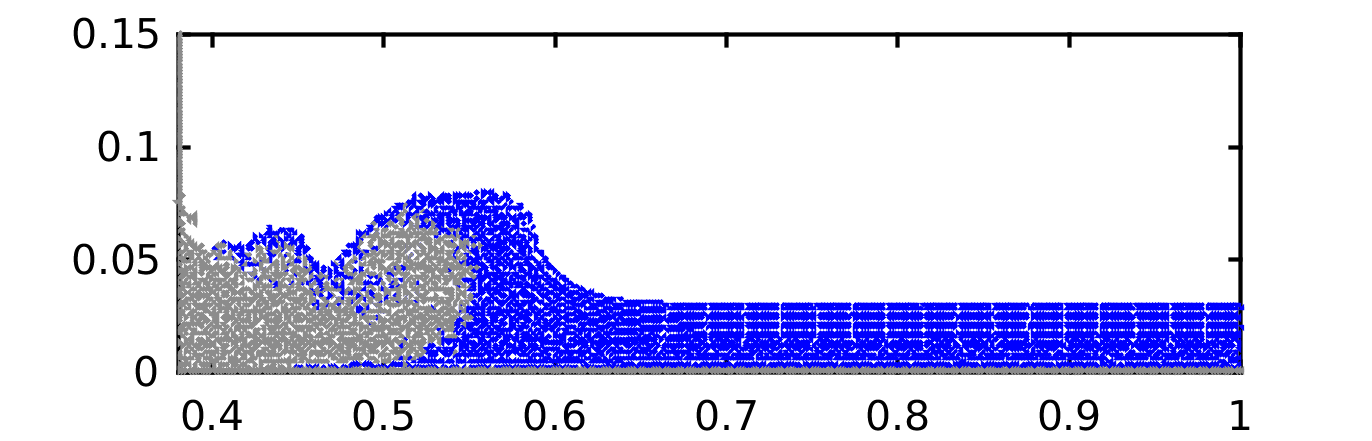}
		
		\vspace{0.1ex}
		
		\includegraphics[width=\textwidth,height=2cm]{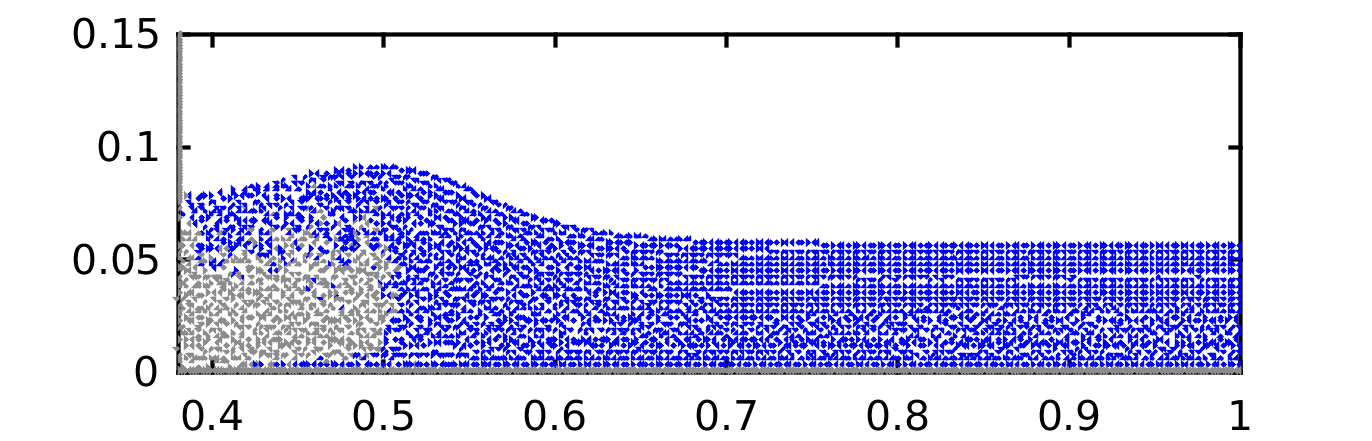}
		\vspace{0.1ex}
		
		\includegraphics[width=\textwidth,height=2cm]{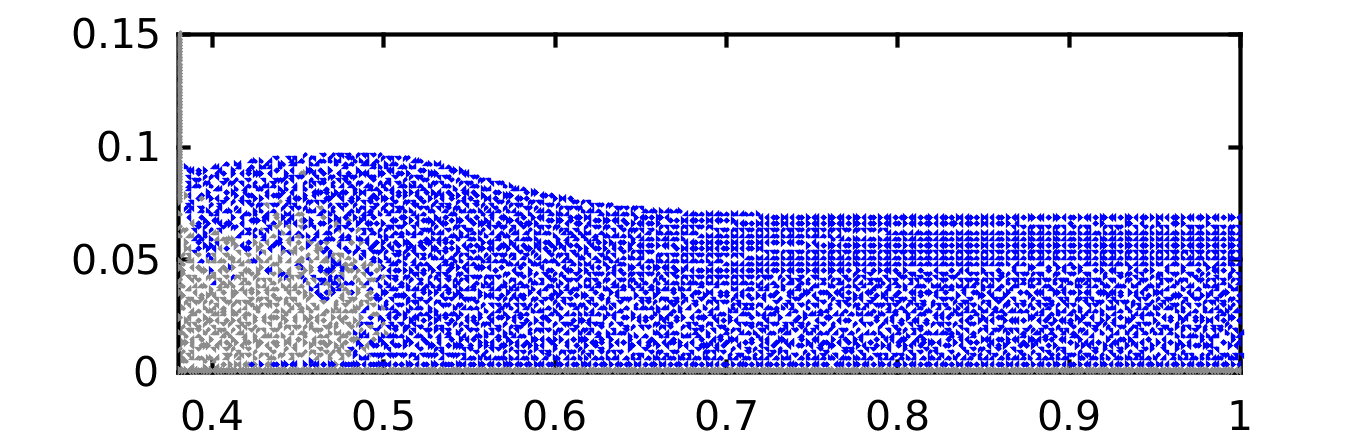}
	\end{subfigure}
	\caption[Effect of ambient layer on mixing at $t=0.3\textup{s}$]{Mixing patterns generated by the release of a dam-break front into a wet bed of increasing depth $d$ (from top to bottom) $0.005\textup{m}$, $0.015\textup{m}$, $0.058\textup{m}$ and $0.070\textup{m}$ at time $t=0.3\textup{s}$(left panel, Janosi et.al. \cite{Janosi2004}) with simulations performed using the proposed SPH$-i$ model (right panel).}
	\label{fig:mixing2a}
\end{figure}

\paragraph{effect of ambient layer:} the presence of a shallow ambient layer of fluid in the downstream channel has an important influence on the flow behavior, even for a very small ambient depth $d$. The next experiment investigates the effect of the wet bed downstream on the flow properties with five ambient depths of  $d=0.005\textup{m}$, $0.015\textup{m}$, $0.058\textup{m}$ and $0.070\textup{m}$. Figure \ref{fig:mixing2a} shows a qualitative comparison of the experimental and numerical predictions at time $t=0.3\textup{s}$. 

For very small ambient depths, the potential energy of the wet bed fluid is much less than that of the fluid upstream. Therefore, as the moving front of the upstream fluid collides with ambient fluid upon release of the gate, due to high kinetic energy, a strong upthrust is recorded. This effect leads to the quick formation of a propagating bore for depths $d=0.005\textup{m}$, $0.015\textup{m}$. For these three depths, it can be further observed that the plunging wave column consists mainly of the upstream fluid, with a comparatively smaller ambient fluid layer. Furthermore, due to the energetic collision between the moving front and the stationary ambient fluid, the mixing interface tilts towards the downstream direction. 
In contrast, for larger ambient depths $d=0.058\textup{m}$ and $0.070\textup{m}$, due to the small difference in potential energy between the upstream and ambient fluid downstream, the moving front collides with the ambient fluid with low kinetic energy. For this case, it takes longer for the waves to fully develop and the free surface breaking phenomena may not occur. Due to the less energetic collision, the mixing interface for larger depths largely remains vertical as the waveform gains height. Clearly, the SPH$-i$ model shows a satisfactory agreement with the experimental results at time $t=0.3\textup{s}$.     

\begin{figure}
	\centering
	\begin{subfigure}[b]{.4\textwidth}
		\includegraphics[width=\textwidth,height=13cm]{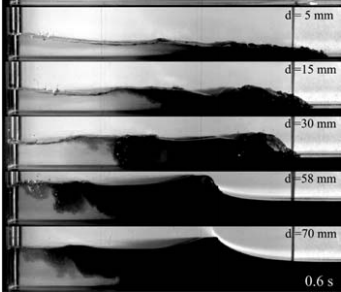}
	\end{subfigure}\qquad
	\begin{subfigure}[b]{.4\textwidth}
		\includegraphics[width=\textwidth,height=2cm]{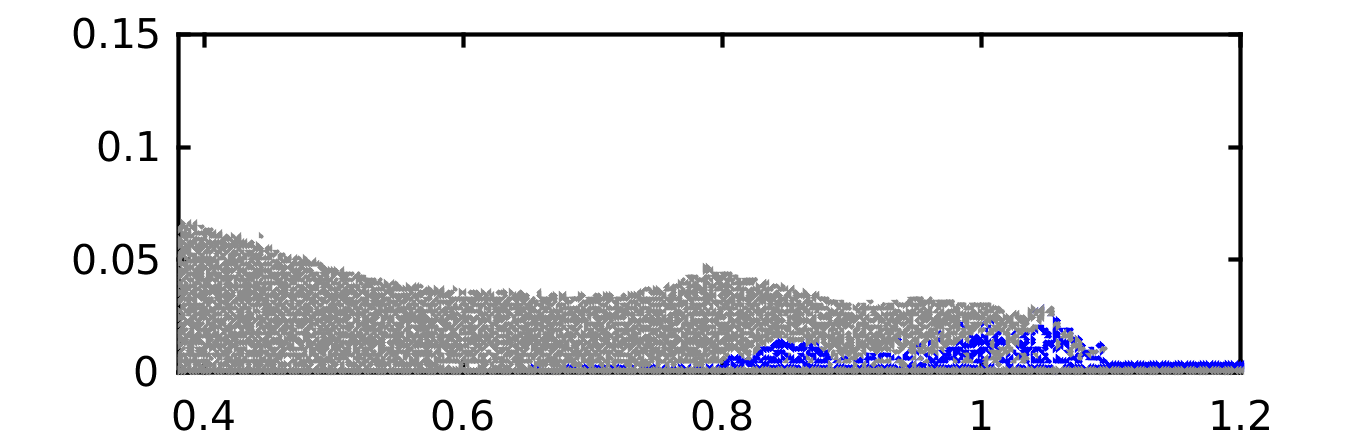}
		\vspace{0.1ex}
		
		\includegraphics[width=\textwidth,height=2cm]{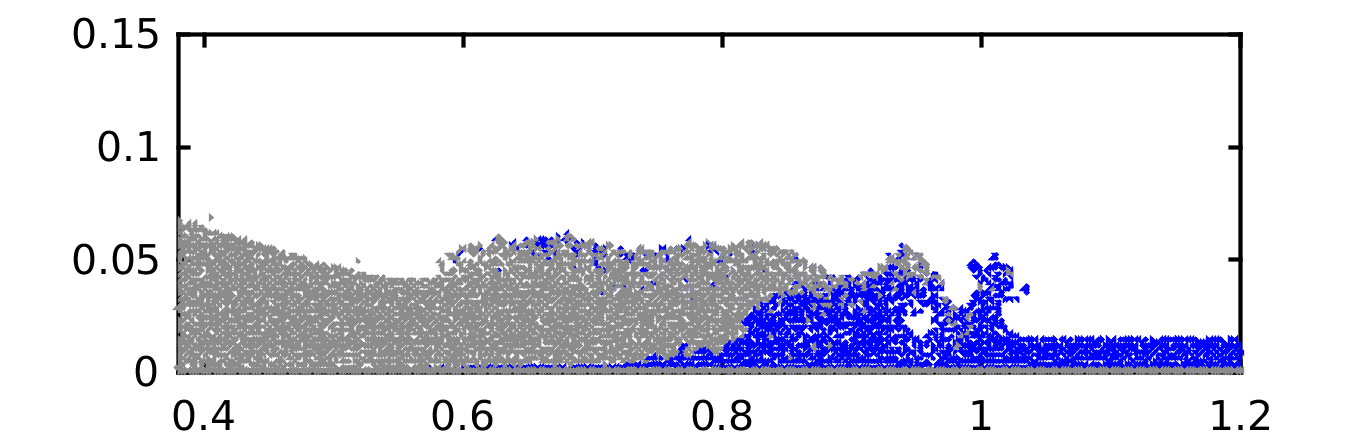}
		\vspace{0.1ex}
		
		\includegraphics[width=\textwidth,height=2cm]{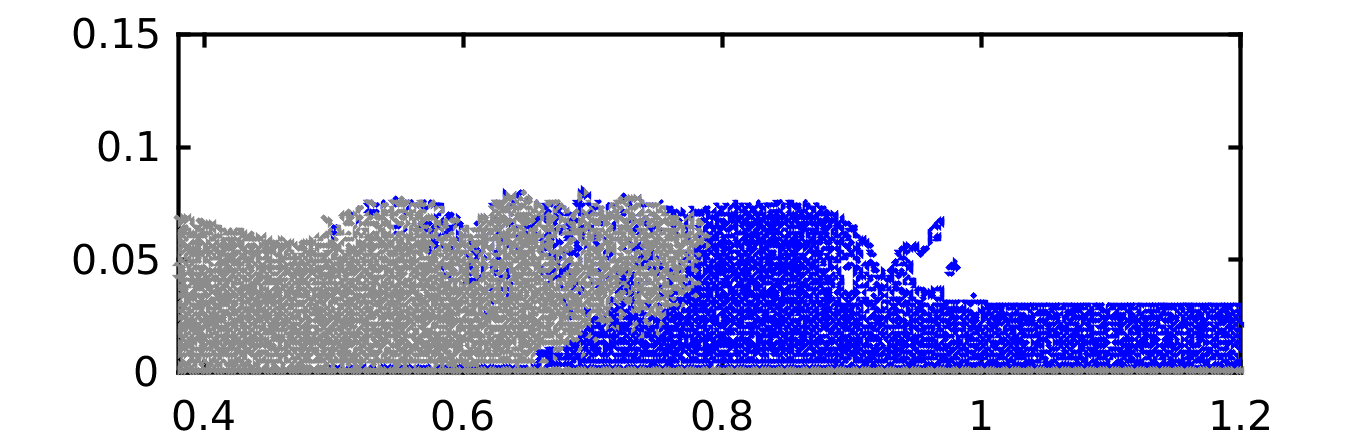}
		
		\vspace{0.1ex}
		
		\includegraphics[width=\textwidth,height=2cm]{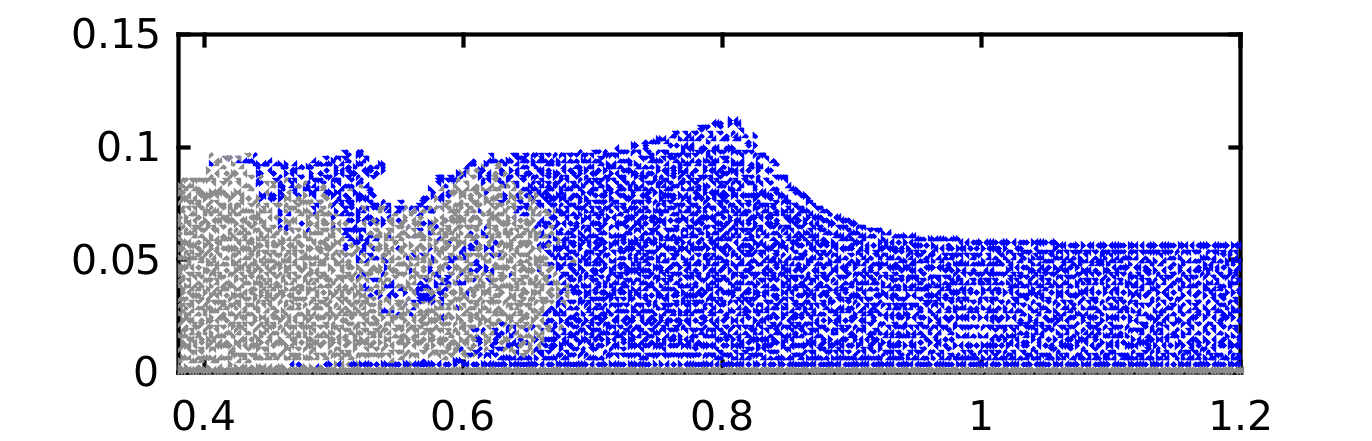}
		\vspace{0.1ex}
		
		\includegraphics[width=\textwidth,height=2cm]{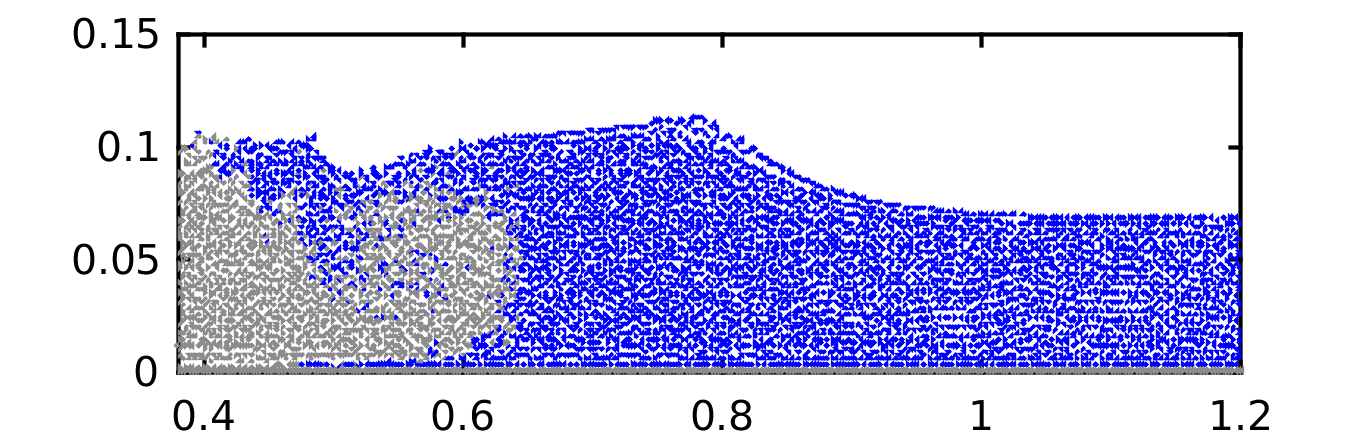}
	\end{subfigure}
	\caption[Mixing process ambient layer downstream]{Mixing patterns generated by the release of a dam-break front into a wet bed of increasing depth $d$ (from top to bottom) $0.005\textup{m}$, $0.015\textup{m}$, $0.058\textup{m}$ and $0.070\textup{m}$ at time $t=0.6\textup{s}$(left panel, Janosi et.al. \cite{Janosi2004}) with simulations performed using the proposed SPH$-i$ model (right panel).}
	\label{fig:mixing2b}
\end{figure}

As the generated turbulent bore propagates further downstream, breaking phenomena occurs, depending on the wet bed depth. This phenomena is captured in figure \ref{fig:mixing2b} at $t=0.6\textup{s}$. Again, for low ambient depths $d=0.005\textup{m}$, $0.015\textup{m}$ the mixing keeps tilting towards the downstream direction whereas for the larger depths $d=0.058\textup{m}$ and $0.070\textup{m}$ the mixing interface remains largely vertical as in the early mixing stages. It can also be observed that the two fluid bodies are quite well mixed by the time.  

Besides experimental studies by Janosi et.al. \cite{Janosi2004}, the mixing process in dam break flows with upstream and downstream fluid bodies has recently been studied using various corrected versions of SPH and MPS models \cite{Khayyer2010,Jian2015}.

\section{Conclusion}
In this paper a version of smoothed particle hydrodynamics that implicitly models turbulence has been subjected to preliminary tests to determine its capability to simulate turbulent, free surface flows in general. The algorithms is comprised of two main steps; a smoothing or filtering process and an un-smoothing or de-filtering process. Unlike standard SPH which evolves the smoothed field and turbulence must be explicitly modeled, the proposed method solves for the disordered field and is thus a coarse-grained direct numerical simulation approach.

Four benchmark tests including; hydrostatic pressure, dam-break on dry bed, dam-break on wet bed and periodic breaking waves on a gentle slope were performed. In all the tests the proposed SPH-$i$ model captures the flow patterns relatively well when compared with experimental results. The pressure field obtained was generally smooth and stable and this may be attributed to a new equation for pressure rather than the simple equation of state commonly used in conventional SPH.

With regards to future research, the SPH-$i$ model will be tested to ascertain how well it can reproduce the energy cascade in turbulence. The numerical examples for such study will include both freely decaying and driven turbulence in a periodic box.

\section*{acknowledgements}
The first author would like to acknowledge Pavel Puchenkov of the OIST Graduate University scientific computation  section for his support with visualization of SPH data.
\newpage



\bibliography{sample}



\end{document}